\documentclass{article}

\usepackage{arxiv}

\usepackage[utf8]{inputenc} 
\usepackage[T1]{fontenc}    
\usepackage[hidelinks]{hyperref}       
\usepackage{url}            
\usepackage{booktabs}       
\usepackage{amsmath}
\usepackage{amsfonts}       
\usepackage{nicefrac}       
\usepackage{microtype}      
\usepackage{lipsum}
\usepackage{graphicx}
\graphicspath{ {./images/} }
\usepackage{subfig}
\usepackage{multirow}
\usepackage{natbib}
\renewcommand\footnotemark{}

\title{Quantifying the Consistency and Characterizing the Confidence of Coronal Holes Detected by Active Contours without Edges (ACWE)}

\author{Jeremy A.~Grajeda, Laura E.~Boucheron \\
  Klipsch School of Electrical and Computer Engineering\\
  New Mexico State University\\
  Las Cruces, NM 88001, USA \\
  \texttt{\{jgra,lboucher\}@nmsu.edu} \\
   \And
  Michael S.~Kirk \\
  Heliophysics Space Division, Goddard Space Flight Center \\
  National Aeronautics and Space Administration \\
  Greenbelt, MD 20771, USA\\
  \texttt{michael.s.kirk@nasa.gov} \\
  \And
  Andrew Leisner \\
  Department of Physics and Astronomy \\
  George Mason University \\
  Fairfax, VA 22030, USA \\
  \texttt{aleisner@gmu.edu} \\
  \And
  C.~Nick Arge \\
  Heliophysics Space Division, Goddard Space Flight Center \\
  National Aeronautics and Space Administration \\
  Greenbelt, MD 20771, USA\\
  \texttt{charles.n.arge@nasa.gov} \\
}

\begin{document}
\maketitle\let\thefootnote\relax\footnotetext{This version of the article has been accepted for publication, after peer review, but is not the Version of Record and does not reflect post-acceptance improvements, or any corrections.  The Version of Record is available online at \url{https://doi.org/10.1007/s11207-023-02228-0}}
\begin{abstract}
Coronal Holes (CHs) are regions of open magnetic field lines, resulting in high speed solar wind. Accurate detection of CHs is vital for space weather prediction. This paper presents an intramethod ensemble for coronal hole detection based on the Active Contours Without Edges (ACWE) segmentation algorithm. The purpose of this ensemble is to develop a confidence map that defines, for all on disk regions of a Solar extreme ultraviolet (EUV) image, the likelihood that each region belongs to a CH based on that region's proximity to, and homogeneity with, the core of identified CH regions. By relying on region homogeneity, and not intensity (which can vary due to various factors including line of sight changes and stray light from nearby bright regions), to define the final confidence of any given region, this ensemble is able to provide robust, consistent delineations of the CH regions.  Using the metrics of global consistency error (GCE), local consistency error (LCE), intersection over union (IOU), and the structural similarity index measure (SSIM), the method is shown to be robust to different spatial resolutions maintaining a median IOU $>0.75$ and minimum SSIM $>0.93$ even when the segmentation process was performed on an EUV image decimated from $4096\times4096$ pixels down to $512\times512$ pixels. Furthermore, using the same metrics, the method is shown to be robust across short timescales, producing segmentation with a mean IOU of 0.826 from EUV images taken at a 1 hour cadence, and showing a smooth decay in similarity across all metrics as a function of time, indicating self-consistent segmentations even when corrections for exposure time have not been applied to the data.  Finally, the accuracy of the segmentations and confidence maps are validated by considering the skewness (i.e., unipolarity) of the underlying magnetic field. 
\end{abstract}


\section{Introduction}
Within the Sun's corona, there are regions of lower temperature, lower density plasma called coronal holes (CHs). These regions have open magnetic field lines, emerging from the photosphere \citep{Altschuler1972,Munro1972,Wang1996}, which results in high-speed solar winds \citep{Wang1990,Wang1996,Antonucci2004,McComas2007}. This relationship has been leveraged to improve our understanding of solar wind behavior. In particular, \cite{Wang1990} demonstrated that CH area provides key information for understanding solar wind speeds at 1~{AU}. In addition to this, further studies into the relationship between CHs and solar wind have revealed additional constraints which have improved space weather prediction \citep{Arge2003,Arge2004}. For this reason, understanding CH behavior is important for space weather prediction, creating a need for accurate and robust segmentation methods.

Current methods of CH detection often rely on extreme ultraviolet (EUV) and x-ray imagery, such as the data produced by the Atmospheric Imaging Assembly (AIA) aboard the National Aeronautics and Space Administration’s (NASA's) Solar Dynamics Observatory (SDO) \citep{lemen2012}. At these wavelengths CHs appear as dark regions \citep{Munro1972}. For this reason segmentation methods often rely on an intensity threshold to find CH regions. A strict intensity threshold, however, may not adequately segment CHs as the intensity of CH regions within solar EUV observations can vary as a result of both intrinsic and extrinsic properties including limb brightening, stray light from nearby regions, and instrument noise \citep{verbeeck2014,Caplan2016}. To mitigate this issue, various solutions have been implemented. \cite{Krista2009}, for example, subdivide the 193~{\AA} observations and determine a threshold for each sub-image. \cite{Lowder2014} also rely on a region-based threshold. \cite{Hamada2018} define a coronal hole based on the majority agreement of segmentations of synoptic maps generated from a region-based threshold applied to 304 \AA, 171 \AA, and 195 \AA~observations, followed by morphological processes to refine the segmentations. Alternatively, \cite{Caplan2016} use two thresholds, representing the initial seed and maximum possible CH area, then rely on a region growing algorithm to refine the initial seed.

In addition to intensity-based thresholds, alternate methods of CH segmentation have also been developed. \cite{verbeeck2014}, for example, rely on fuzzy c-means clustering to identify CHs, active regions, and quiet Sun. \cite{Jarolim2021} rely on a supervised machine learning method which uses all AIA EUV observations and Helioseismic and Magnetic Imager (HMI) magnetograms \citep{scherrer2012}. This method was trained using a hand-curated dataset derived from a modified version of the algorithm of \cite{verbeeck2014} and generates segmentations or probabilistic maps at a resolution of $512\times512$ pixels. To our knowledge the network developed by \cite{Jarolim2021} is the only other automated segmentation method that generates a probabilistic (or confidence) map of CH locations and boundaries, which may indicate that our version of ACWE is the first classical approach to CH confidence map generation.

In a study of nine automated CH segmentation methods, \cite{Reiss2021} note that segmentations generated by automated methods can vary significantly, with the difference in area for one CH varying by a factor of 4.5 between methods. These differences may be due, in part, to how each algorithm accounts for the fact that the intensity of CH regions within solar EUV observations can vary.

\cite{boucheron2016segmentation} introduced the use of Active Contours Without Edges (ACWE) as a method for CH detection. Like the aforementioned threshold-based methods, \cite{boucheron2016segmentation} identify candidate CHs using an intensity threshold. These initial candidates, however, are then refined in order to maximize the homogeneity of the CH regions. By using region homogeneity as the criterion for defining CH boundaries, ACWE was able to overcome the effects of limb brightening and stray light from nearby regions and include portions of the identified CHs that were excluded by the initial threshold, generating more robust CH segmentations. The work in this paper expands on \cite{boucheron2016segmentation} with two investigations applied to a larger dataset. The first characterizes the consistency of the ACWE algorithm across spatial resolutions, intensity resolutions, and short timescales. The second investigation further accounts for uncertainty in detection of CH regions by developing a “confidence map” via an intramethod ensemble that correlates region homogeneity (compared to the core of a CH region) with the likelihood that any particular region is part of a CH.

The remainder of the paper is organized as follows. Section \ref{sec:background} provides an overview of the ACWE algorithm and the adaptation of ACWE to CH detection. Section \ref{sec:data} describes the dataset and provides an overview of the metrics used for evaluation of the segmentations produced by ACWE. Section \ref{sec:robustness} characterizes the invariability of CH segmentations generated by ACWE as a function of spatial resolution, comparing the effects of performing ACWE on images decimated by $8\times$ in each dimension, which is the default from \cite{boucheron2016segmentation}, to the effects of performing ACWE at higher resolutions.  Section \ref{sec:robustness} also characterizes the invariability of CH segmentations generated by ACWE across short time scales, where CH evolution is expected to be minimal, to determine the consistency in identifying and segmenting CH regions. The development of a confidence map via an intramethod ensemble is outlined in Section \ref{sec:conMap}. This method is then evaluated in Section \ref{sec:val} by comparing the ensemble to underlying magnetic field as expressed in HMI magnetograms. Section \ref{sec:conclustion} provides a conclusion and discusses future work.  Appendix~\ref{sec:intensity} contains an additional study of the performance of ACWE as a function of different intensity resolutions and dynamic ranges, comparing the resulting segmentations to the corresponding default segmentation generated from the EUV at the original dynamic range. These results are included as they indicate that preserving the dynamic range between on-disk features is crucial for accurate CH detection.

\section{Active Contours Without Edges}
\label{sec:background}

\subsection{General Implementation}
Developed by \cite{chan2001}, Active Contours without Edges (ACWE) uses one or more enclosed shapes to separate an image into the foreground (or region enclosed within the shapes) and background (the remainder of the image). Collectively these shapes are defined by the contour $C$ which separates the foreground and background, where $C_i$ denotes the foreground and $C_o$ denotes the background. In order to ensure that the segmentation represents a meaningful separation of the input image into foreground and background, the contour is manipulated in an iterative fashion according to a series of ``forces'' derived from the characteristics of the image. In the case of ACWE, the goal of these forces is to ensure that both the foreground and background have a relatively narrow collection of intensities within them, creating visually homogeneous regions. With each iteration every pixel along the boundary of $C$ is evaluated to determine if it is more similar to the pixels within the foreground or background, and the contour is redrawn accordingly. This process is achieved through the definition of an energy functional which summarizes the image forces and any additional constraints applied to $C$.

The ACWE energy functional is defined as
\begin{equation}
  \begin{aligned}
    F(m_i,m_o,C)=\mu \ell(C)&+\lambda_i\int_{C_i}|I(x,y)-m_i|^2dxdy\\
    &+\lambda_o\int_{C_o}|I(x,y)-m_o|^2dxdy,
  \end{aligned}
 \label{eq:acwe}
\end{equation}
and can be separated into three primary elements representing the ``forces'' acting on $C$ in order to create visually homogeneous regions. The first term, $\mu \ell(C)$, seeks to minimize the length of the contour boundary $\ell(C)$; it is weighted by user-defined weight $\mu$. The term $\lambda_i\int_{C_i}|I(x,y)-m_i|^2dxdy$ seeks to maximize the homogeneity of the foreground by comparing the region of the input image $I$ inside of $C$ to the mean intensity of that region, $m_i$; it is weighted by user-defined weight $\lambda_i$. The final term, $\lambda_o\int_{C_o}|I(x,y)-m_o|^2dxdy$, seeks to maximize the homogeneity of the background by comparing the region of $I$ outside of $C$ to the mean intensity of the background, $m_o$; it is weighted by user-defined weight $\lambda_o$. By manipulating weights $\mu$, $\lambda_i$, and $\lambda_o$, the user can prioritize a shorter (e.g., less fractal) contour boundary, a foreground with a narrow or homogeneous distribution of intensities, or a background with a narrow or homogeneous distribution of intensities.

\subsection{Segmenting Coronal Holes via ACWE}
\label{sec:ACWEevolve}
In adapting ACWE for CH segmentation, \cite{boucheron2016segmentation} developed a seeding algorithm for defining the initial contour, developed methods for constraining ACWE to on-disk areas, defined a set of stopping criteria to account for finite precision due to input image resolution, and identified an optimal range of parameters for CH segmentation. The segmentation method outlined by \cite{boucheron2016segmentation} is preceded by the decimation of the spatial resolution of the EUV images by $8\times$, creating a copy of the EUV image that is $512\times512$ pixels, and the correction for limb brightening outlined in \cite{verbeeck2014}. To create the initial contour, \cite{boucheron2016segmentation} utilize a circular mask on the decimated image to identify and extract a 100-bin intensity histogram of the on-disk region of the image. The mean intensity of the quiet Sun (QS) (low-activity) region, $m_{QS}$, is estimated from the histogram by calculating the mean intensity of the bin with the most pixels. Since CHs are expected to be darker than QS, the initial seed is defined as all pixels with an intensity $\leq\alpha\times m_{QS}$ where $\alpha$ is a user-defined parameter $<1$.

When performing ACWE \cite{boucheron2016segmentation} constrain evolution to on-disk areas by performing ACWE on a copy of the decimated EUV image where all off-disk areas are set to the mean intensity of the non-CH region. ACWE is then performed with $\mu=0$, which causes ACWE to ignore the length constraint, and the remaining parameters, $\lambda_i$ and $\lambda_o$, are co-defined through a ratio $\lambda_i/\lambda_o$, which describes how much more homogeneous CHs are expected to be compared to the aggregate of all remaining on-disk features. To account for cases where the boundary between foreground and background lies within a pixel instead of at a pixel boundary, which will result in pixels alternating between foreground and background without converging, evolution of the contour is halted when the only evolution that occurs between iterations consists of pixels along the boundary that alternate between foreground and background. \cite{boucheron2016segmentation}
utilize the ratio $\lambda_i/\lambda_o=50$, with a length constraint $\mu=0$, and an initial seeding parameter $\alpha=0.3$, for method validation.

The work presented in this paper uses the ACWE CH segmentation algorithm as described in \cite{boucheron2016segmentation} and expands on that work in two ways.  First, the robustness of the ACWE CH algorithm to changes in spatial resolution is characterized in Section~\ref{sec:spatial} and to small temporal changes is characterized in Section~\ref{sec:temporal}.  Second, a confidence map is generated via an intraensemble method to correlate region homogeneity to likelihood a region belongs to a CH.  The confidence maps are validated by considering the unipolarity (skewness) of the underlying magnetic field. An Appendix is also provided which explores the robustness of ACWE CH algorithm to intensity resolution.  The GitHub repository at \url{https://github.com/DuckDuckPig/CH-ACWE} contains code to download the dataset used in this paper, the base ACWE segmentation code, and code to replicate all experiments described herein.

It should be noted that the code in this repository will, by default, save three elements for each segmentation: the header of the original EUV observation, an ``ACWE header'' which contains both a summary of the pre-processing steps performed and the ACWE parameters themselves, and the final segmentation or group of segmentations (in the case of a confidence map). The code in the GitHub repository also includes functions for resizing both single segmentations and confidence maps to match the resolution of the EUV observation. The format in which the original EUV header is saved and the resized segmentations can be directly ingested by the tools provided by the Python package SunPy~\citep{sunpy_community2020}, which ensures that crucial position, orientation, and coordinate data can be taken directly from the segmentation using the tools already developed by the broader community and that existing reprojection tools, which take into account the rotation of the Sun and motion of the observer, can be applied to segmentations. These features are used in Section \ref{sec:temporal} to align segmentations taken at different times to each other, in order to gauge the consistency of the ACWE segmentation process, and in Section \ref{sec:val} to facilitate the alignment of the HMI magnetograms.

\section{Data and Metrics}
\label{sec:data}
\subsection{Dataset}
The dataset used in this work consists of 2381 observations taken from Carrington rotations (CRs) 2099, 2100, 2101 and 2133. To produce this dataset, image groups consisting of AIA Level 1 EUV images at 94, 131, 171, 193, 211, 304, and 335 {\AA} as well as the corresponding 720s HMI magnetograms were collected at a one hour cadence from the aforementioned CRs using the \verb|drms| library \citep{Glogowski2019}. This dataset was then reduced by eliminating image groups where at least one image did not have a \verb+QUALITY+ key of \verb+0+, resulting in the 2381 observations. It should be noted that one gap of 2 days and 2 hours exists in CR 2099, and gaps of $\leq$ 7 hours exist in the remaining CRs. Prior to performing ACWE, the Level 1 EUV images are converted into level 1.5 data products using \verb+aiapy.calibrate.update_pointing+ as described in \cite{Barnes2020} to ensure the header is correct and \verb+aiapy.calibrate.register+ to align the image to solar north and center the image so that solar center lines up with image center. ACWE is performed on the 193 {\AA} observation only. The remaining observations were collected to facilitate future work with this dataset.

Each experiment is initially performed on CRs 2099-2101, which contain data from 13 July – 3 October 2010, and have a total of 1763 observations. The process is then repeated on CR 2133, which contains 618 observations from 25 January – 22 February 2013, in order to verify that the behavior of the algorithm is consistent across different time frames and points within the solar cycle.  

\subsection{Metrics}
Across all experiments the consistency of ACWE is determined by comparing segmentations to each other using the following four metrics: intersection over union (IOU), structural similarity index measure (SSIM), global consistency error (GCE), and local consistency error (LCE).

\subsubsection{Intersection Over Union}
The Jacard Index or Intersection Over Union (IOU) is defined as the area of agreement normalized by the total area. This metric was first introduced in \cite{jacard1912}. For binary segmentations (Section \ref{sec:robustness}) the IOU for two segmentations  $S_1$ and $S_2$ will be defined as 
\begin{equation}
    IOU(S_1,S_2) = \frac{|S_1 \cap S_2|}{|S_1 \cup S_2|},
\end{equation}
where $S_1\cap S_2$ is the intersection of the two segmentations, $S_1\cup S_2$ is the union of the two segmentations, and $|\cdot|$ is the cardinality of the resulting region. When comparing maps that report probability or likelihood, such as the confidence maps (Section \ref{sec:conMap}), the weighted IOU (wIOU) for the maps $P_1$ and $P_2$ will be determined by comparing the value (confidence) of each each pixel $i$ in $P_1$ to the corresponding pixel in $P_2$. It is defined as 
\begin{equation}
    wIOU(P_1,P_2) = \frac{\sum_{i}\min(P_1(i),P_2(i))}{\sum_{i}\max(P_1(i),P_2(i))},
\end{equation}
where the summations over $i$ denote a summation over all pixels contained in $P_1\cup P_2$.  When two segmentations are compared using IOU or wIOU, the resulting score will be bound to a range of [0,1] where 1 indicates that the two segmentations are identical. IOU and wIOU are sensitive to changes in orientation, placement, shape, and scale.

\subsubsection{Structural Similarity Index}
\cite{wang2004image} define the Structural SIMilarity index measure (SSIM) for two images $x$ and $y$ as 
\begin{equation}
 SSIM(x,y) =\frac{(2\mu_x\mu_y + C_1) (2\sigma_{xy} + C_2) }{(\mu_x^2 + \mu_y^2 + C_1)(\sigma_{x}^2 + \sigma_{y}^2 + C_2)},
\end{equation}
where $\mu_x$ is the mean intensity of $x$,  $\mu_y$ is the mean intensity of $y$, $\sigma_x$ is the standard deviation of $x$, $\sigma_y$ is the standard deviation of $y$, $\sigma_{xy}$ is the covariance between $x$ and $y$, and $C_1$ and $C_2$ are small constants that prevent instability when $(\mu_x^2+\mu_y^2)$ or $(\sigma_{x}^2+\sigma_{y}^2)$ are close to zero. SSIM provides an evaluation of the quality of the structure within a pair of images or segmentations in a manner that is considered more consistent with the human visual system \citep{wang2004image}. In order to match the implementation of \cite{wang2004image}, $C_1=(K_1L)^2$ and $C_1=(K_2L)^2$ where $K_1=0.01$, $K_2=0.03$, and $L$ is the dynamic range of the segmentations. SSIM is bound to the range $(-1,1]$~\citep{nilsson2020understanding}, where 1 indicates that the two segmentations are identical, 0 indicates no similarity, and -1 indicates anti-correlation. SSIM is most sensitive to the structure present within the images or segmentations~\citep{wang2004image}. 

\subsubsection{Global and Local Constancy Error}
Global Consistency Error (GCE) and Local Consistency Error (LCE) are introduced in \cite{martin2001} as methods for evaluating segmentation accuracy while accounting for the variability that exists within human-segmented data. In particular, \cite{martin2001} note that, in human-generated segmentations, it is possible that one segmentation may be a refinement of another, either by further subdividing a region of interest or by providing finer granularity at image boundaries. To account for this, both GCE and LCE rely on an intermediate metric called local error $E$, which is evaluated on a per-pixel basis. For pixel $i$ present in segmentations $S_1$ and $S_2$, local error is defined as
\begin{equation}
    E(S_1,S_2,i)=\frac{|R(S_1,i)-R(S_2,i)|}{|R(S_1,i)|},
\end{equation}
where $-$ denotes the set difference, $R(S_1,i)$ is the region in $S_1$ which contains pixel $i$, $R(S_2,i)$ is the region in $S_2$ which contains pixel $i$, and $|\cdot|$ is the cardinality of the region. This metric returns a value of zero any time $R(S_1,i)$ is fully contained in $R(S_2,i)$ to account for subdivision of regions, and a small value when only a few pixels of $R(S_1,i)$ are not contained in $R(S_2,i)$, such as when the region $R(S_1,i)$ is a refinement of $R(S_2,i)$.

For a segmentation of an image with $n$ pixels, GCE is defined as
\begin{equation}
    GCE(S_1,S_2)=\frac{1}{n}\min\Big\{\sum_{i}E(S_1,S_2,i),\sum_{i}E(S_2,S_1,i)\Big\},
\end{equation}
which forces GCE to be most sensitive to cases where one segmentation is a refinement of another. LCE is defined as
\begin{equation}
    LCE(S_1,S_2)=\frac{1}{n}\sum_{i}\min\{E(S_1,S_2,i),E(S_2,S_1,i)\},
\end{equation}
which allows LCE to be sensitive to cases where some portions of each segmentation are refinements of the other. For this reason, LCE is a less strict error measure than GCE. Both GCE and LCE are bound to a range of [0,1] with 0 indicating no error.  

\section{Sensitivity of the ACWE Algorithm}
\label{sec:robustness}

Prior to developing an intramethod ensemble for CH detection, the consistency of ACWE when applied to CH segmentation is first evaluated. These experiments characterize the expected degradation in the accuracy of CH segmentations when reducing the spatial resolution of the input image to optimize the process of generating the segmentation and the consistency of segmentations across short timescales where CH evolution is expected to be minimal.

\subsection{Spatial Resolution Effects}
\label{sec:spatial}
ACWE is an iterative process, requiring successive manipulations of the contour that defines a region of interest to minimize the energy functional (Equation (\ref{eq:acwe})). To help reduce the computation time needed to calculate the energy functional in each iteration, and, by extension, reduce the computation time needed to generate a contour map, ACWE can be performed on a copy of the target image which has a reduced spatial resolution. \cite{boucheron2016segmentation} demonstrate that this process may be viable for improving computational efficiency. In particular, \cite{boucheron2016segmentation} note that segmentations remain qualitatively similar across spatial resolutions, and that the primary difference between the reduced resolution segmentations, generated from images with a resolution of $512\times512$ pixels, and the full-scale  ($4096\times4096$ pixels) segmentations is the absence of smaller regions that, due to pixel removal and lowpass filtering, are no longer dark enough to be included in the initial seed.

\begin{figure}
    \centering
    \subfloat[Intersection Over Union]{\includegraphics[width=.49\textwidth]{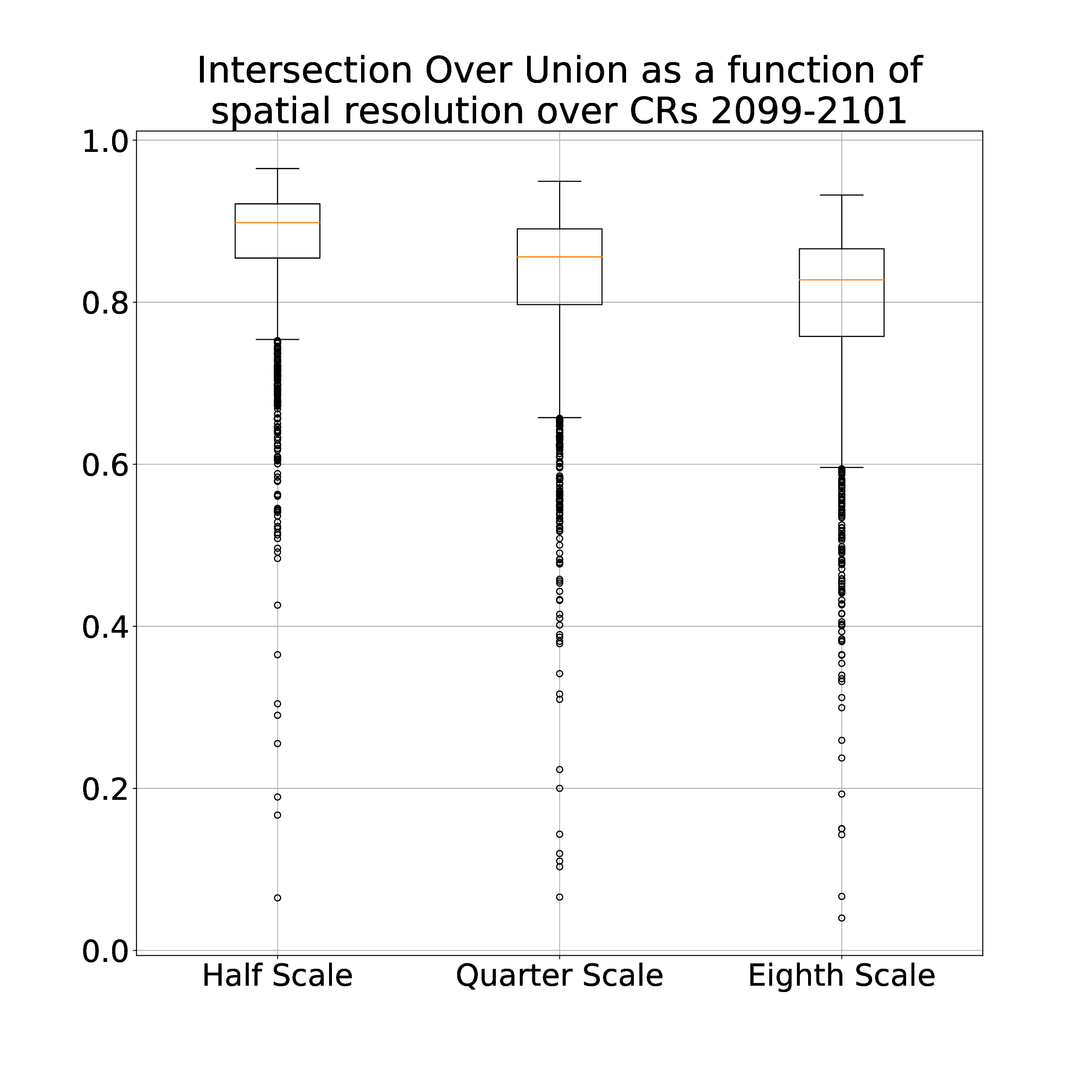}}~~
    \subfloat[Structural Similarity]{\includegraphics[width=.49\textwidth]{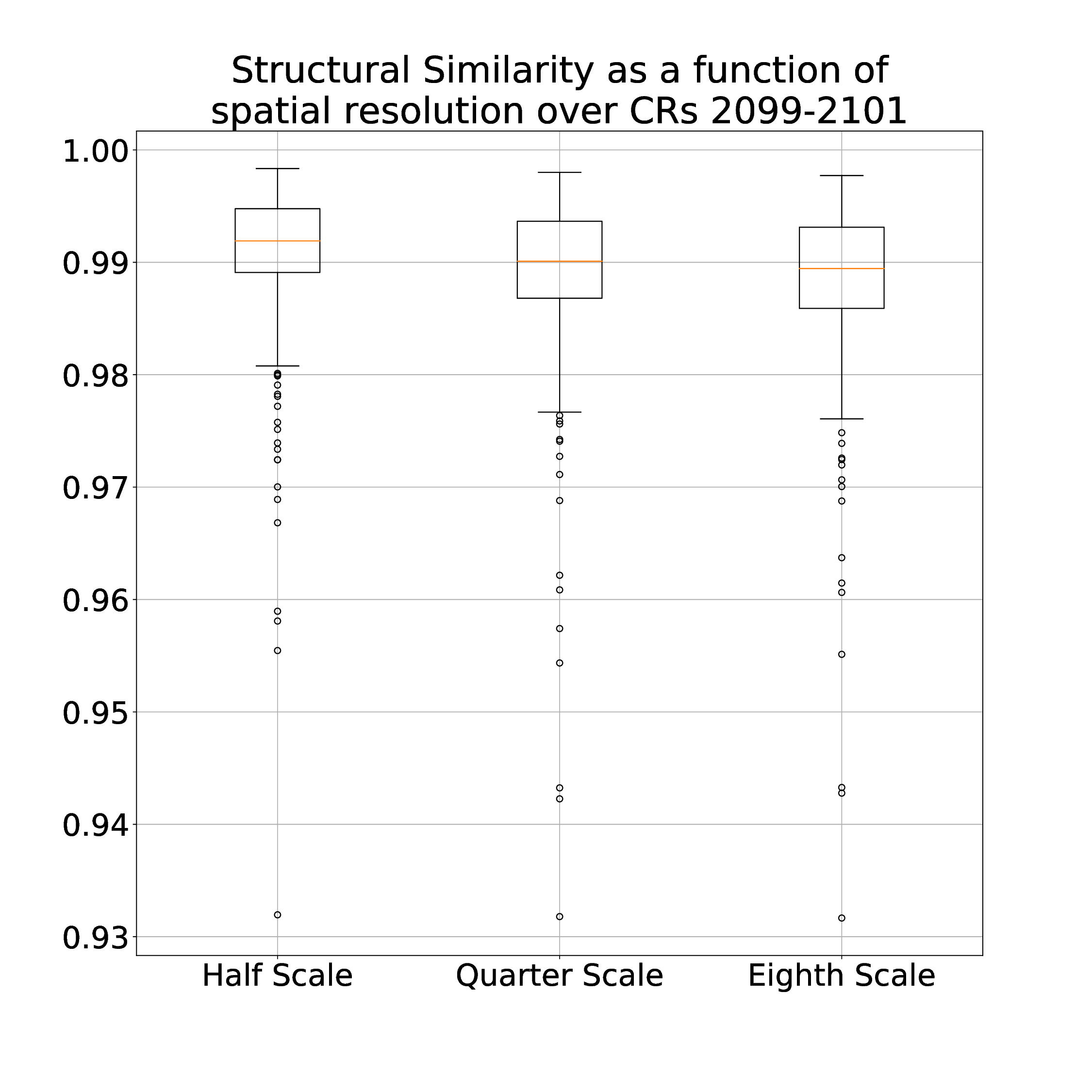}}\\
    \subfloat[Global Consistency Error]{\includegraphics[width=.49\textwidth]{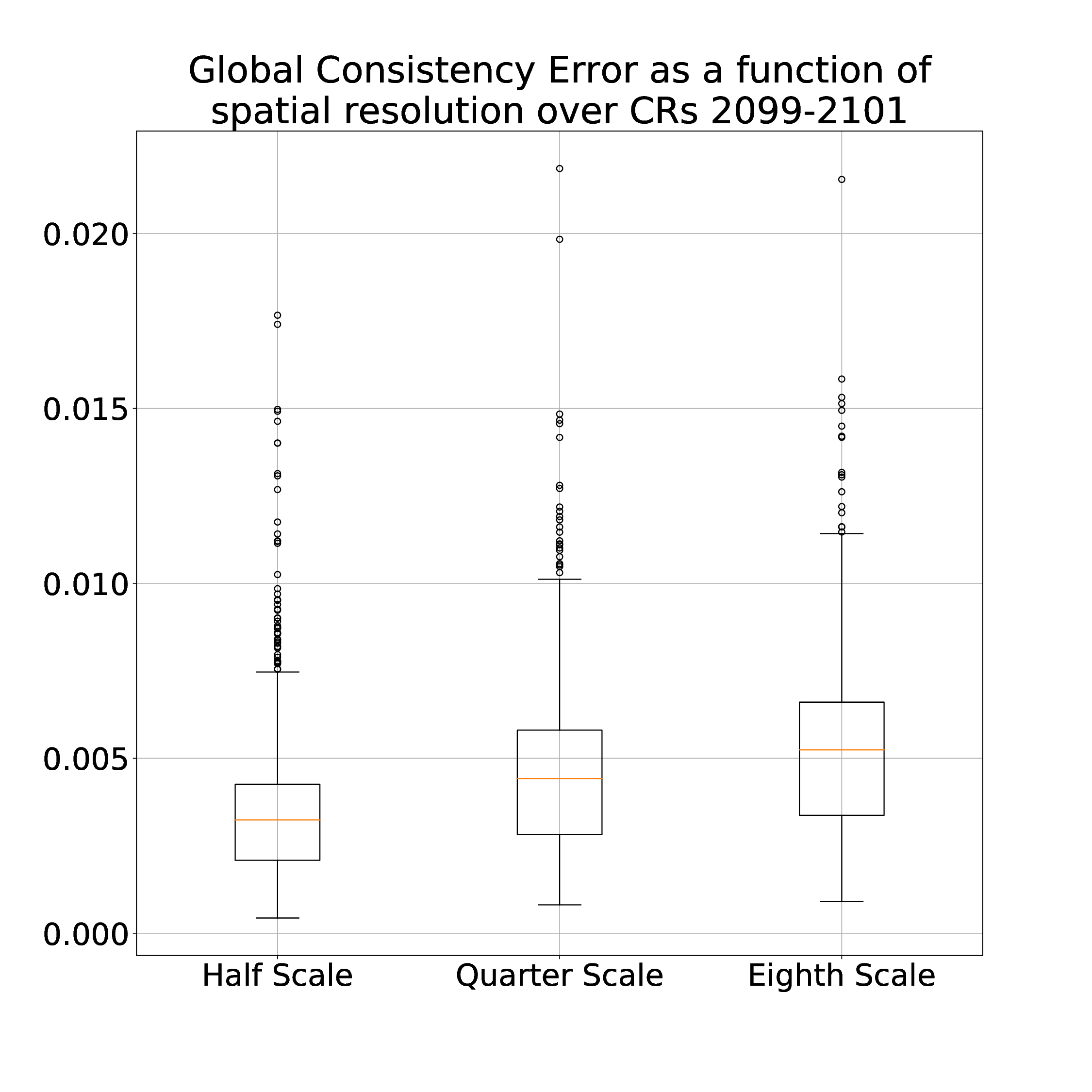}}~~
    \subfloat[Local Consistency Error]{\includegraphics[width=.49\textwidth]{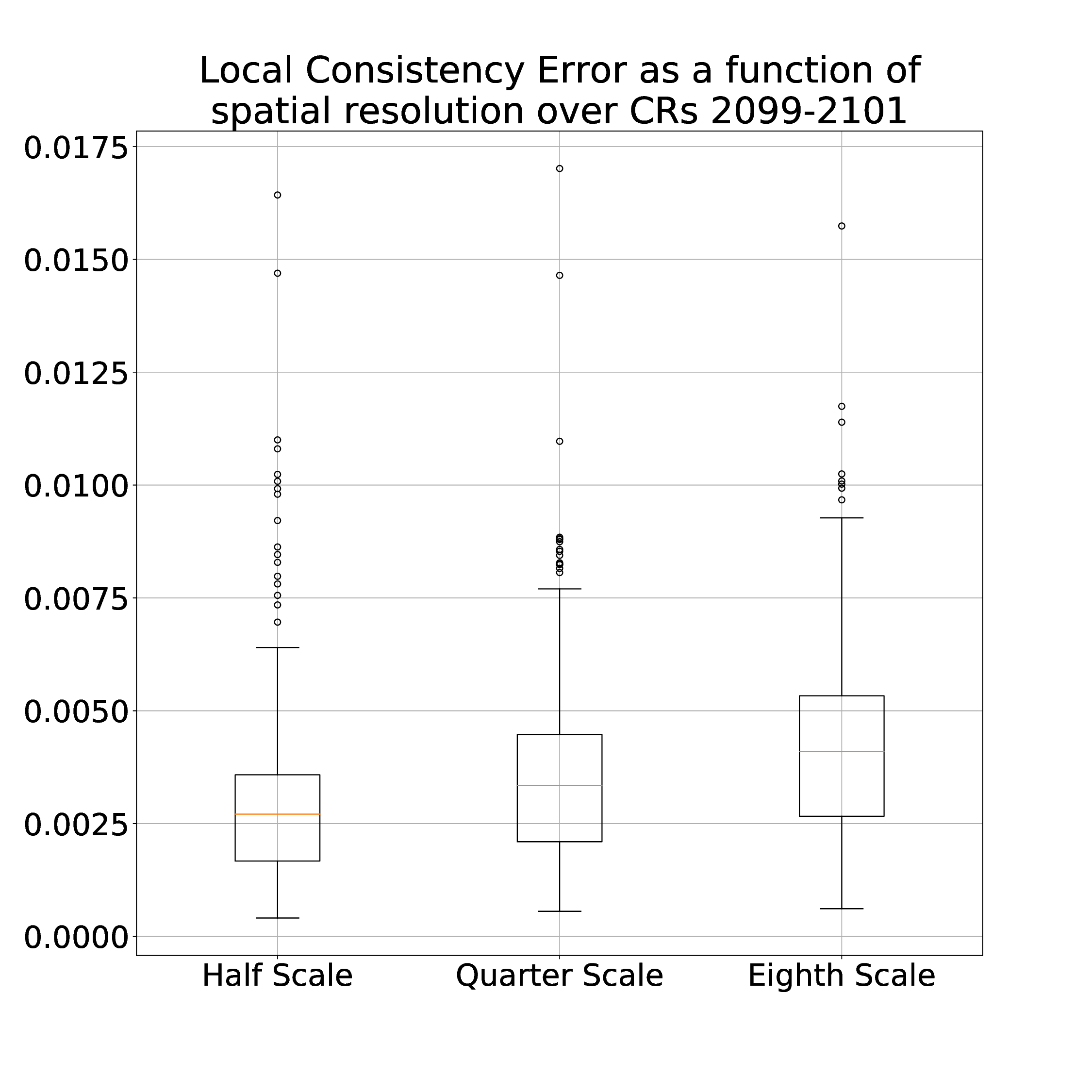}}\\
    \caption{Effects of decimation in spatial resolution for CRs 2099, 2100, and 2101 using bilinear interpolation. In this figure the orange line is the median value, the box represents the range between the first (Q1) and third (Q3) quartile, the whiskers represent 1.5 times the interquartile range above Q3 or below Q1, and circles represent outliers.}
    \label{fig:Scale_CR2099_2101}
\end{figure}

Here, a study of the effects of down-sampling the original AIA 193 {\AA} image on final segmentation was performed by generating ACWE segmentations for each image within the dataset at original resolution  ($4096\times4096$ pixels), one-half scale (decimation by $2\times$ in each dimension  or $2048\times2048$ pixels), one-quarter scale ($1024\times1024$ pixels), and one-eighth scale ($512\times512$ pixels). In each instance correction for limb brightening, as implemented by \cite{verbeeck2014}, was applied to the image after decimation. For the initial seeding of the algorithm, the seeding process was performed through analysis of the down-sampled image, with threshold $\alpha$ of 0.3. Across all spatial resolutions, evolution of the segmentation was performed using a homogeneity ratio $\lambda_i/\lambda_o$ of 50 and a length constraint $\mu$ of 0. The segmentations generated from the decimated images were then upscaled to $4096\times4096$ pixels to match the resolution of the segmentation generated from the original image and directly compared.

\begin{figure}
    \centering
    \subfloat[Intersection Over Union]{\includegraphics[width=.49\textwidth]{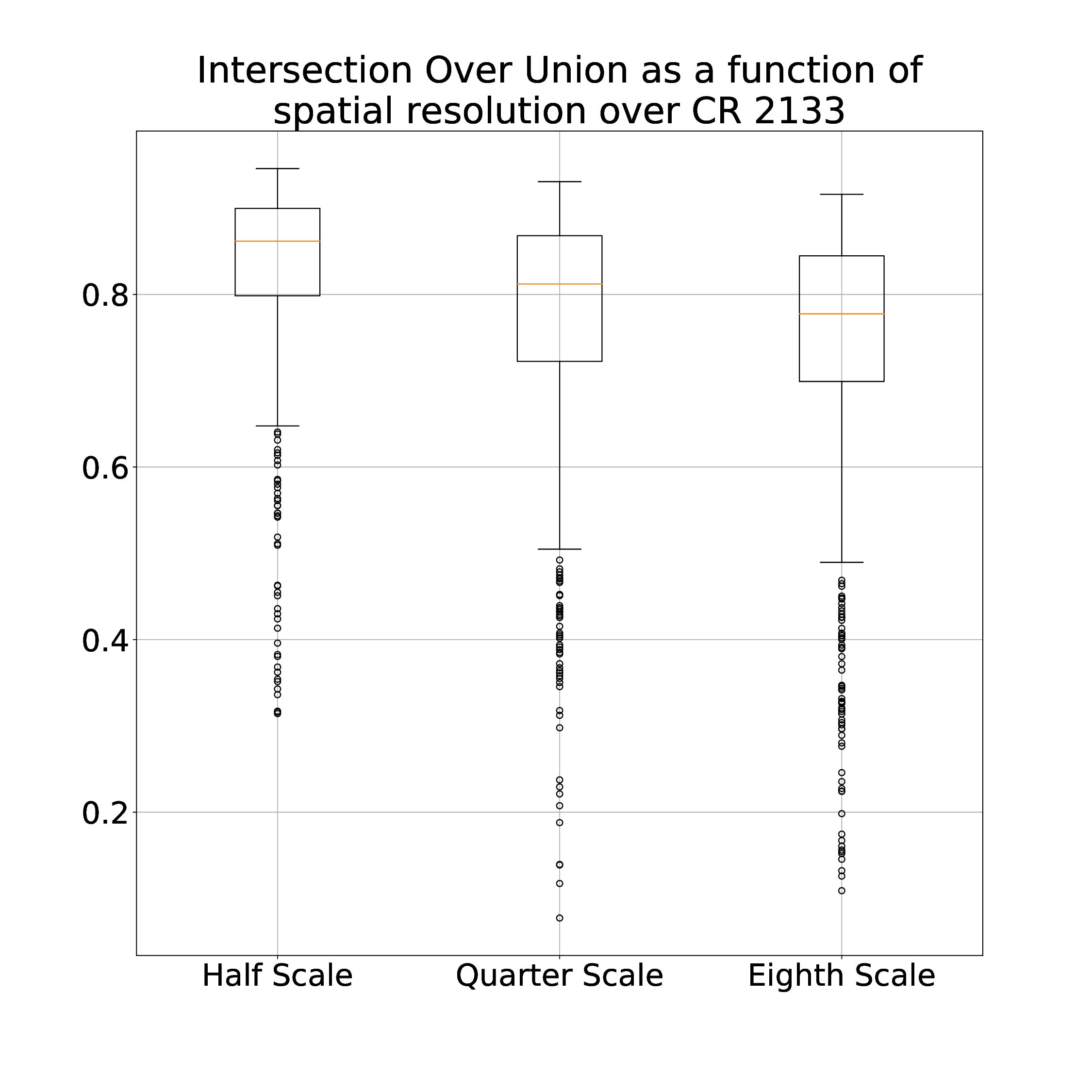}}~~
    \subfloat[Structural Similarity]{\includegraphics[width=.49\textwidth]{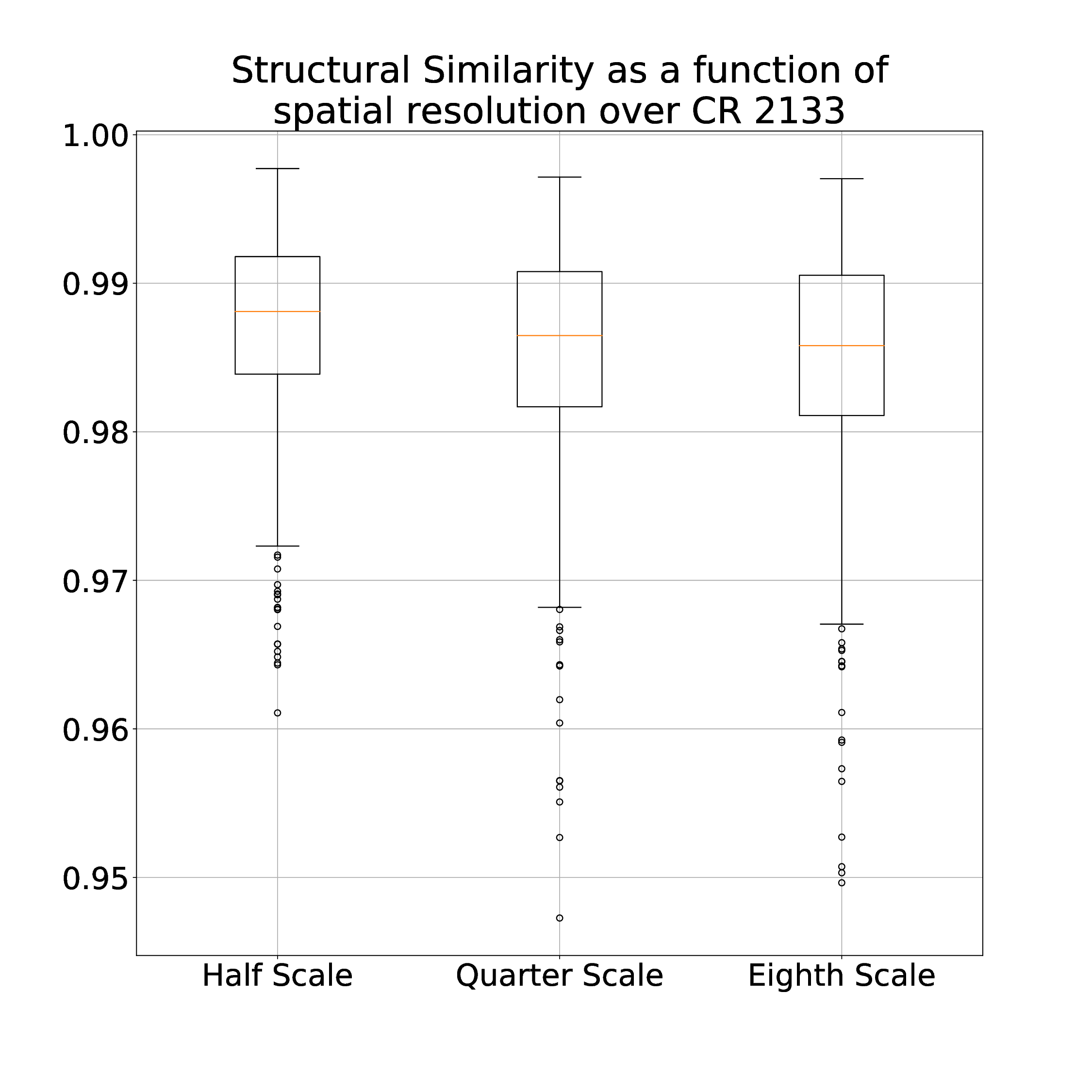}}\\
    \subfloat[Global Consistency Error]{\includegraphics[width=.49\textwidth]{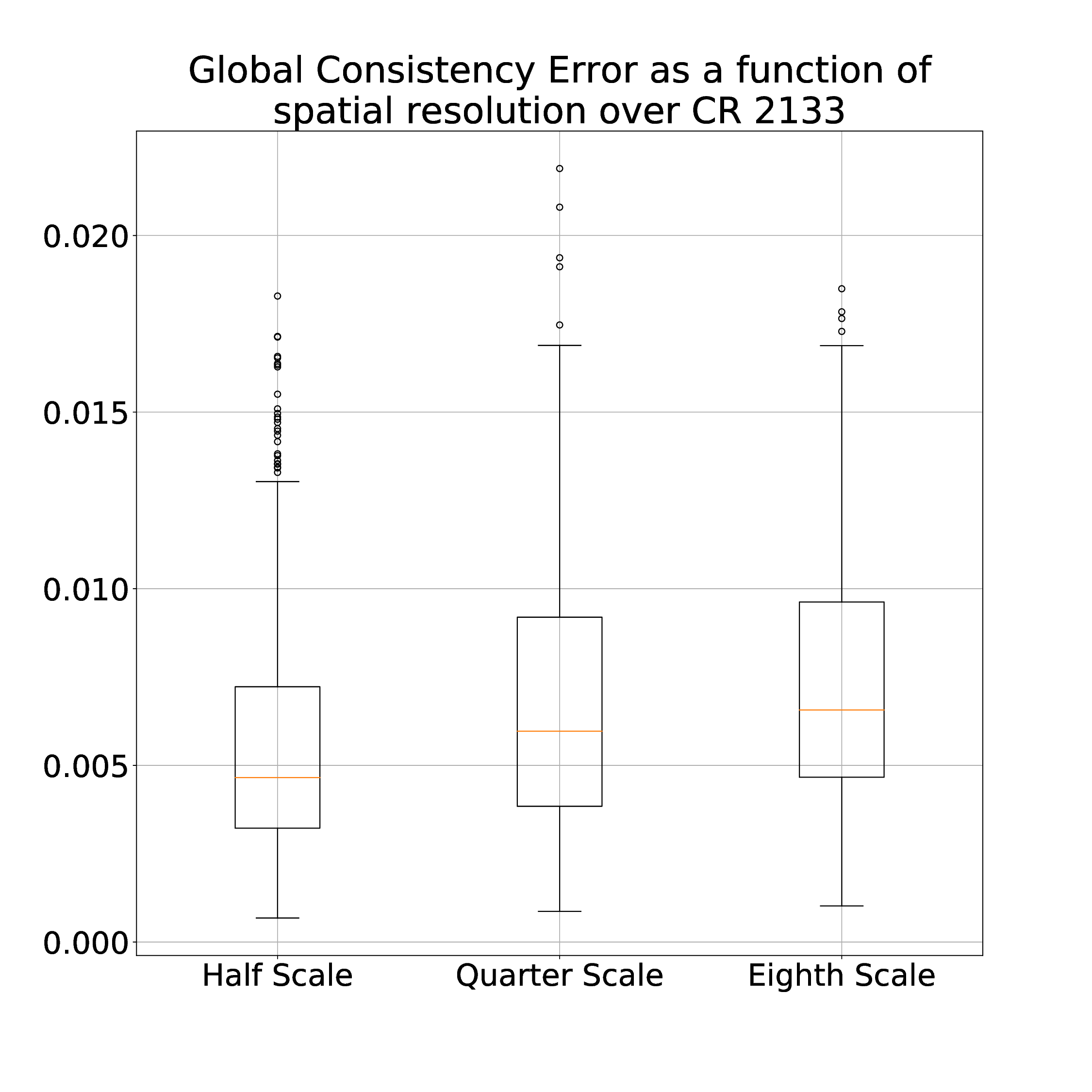}}~~
    \subfloat[Local Consistency Error]{\includegraphics[width=.49\textwidth]{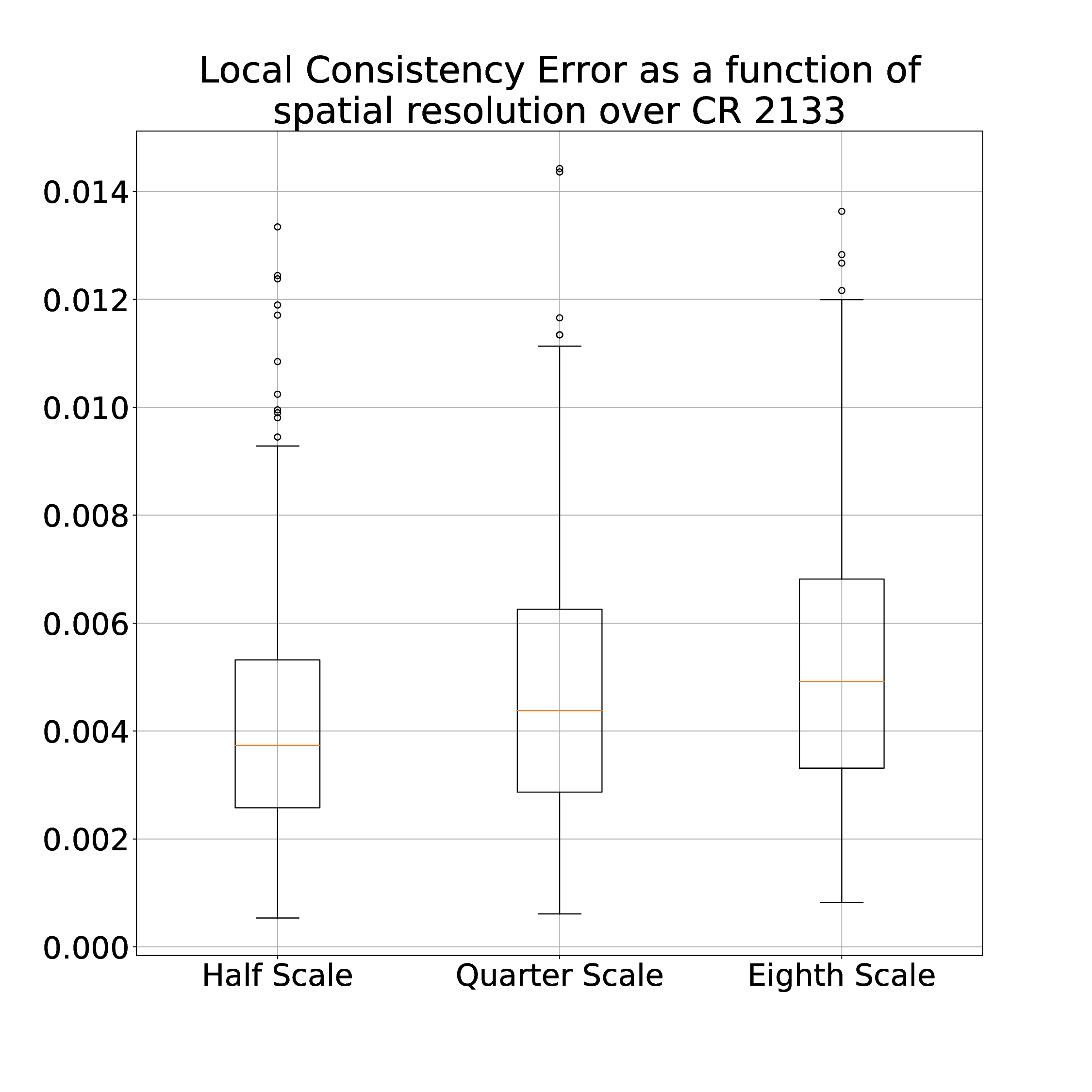}}\\
    \caption{Effects of decimation in spatial resolution for CR 2133 using bilinear interpolation. In this figure the orange line is the median value, the box represents the range between the first (Q1) and third (Q3) quartile, the whiskers represent 1.5 times the interquartile range above Q3 or below Q1, and circles represent outliers.}
    \label{fig:Scale_CR2133}
\end{figure}

\begin{figure}
    \centering
    \subfloat[CRs 2099, 2100, and 2101]{\includegraphics[width=\textwidth]{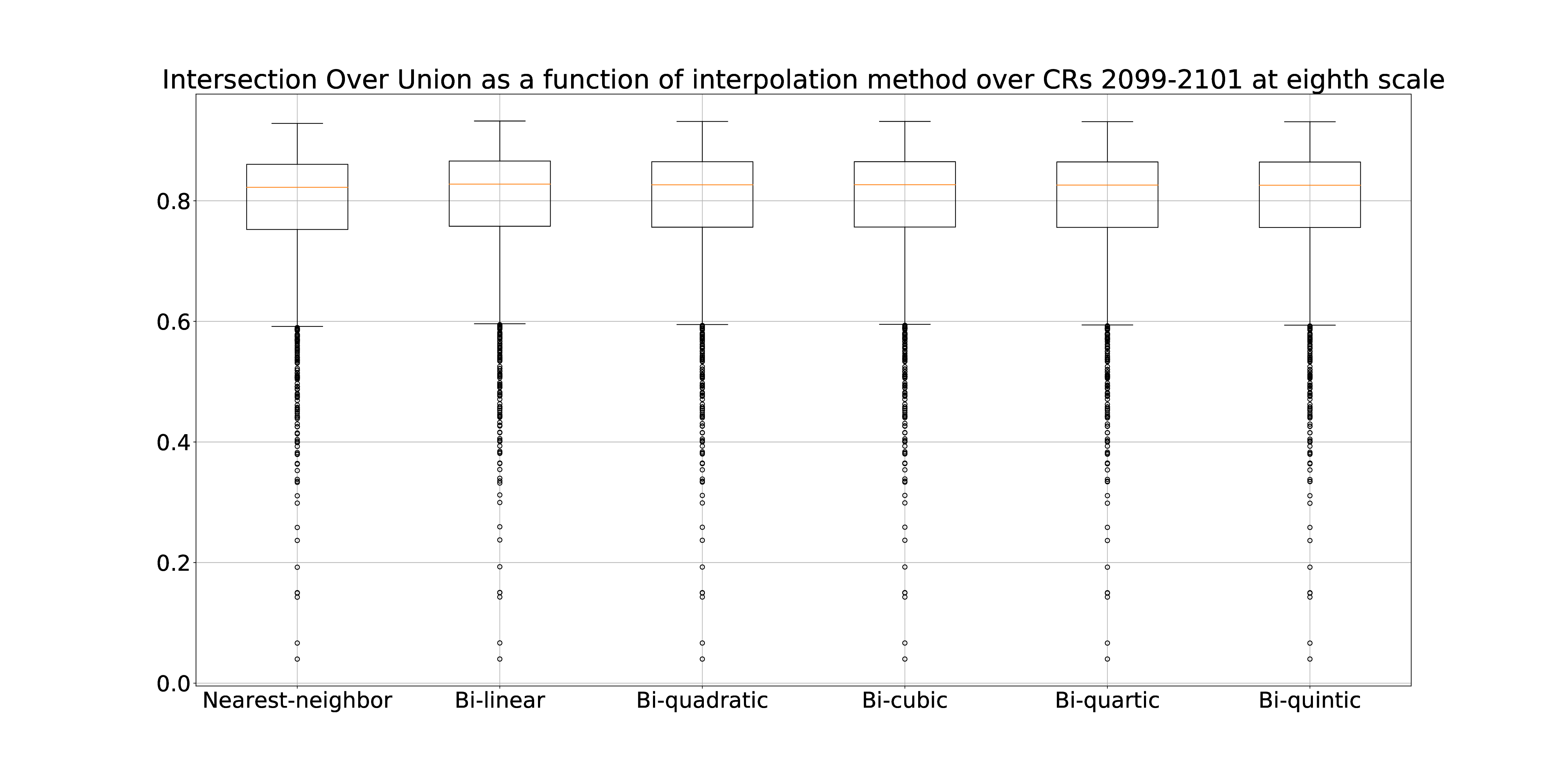}}\\
    \subfloat[CR 2133]{\includegraphics[width=\textwidth]{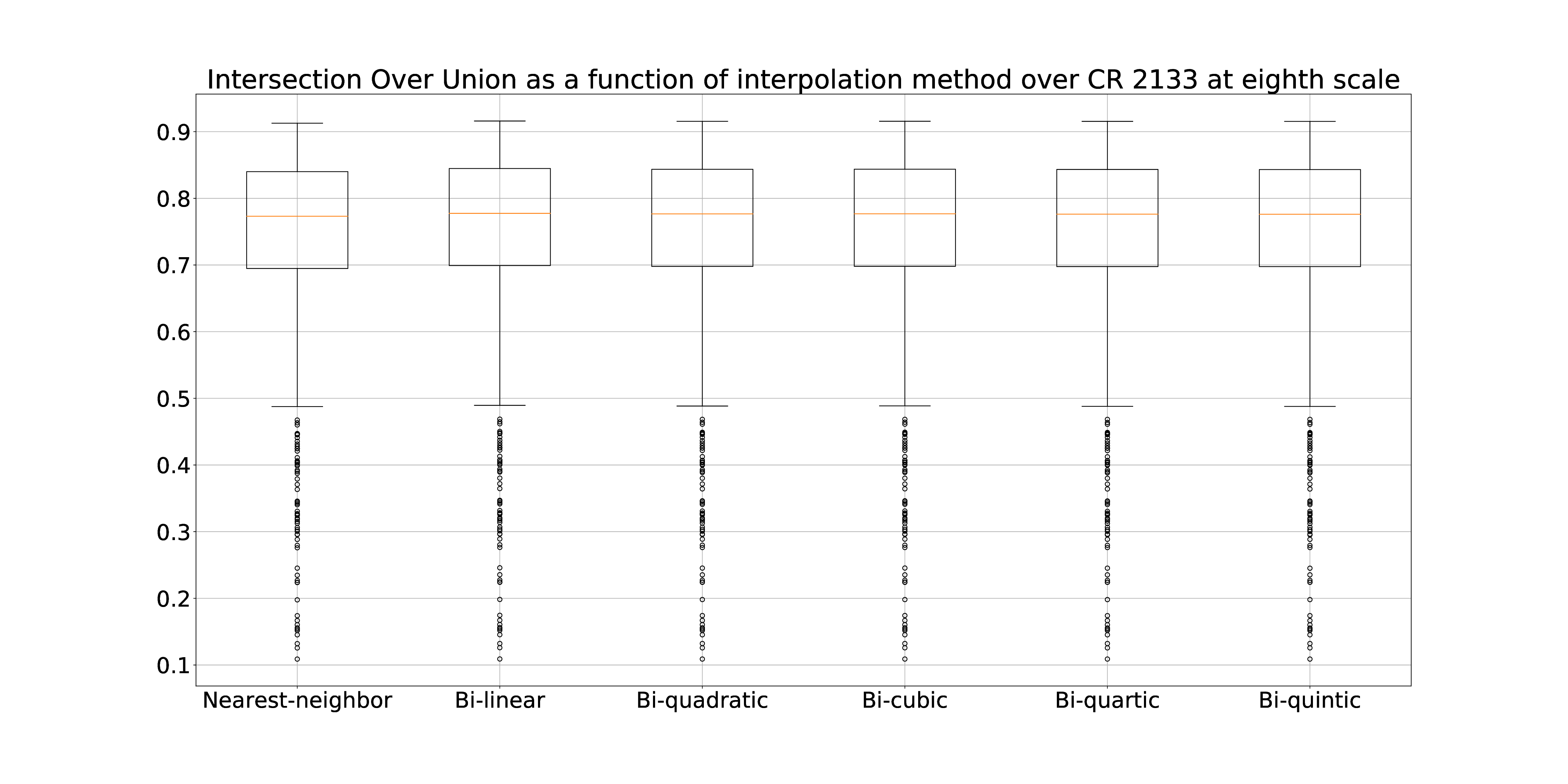}}
    \caption{Effects of interpolation method on similarity between segmentation generated at one-eighths scale and full resolution ($512\times512$ and $4096\times4096$ pixels, respectively). In this figure the orange line is the median value, the box represents the range between the first (Q1) and third (Q3) quartile, the whiskers represent 1.5 times the interquartile range above Q3 or below Q1, and circles represent outliers.}
    \label{fig:Scale_eights_all}
\end{figure}

The effects that reducing spatial resolution had on the final segmentation for CRs 2099 through 2101 are presented in Figure \ref{fig:Scale_CR2099_2101}. Figure \ref{fig:Scale_CR2133} presents the effects for CR 2133. Both figures show the resulting similarity after applying bi-linear interpolation to upscale the image and converting the result back to a binary mask by setting all values $>0.5$ to $1$ and setting all values $\leq0.5$ to $0$. The results in Figure \ref{fig:Scale_CR2099_2101} and Figure \ref{fig:Scale_CR2133} suggest that general structures in the segmentations are preserved, even when decimated by $8\times$, resulting in a high IOU and SSIM, and low GCE and LCE in both time frames. It should be noted that this process was repeated with up scaling performed using nearest-neighbor, bi-quadratic, bi-cubic, bi-quartic, and bi-quintic interpolation. Figure \ref{fig:Scale_eights_all} shows the IOU between the one-eighth scale segmentations, upscaled using each of the described interpolation methods, and the corresponding full scale segmentation. These results indicate that interpolation method has minimal effect on the similarity of final segmentation.

\begin{figure}
    \centering
    \includegraphics[trim=0in .25in 0in 0in,clip,width=\textwidth]{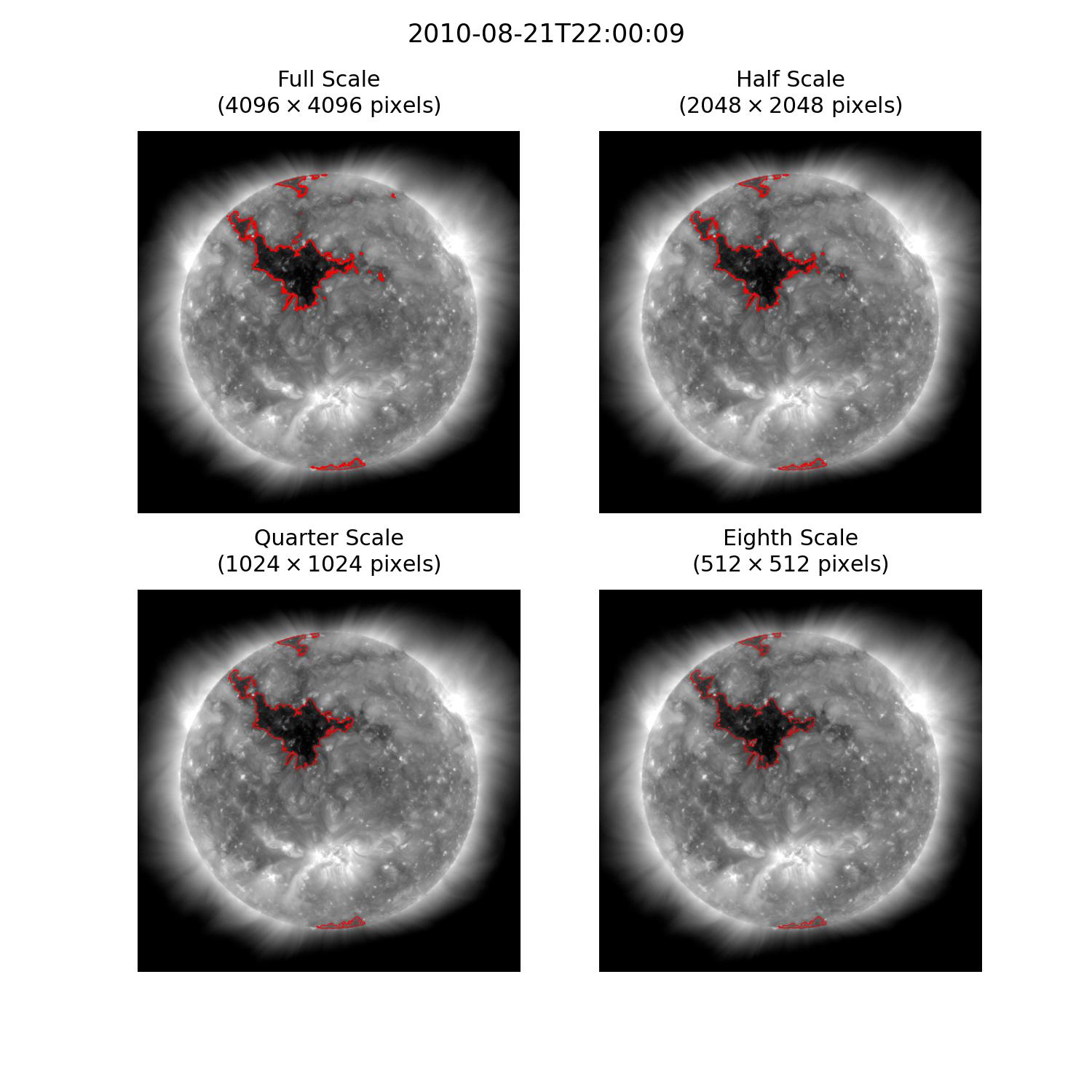}
    \caption{Comparison of segmentations of the same observation performed at different spatial resolutions.}
    \label{fig:ScaleSample}
\end{figure}

Visual examination of ACWE segmentations revealed three primary sources of discrepancy between segmentations performed at the original resolution and segmentations performed on decimated images. First, segmentations generated from decimated images sometimes excluded  smaller CH regions. This is consistent with the observations of \cite{boucheron2016segmentation}. Second, spurious bright regions within larger CHs may be excluded or included at different spatial resolutions. Finally, larger CH regions show a gradual reduction in the granularity of the segmentation boundary as spatial resolution decreases. All three effects can be seen in Figure \ref{fig:ScaleSample}.

\begin{table}[t]
\centering
    \caption{Minimum, mean, and maximum computation time in seconds for ACWE over CR 2099 as a function of spatial resolution.}
    \label{tab:spaceTime}
     {\begin{tabular}{ccccc}
    \hline
    \textbf{} & \textbf{Full} & \textbf{Half} & \textbf{Quarter} & \textbf{Eighth}\\
    \textbf{} & \textbf{($4096\times4096$)}  & \textbf{($2048\times2048$)}  & \textbf{($1024\times1024$)}  & \textbf{($512\times512$)} \\\hline
    Min & $343.855$ & $54.851$ & $7.936$ & $1.491$\\
    Mean & $600.698$ & $86.174$ & $12.260$ & $2.011$\\
    Max & $2021.239$ & $270.058$ & $24.975$ & $4.285$\\\hline
    \end{tabular}}
\end{table}

To demonstrate the computational efficiency provided by segmenting at a reduced resolution, the segmentation process for CR 2099 was performed on an Intel Core i7-8700K running at base clock speed (3.70~{GHz}), and the time needed to generate each individual segmentation was recorded. Table \ref{tab:spaceTime} outlines the minimum, maximum, and mean time in seconds to generate a segmentation at each spatial resolution, aggregated for all images in CR 2099. A direct comparison of computation time as a function of spatial resolution (also performed over CR 2099) revealed that, for the same EUV observation, performing ACWE on an image decimated to $2048\times2048$ pixels will result in a segmentation an average of $7.025\pm1.561$ times faster than operating on the original $4096\times4096$ pixel image. Reducing the resolution to $1024\times1024$ pixels results in a segmentation an average of $49.050\pm13.004$ times faster than operating on the original image. At $512\times512$ pixels, the segmentation is generated an average of $298.126\pm87.214$ times faster than at full resolution.

\subsection{Effects Across Small Temporal Changes}
\label{sec:temporal}

In order to gauge the consistency of ACWE segmentations, a study of the similarity of segmentations across small time spans was performed on all four CRs. This process was performed using the the one-eighth scale ($512\times512$ pixel) segmentations as described in Section~\ref{sec:spatial} in order to characterize the behavior of the default implementation of ACWE. These images were upscaled to the original $4096\times4096$ pixel resolution using bi-linear interpolation to allow for the reprojection process. For each segmentation in the dataset, the preceding twelve hours of segmentations and the succeeding twelve hours of segmentations were reprojected to account for the rotation of the Sun and then compared. The reprojection was achieved using the \verb+sunpy.coordinates+ tools \verb+Helioprojective+, \verb+RotatedSunFrame+, and \verb+transform_with_sun_center+ described in \cite{sunpy_community2020} to define the reprojection and \verb+reproject.reproject_interp+ described in \cite{repoject_2020} to perform the reprojection.

\begin{figure}
    \centering
    \subfloat[Intersection Over Union]{\includegraphics[width=.49\textwidth]{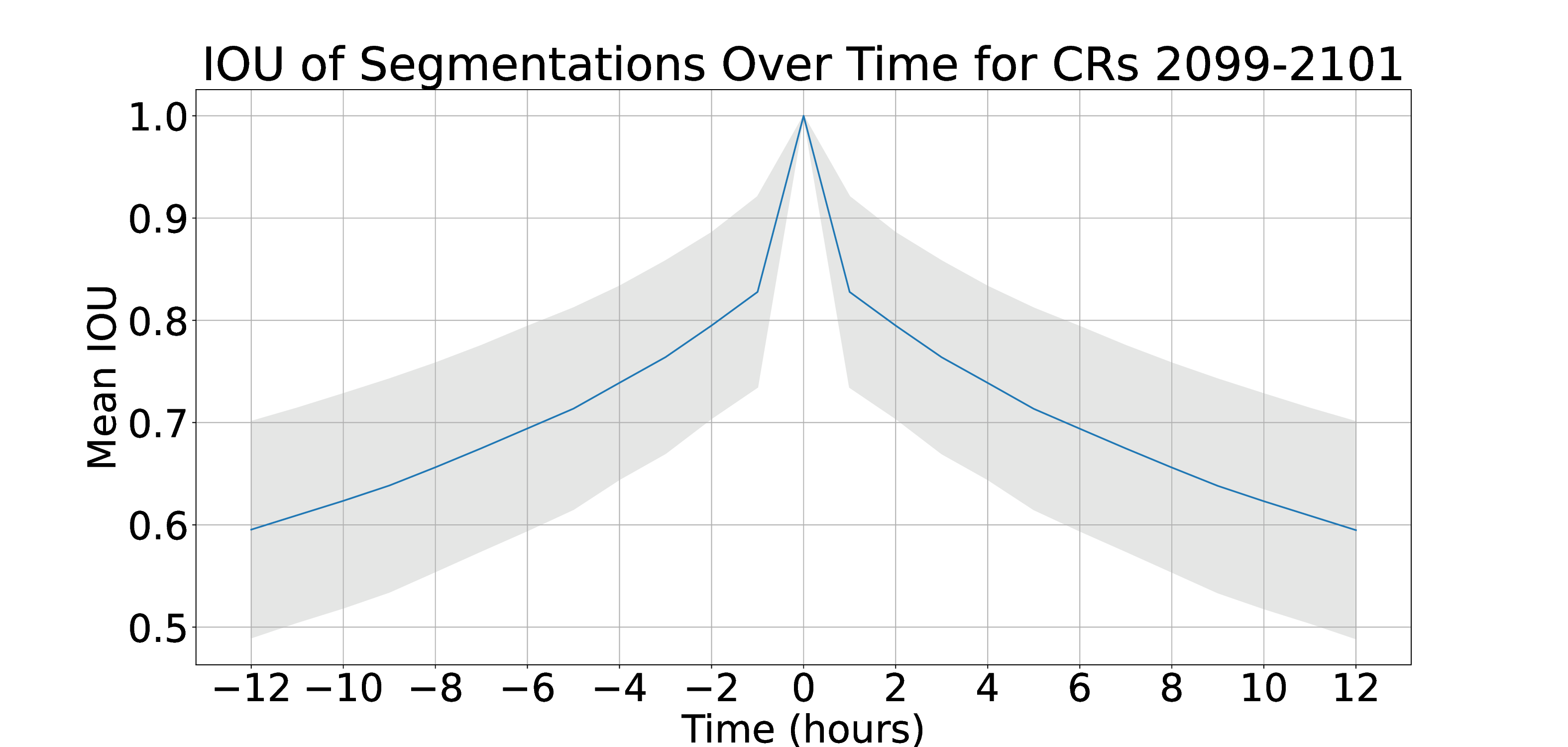}}~~
    \subfloat[Structural Similarity]{\includegraphics[width=.49\textwidth]{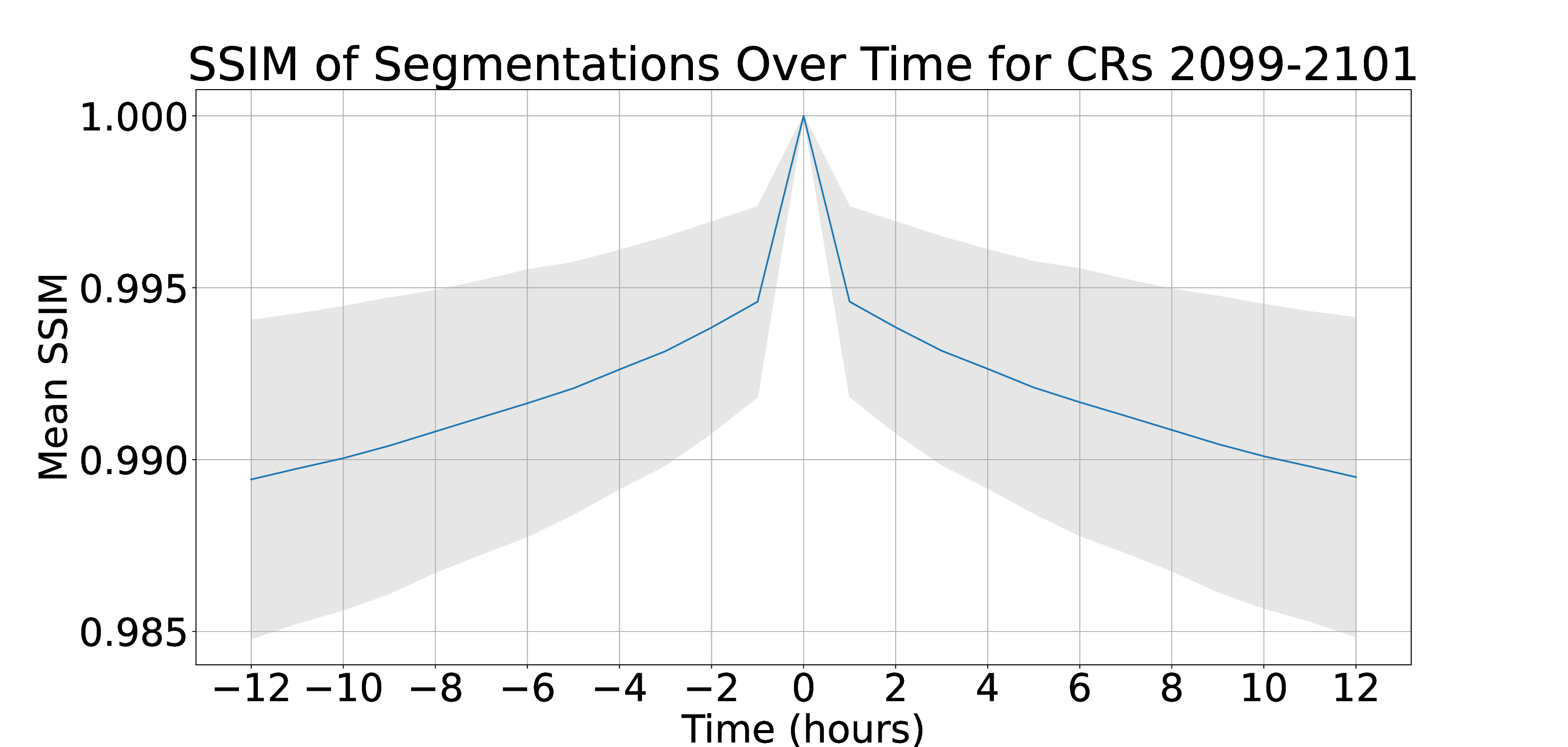}}\\
    \subfloat[Global Consistency Error]{\includegraphics[width=.49\textwidth]{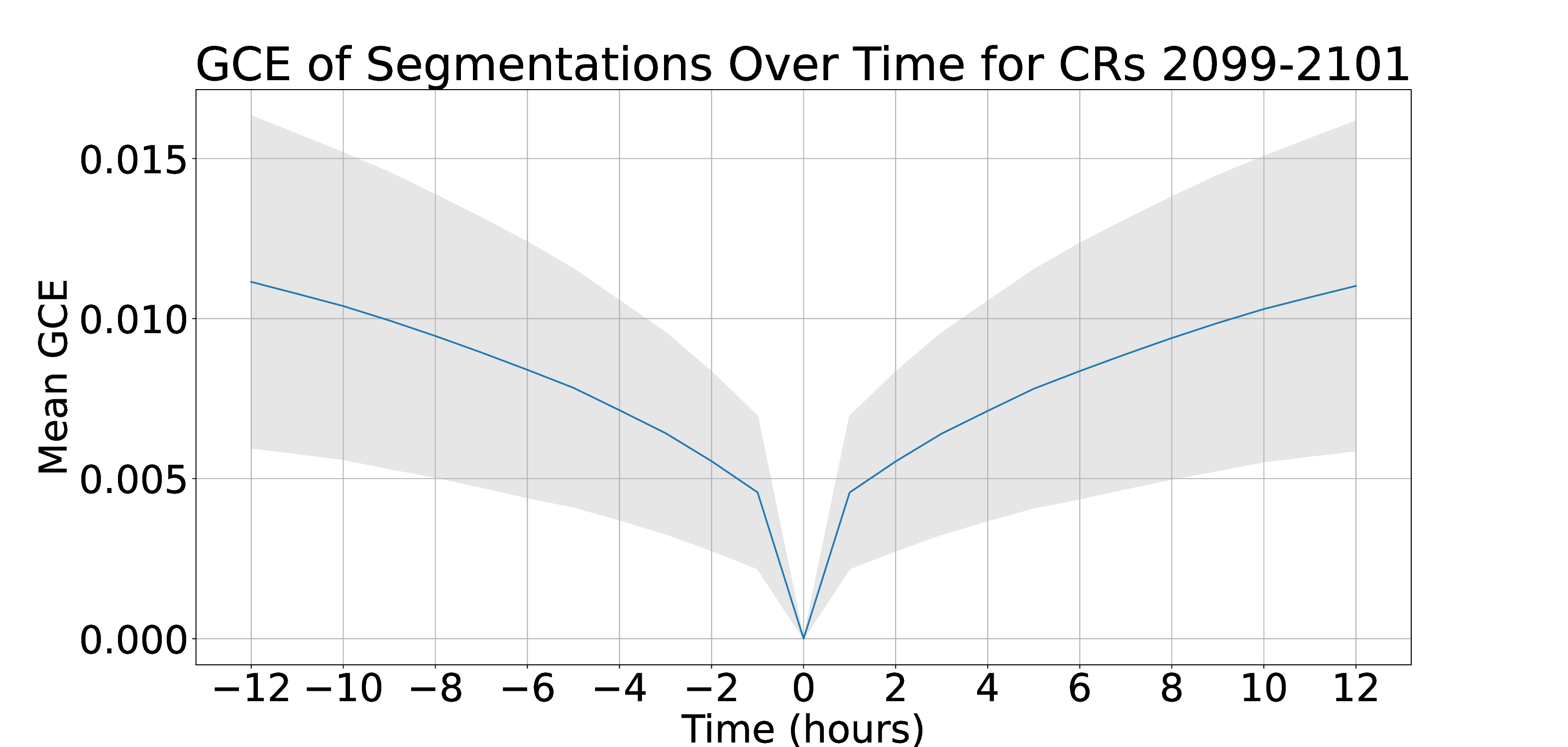}}~~
    \subfloat[Local Consistency Error]{\includegraphics[width=.49\textwidth]{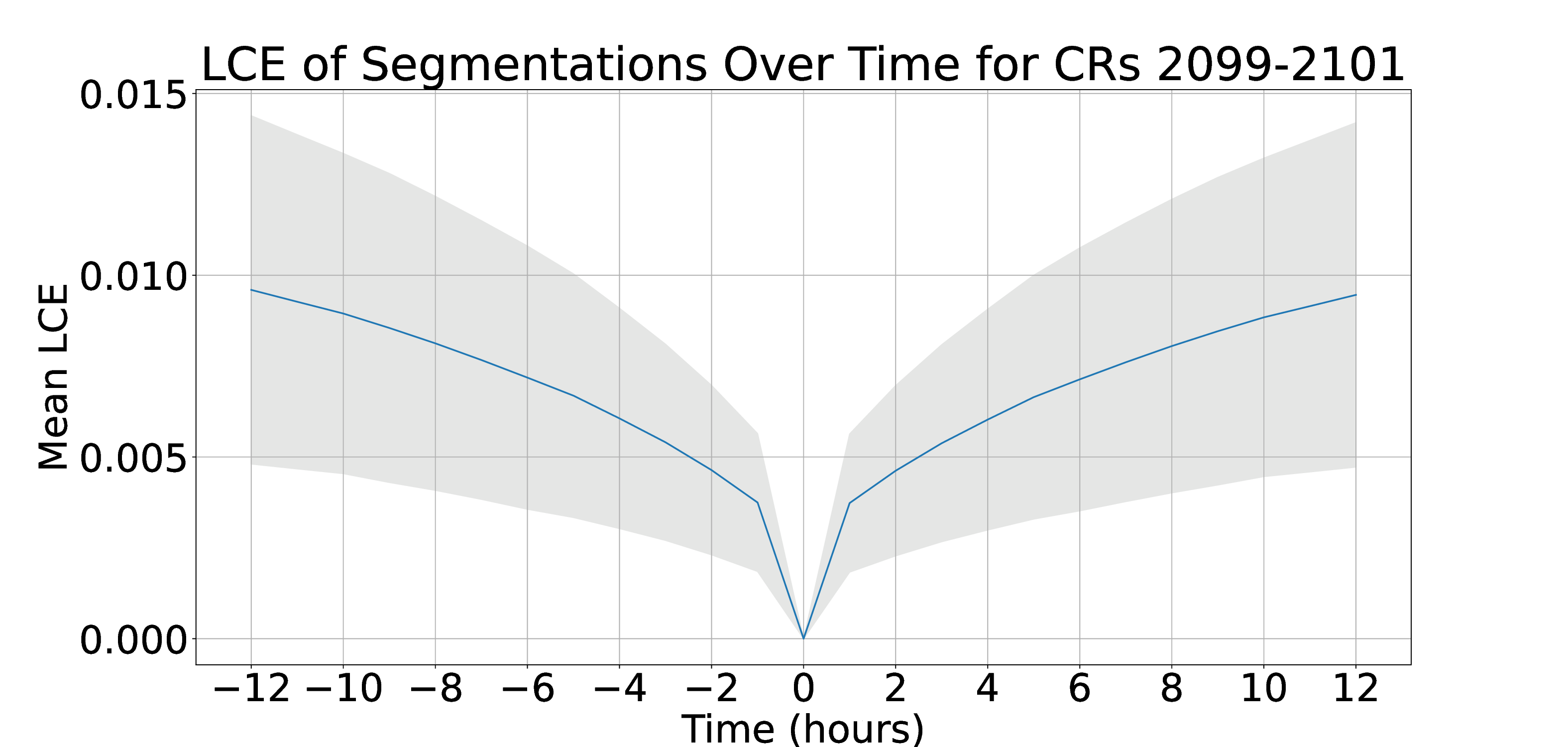}}\\
    \caption{Similarity of segmentation as a function of time across CRs 2099, 2100, and 2101. The blue line represents the mean similarity or error between a segmentation and its successor or predecessor as a function of the time difference between the two EUV images while the gray shaded region is $\mu\pm\sigma$ where $\mu$ is the mean and $\sigma$ is the standard deviation.}
    \label{fig:Temporal_CR2099_2101}
\end{figure}

\begin{figure}
    \centering
    \subfloat[Intersection Over Union]{\includegraphics[width=.49\textwidth]{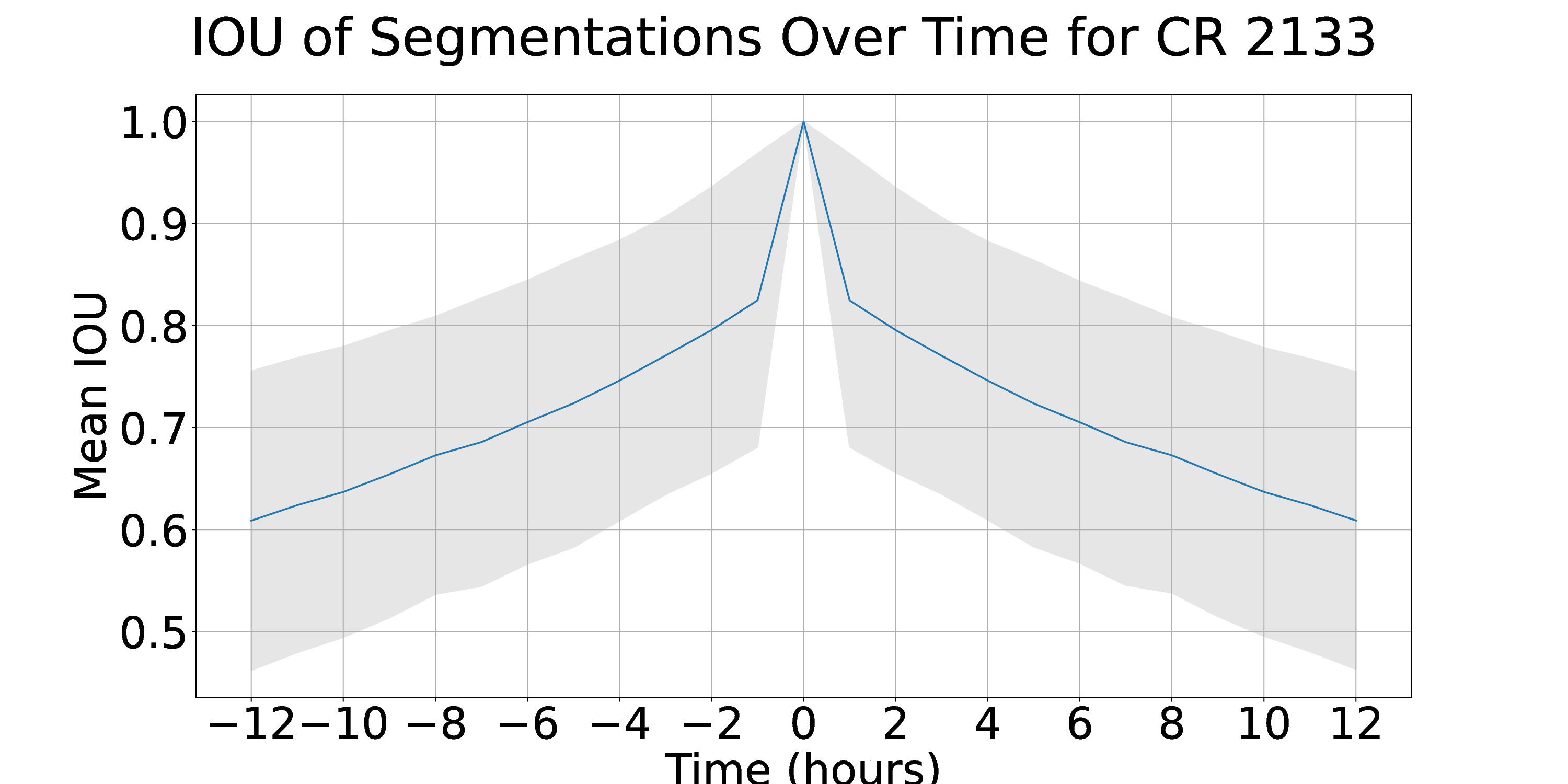}}~~
    \subfloat[Structural Similarity]{\includegraphics[width=.49\textwidth]{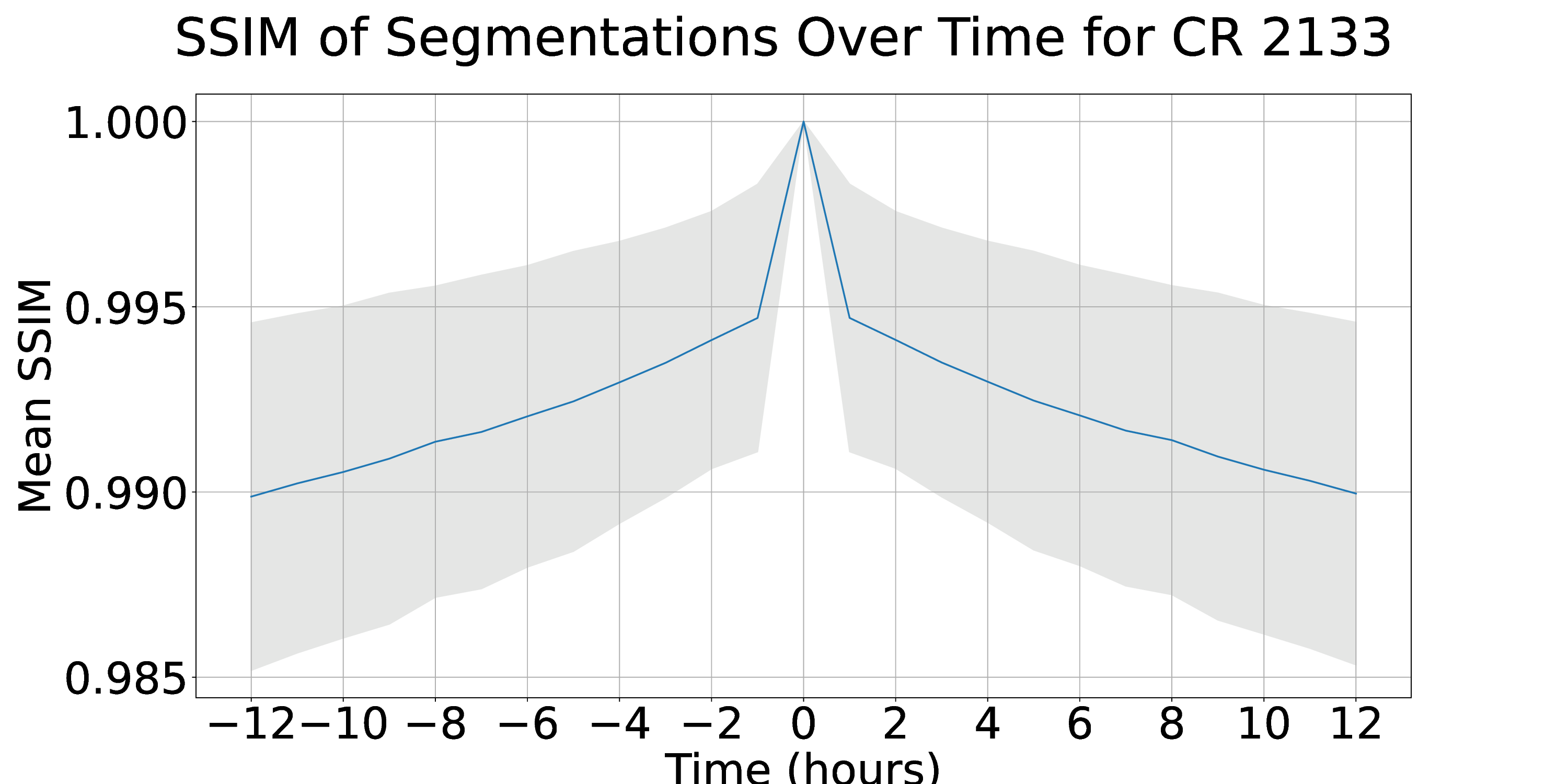}}\\
    \subfloat[Global Consistency Error]{\includegraphics[width=.49\textwidth]{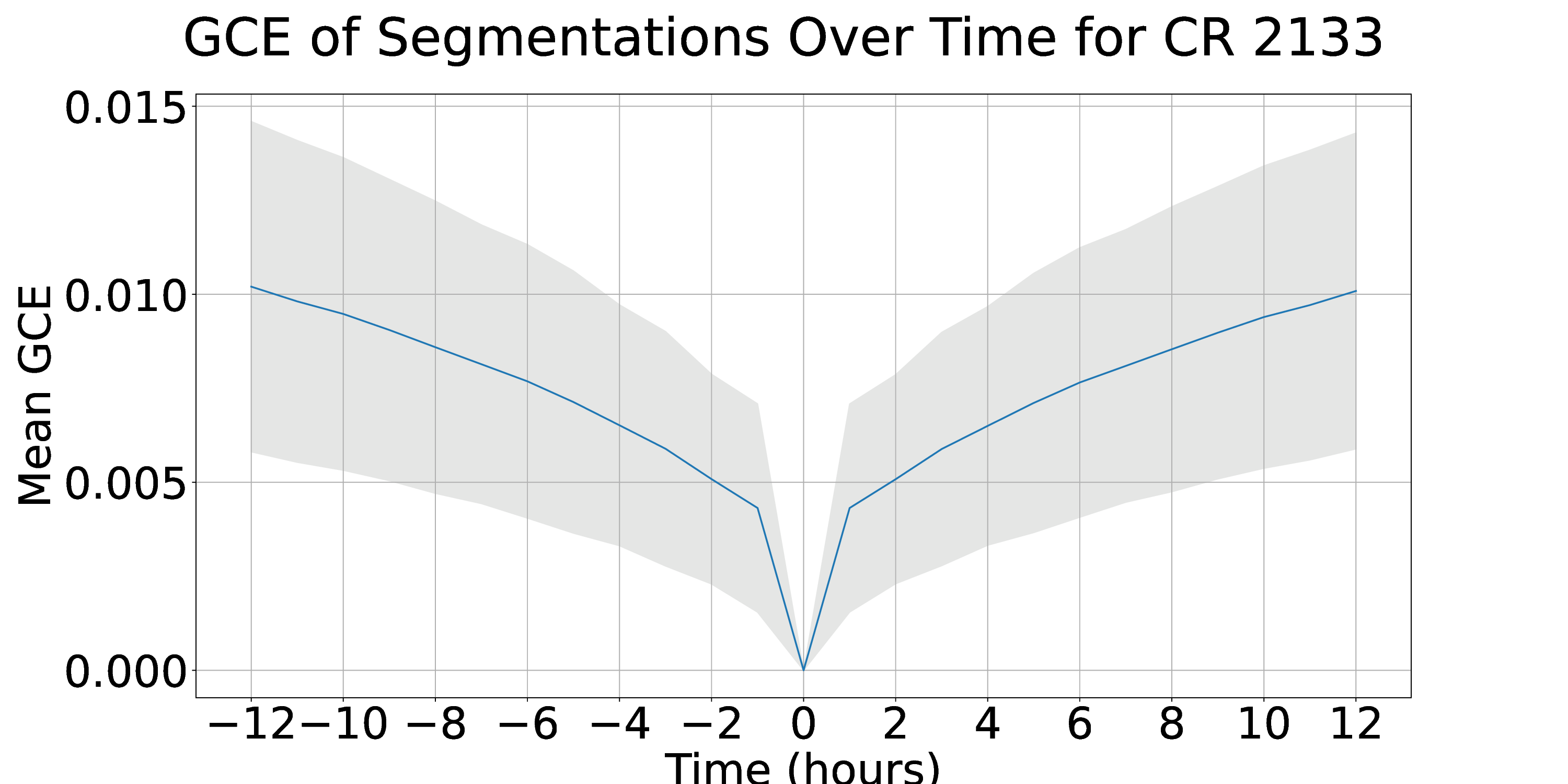}}~~
    \subfloat[Local Consistency Error]{\includegraphics[width=.49\textwidth]{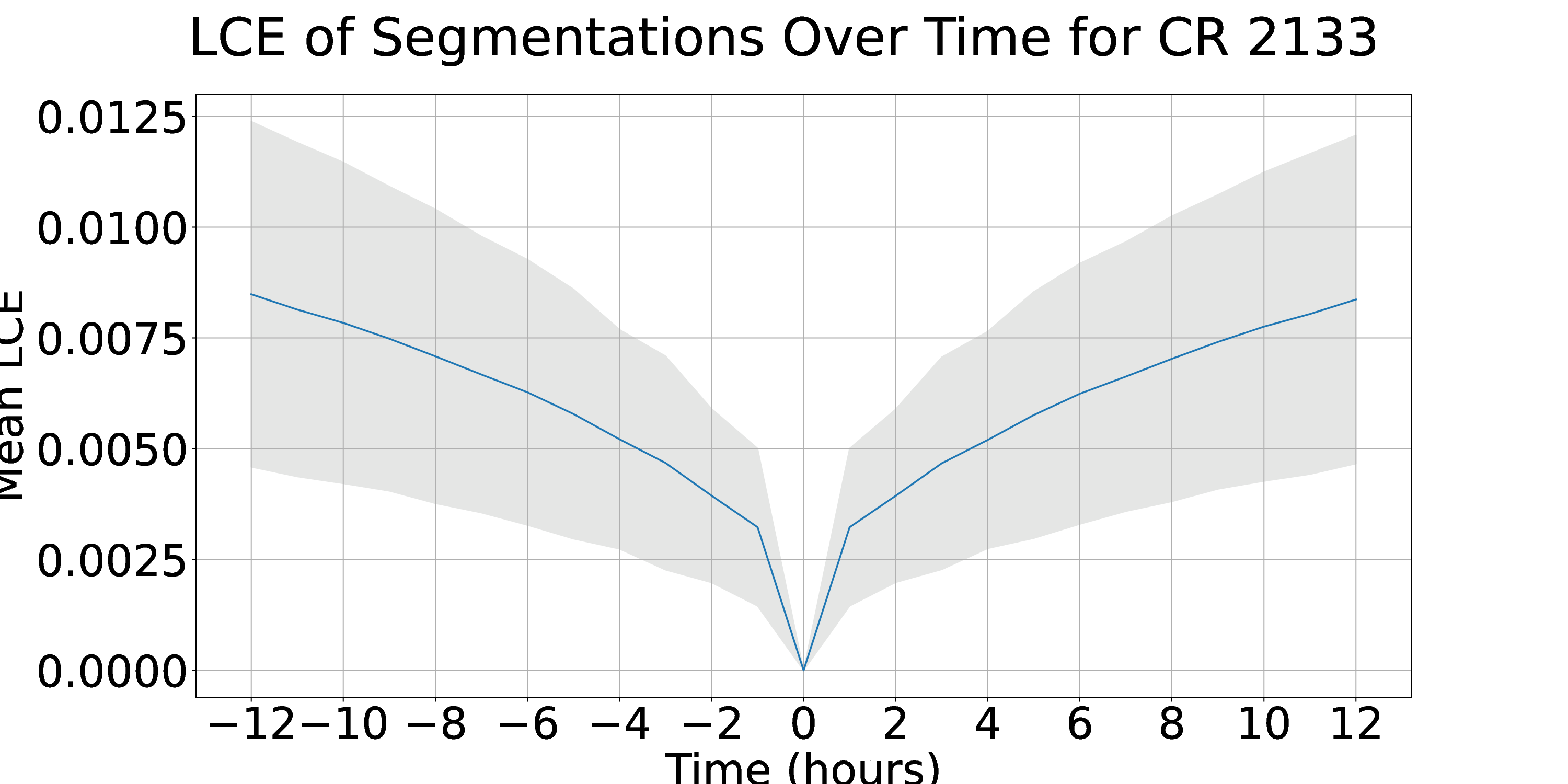}}\\
    \caption{Similarity of segmentation as a function of time across CR 2133. The blue line represents the mean similarity or error between a segmentation and its successor or predecessor as a function of the time difference between the two EUV images while the gray shaded region is $\mu\pm\sigma$ where $\mu$ is the mean and $\sigma$ is the standard deviation.}
    \label{fig:Temporal_CR2133}
\end{figure}

Segmentation similarity as a function of the difference in time for CRs 2099, 2100, and 2101 is presented in Figure \ref{fig:Temporal_CR2099_2101}. Segmentation similarity for CR 2133 is presented in Figure \ref{fig:Temporal_CR2133}. 
In both time frames, segmentation similarity as measured by all metrics decays smoothly as a function of time, with the most dramatic change in similarity occurring within the first hour. In addition to this, a high SSIM and low GCE and LCE are maintained, even with 12 hours difference between segmentations.  These results demonstrate the robustness of the ACWE algorithm across time, indicating that the differences in segmentation are minimal across time scales shorter than those expected for CH evolution.  

\section{Confidence Map Generation from an ACWE Ensemble}
\label{sec:conMap}

\begin{figure}[tp]
    \centering
    \includegraphics[width=\textwidth]{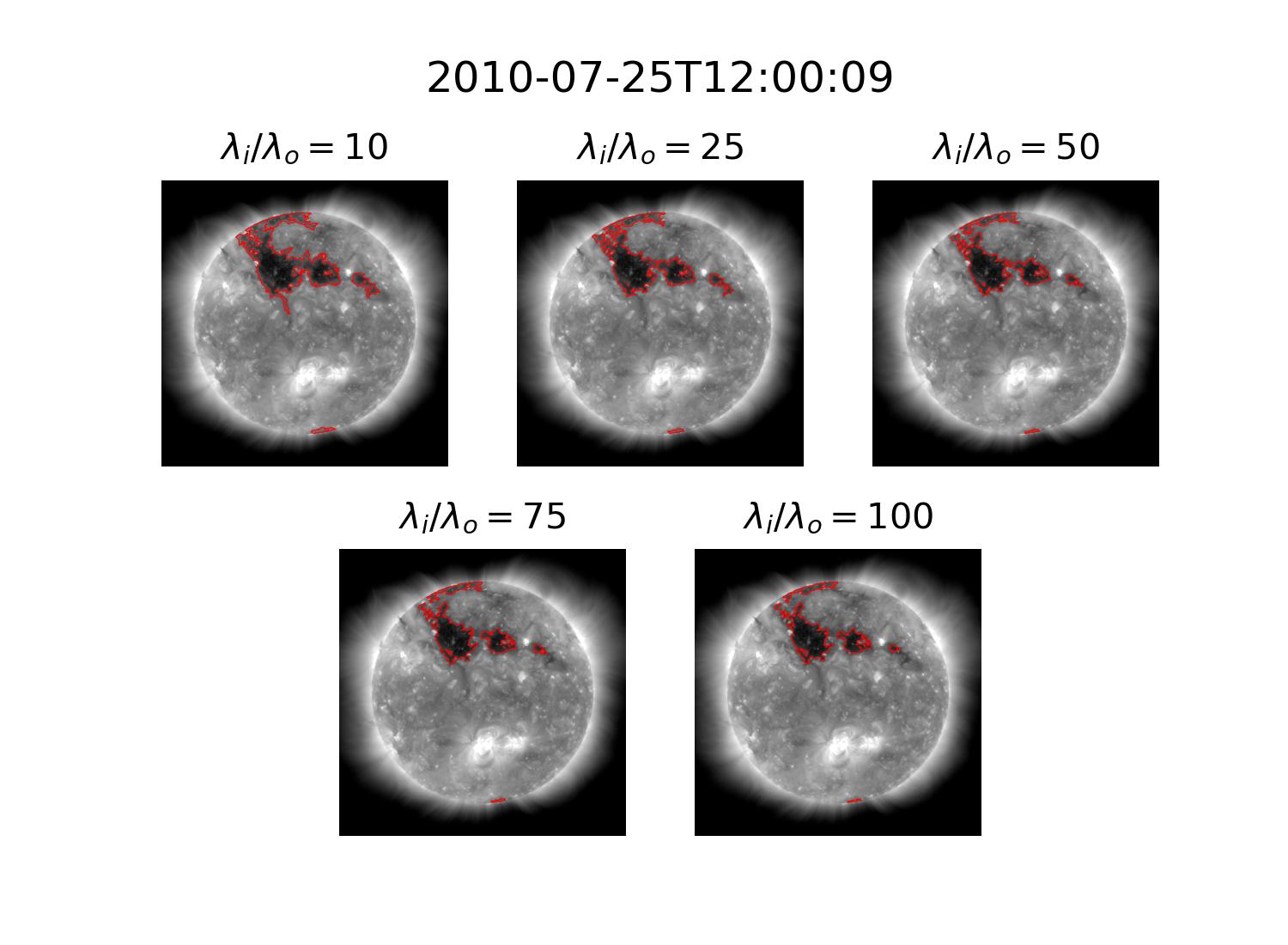}
    \caption{Example of slowly varying segmentation versus homogeneity ratio.}
    \label{fig:Hslow}
\end{figure}

\begin{figure}[tp]
    \centering
    \includegraphics[width=\textwidth]{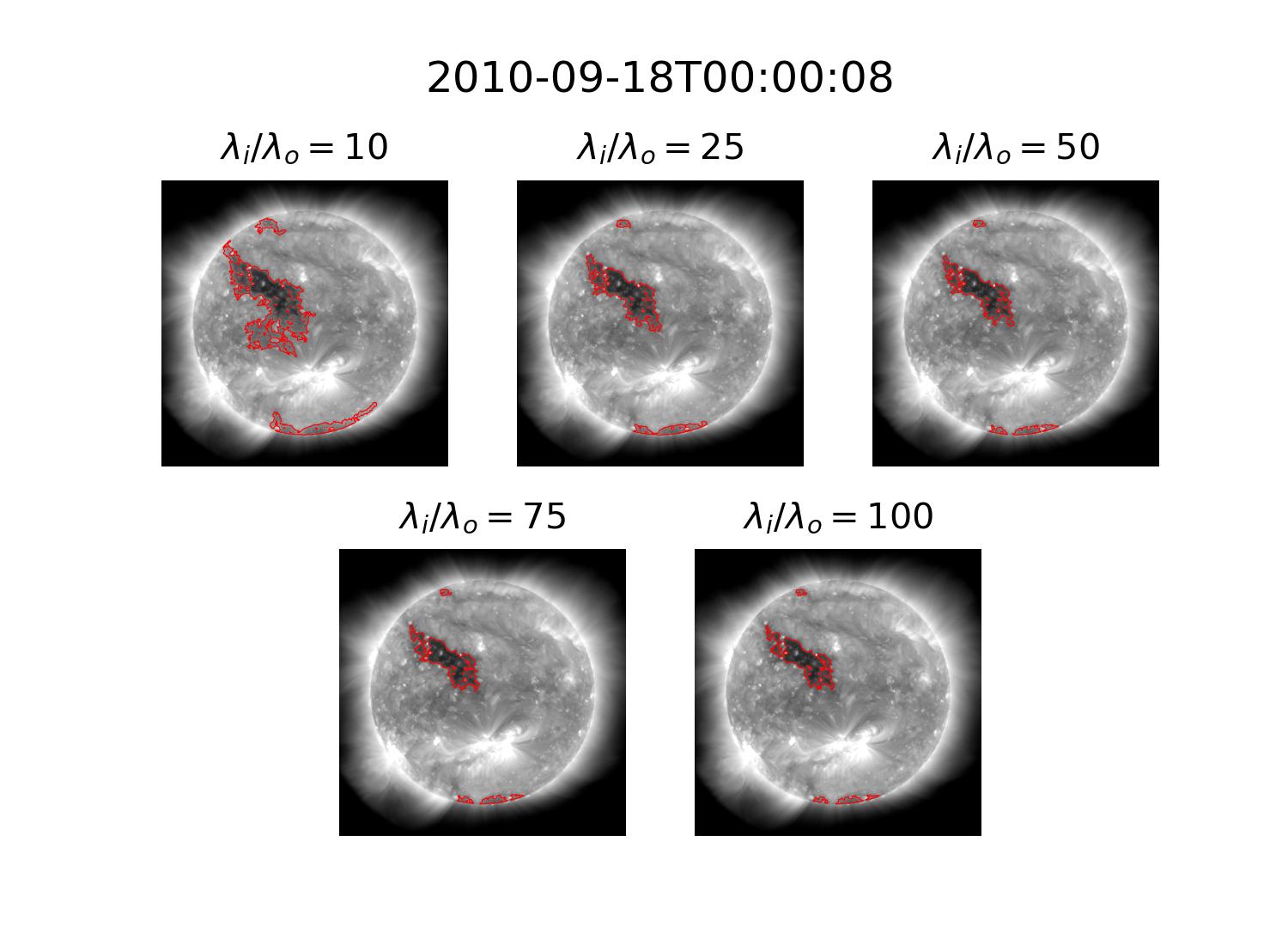}
    \caption{Example of quickly varying segmentation versus homogeneity ratio for low homogeneity ratios.}
    \label{fig:Hfast}
\end{figure}

\begin{figure}[tp]
    \centering
    \includegraphics[width=\textwidth]{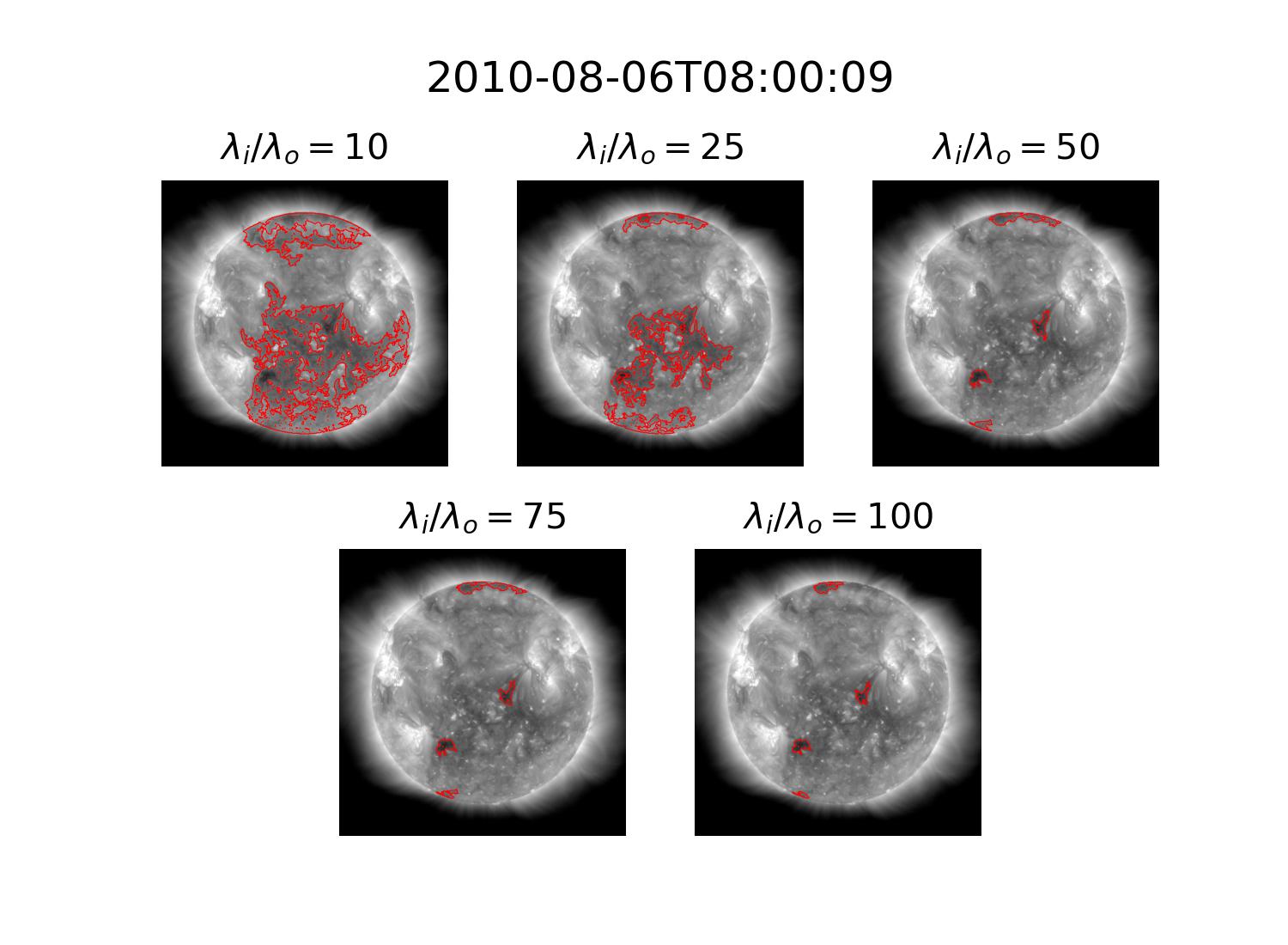}
    \caption{Example of change of target at low homogeneity ratios.}
    \label{fig:Hchange}
\end{figure}

By using region homogeneity to define CHs within a Solar EUV image, ACWE provides the ability to account for intensity variations within a coronal hole region, allowing for a more robust definition of the CHs. This method, however, is affected by all on-disk features, which control the relative homogeneity of the collective “background” or non-CH region.  \cite{boucheron2016segmentation} suggest that homogeneity ratios $\lambda_i/\lambda_o\geq10$ result in “a reasonable segmentation of the CHs and that the segmentation varies very slowly with respect to $\lambda_i/\lambda_o$,” such as the results seen in Figure \ref{fig:Hslow}. While this holds true for the majority of images within the dataset, two other behaviors were noted in a minority of cases. The first behavior, seen in Figure \ref{fig:Hfast}, consists of examples where segmentation area decreases drastically as the homogeneity ratio increases for lower homogeneity ratios. The second behavior, seen in Figure \ref{fig:Hchange} at $\lambda_i/\lambda_o = 25$, consists of cases where some segmentations with small homogeneity ratios, typically $\lambda_i/\lambda_o < 50$, exclude the darkest portions of the initial seed. These segmentations contain large portions of QS, suggesting that the algorithm captured enough QS to change the target of ACWE from CH to QS. Both the case where lower homogeneity ratios result in large area and where a change of target occurs appear in groups throughout the dataset with small time windows between cases. This further highlights the effect that the makeup of the non-CH region has on the final segmentation.

\subsection{Development of Confidence Maps}
\label{sec:makeconmap}

\begin{figure}
    \centering
    \subfloat[Example of slowly varying segmentation versus homogeneity ratio.]{\includegraphics[trim=0in 0in 0in 0in,clip,width=\textwidth]{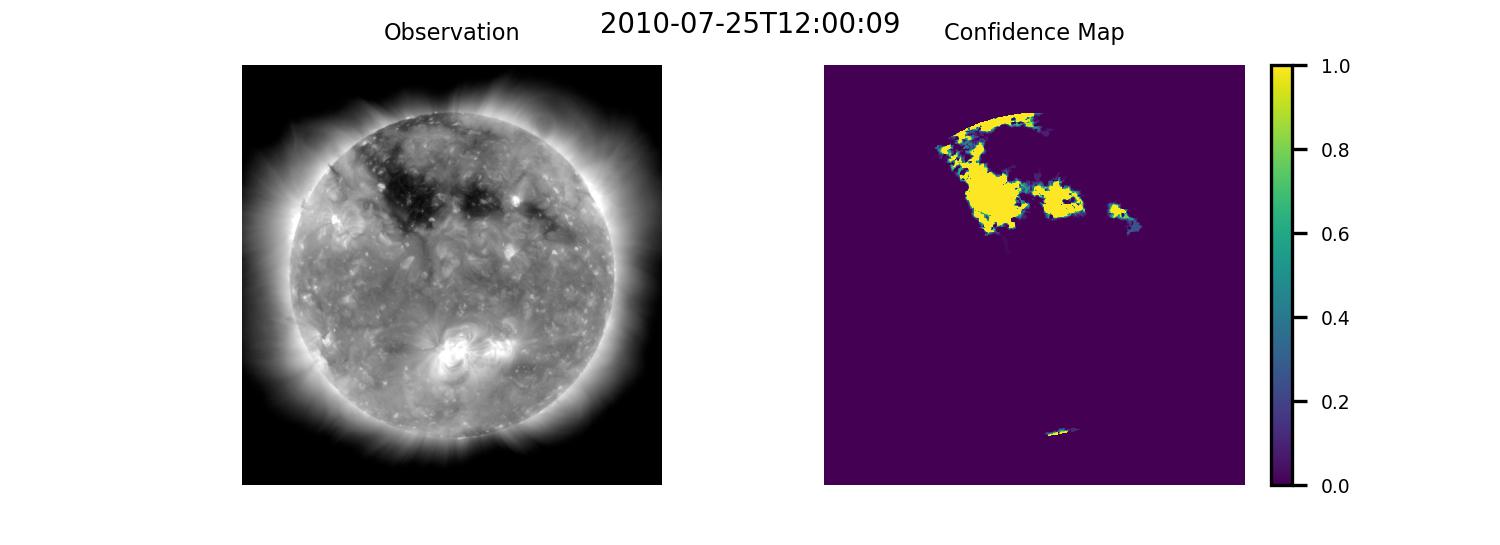}\label{fig:CMslow}}\\
    \subfloat[Example of quickly varying segmentation versus homogeneity ratio for low homogeneity ratios.]{\includegraphics[trim=0in 0in 0in 0in,clip,width=\textwidth]{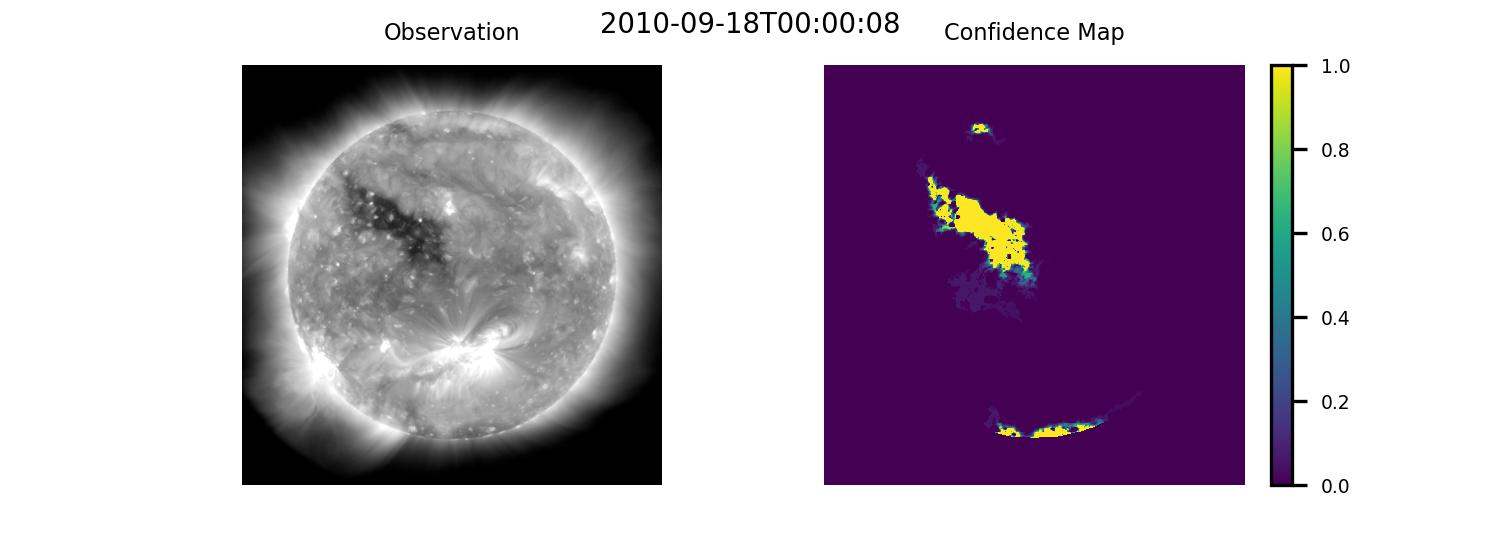}\label{fig:CMfast}}\\
    \subfloat[Example of change of target at low homogeneity ratios.]{\includegraphics[trim=0in 0in 0in 0in,clip,width=\textwidth]{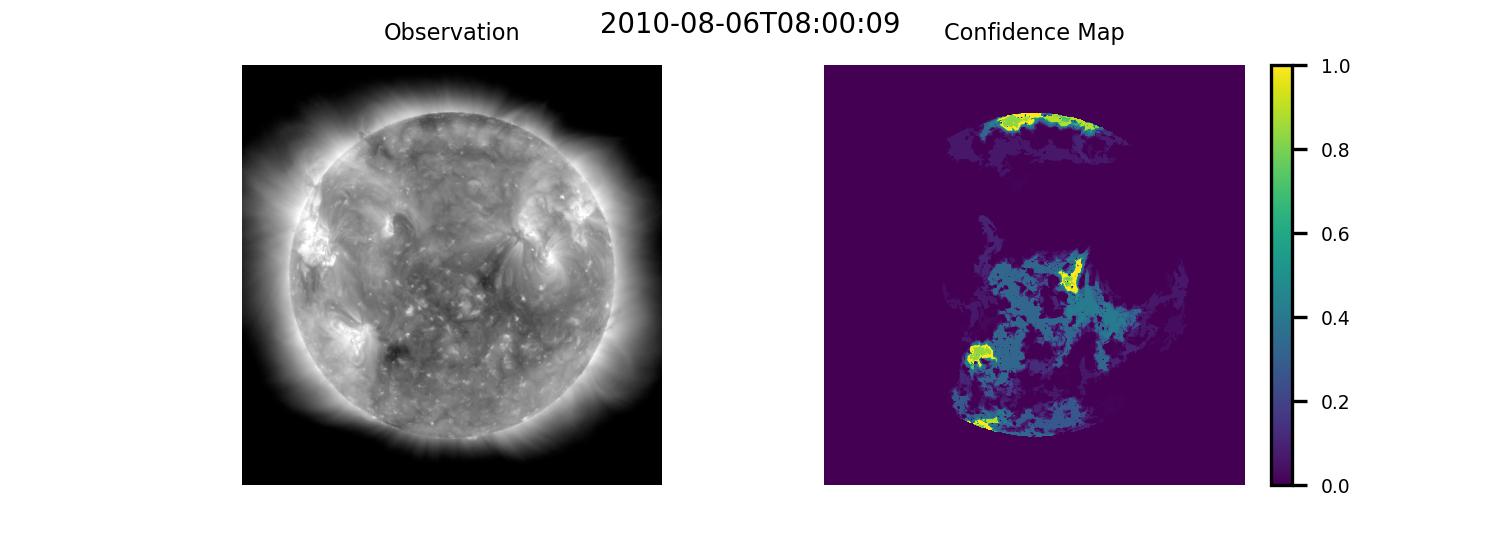}\label{fig:CMchange}}\\
    \caption{Normalized summation of segmentations at homogeneity ratios $\lambda_i/\lambda_o\in[10,100]$. The color bar to the right of each map shows the proportion of the segmentations in which a region in the original image was identified by the segmentation group as belonging to a CH. A value of 1 (yellow) indicates that all segmentations identified that region as a CH region. A value of 0 (purple) indicates that none of the segmentations identified the region as a CH region.}
    \label{fig:ConMapOld}
\end{figure}

\begin{figure}
    \centering
    \includegraphics[trim=.25in 0in .25in .0in,clip,width=\textwidth]{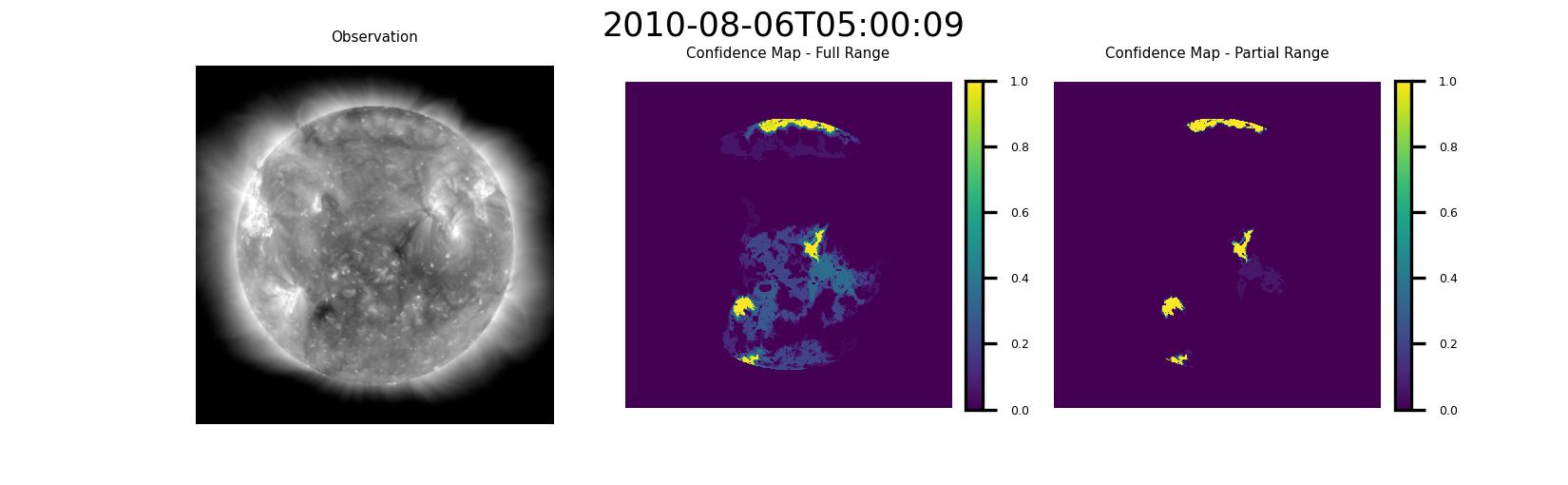}
    \caption{Normalized summation of segmentations at homogeneity ratios $\lambda_i/\lambda_o\in[36,100]$ for the segmentation group in Figure \ref{fig:CMchange}. The color bar to the right of each map shows the proportion of the segmentations in which a region in the original image was identified by the segmentation group as belonging to a CH. A value of 1 (yellow) indicates that all segmentations identified that region as a CH region. A value of 0 (purple) indicates that none of the segmentations identified the region as a CH region.}
    \label{fig:ConMapOldSpecialRange}
\end{figure}

The effects of homogeneity on final segmentation were studied by generating multiple ACWE segmentations at one-eighth spatial resolution ($512\times512$ pixels) for each EUV image. Following the procedure outlined in Section~\ref{sec:spatial}, correction for limb brightening and seeding with $\alpha=0.3$ were performed after resizing the EUV image. Each segmentation was generated independently, starting at the initial seed, and evolving to one $\lambda_i/\lambda_o$ ratio in the range $[10,100]$ (with $\mu=0$ in all cases). This study found that for any collection of segmentations generated from the same EUV image (hereafter referred to as a segmentation group), including those which show a change of target at lower homogeneity ratios, the majority of segmentations converge to a region of high homogeneity surrounding the initial seed. This can be seen in Figure \ref{fig:ConMapOld}, which shows a normalized summation of all segmentations at homogeneity ratios $\lambda_i/\lambda_o$ in the range $[10,100]$ for the observations in Figure \ref{fig:Hslow}, Figure \ref{fig:Hfast}, and Figure \ref{fig:Hchange}. In addition to this, for change of target cases, aggregating only those segmentations with a homogeneity ratio above the change of target point, results in a behavior that matches the rest of the dataset. This can been seen in Figure \ref{fig:ConMapOldSpecialRange}, which was generated from the same segmentation group as Figure \ref{fig:CMchange}, by eliminating all segmentations with a $\lambda_i/\lambda_o\leq35$, the point where the darkest pixel in the initial seed was no longer present in the final segmentation.

Prior to change of target (if it occurs), regions of higher homogeneity are contained within the regions of lower homogeneity. For this reason, the aggregated maps generated from an ensemble of homogeneity ratios above the change of target point (i.e., Figure \ref{fig:CMslow}, Figure \ref{fig:CMfast}, and Figure \ref{fig:ConMapOldSpecialRange}) can be interpreted as confidence maps wherein the confidence that a region belongs to a CH is directly correlated with the similarity of that region to the core region of that CH. 

The confidence maps herein developed are generated in a two-step process. First, the segmentation group, which contains segmentations for all $\lambda_i/\lambda_o$ ratios in the range $[10,100]$, is created  from the one-eighth scale observations. Second, the confidence map is generated from the ensemble by summing the individual maps within the segmentation group and normalizing the result by the total number of maps within that group.

\subsection{Optimization of Ensemble Generation}
\label{sec:levelset}

The generation of segmentation groups can be made more computationally efficient by relying on the deterministic nature of ACWE evolution. In particular, for a given initial seed and input image, the evolution of ACWE will follow a set progression. The stopping criteria, represented by the $\lambda_i/\lambda_o$ ratios, identify points along this evolution where the user wishes to record the results of this process. To generate the segmentation group using this optimized approach, the 193~{\AA} image is resized to one-eighth resolution, correction for limb brightening is applied and the initial seed (using $\alpha=0.3$) is generated. ACWE is then used to evolve the initial seed to the largest $\lambda_i/\lambda_o$ ratio and the resulting segmentation is saved. Evolution then continues, from the current segmentation, targeting the next largest $\lambda_i/\lambda_o$ ratio. This is repeated, recording the result when each $\lambda_i/\lambda_o$ ratio is reached, until a segmentation exists for each $\lambda_i/\lambda_o$ ratio in the range [10,100]. 

\begin{figure}
    \centering
    \subfloat[Example of slowly varying segmentation versus homogeneity ratio.]{\includegraphics[trim=.25in 0in .25in 0in,clip,width=\textwidth]{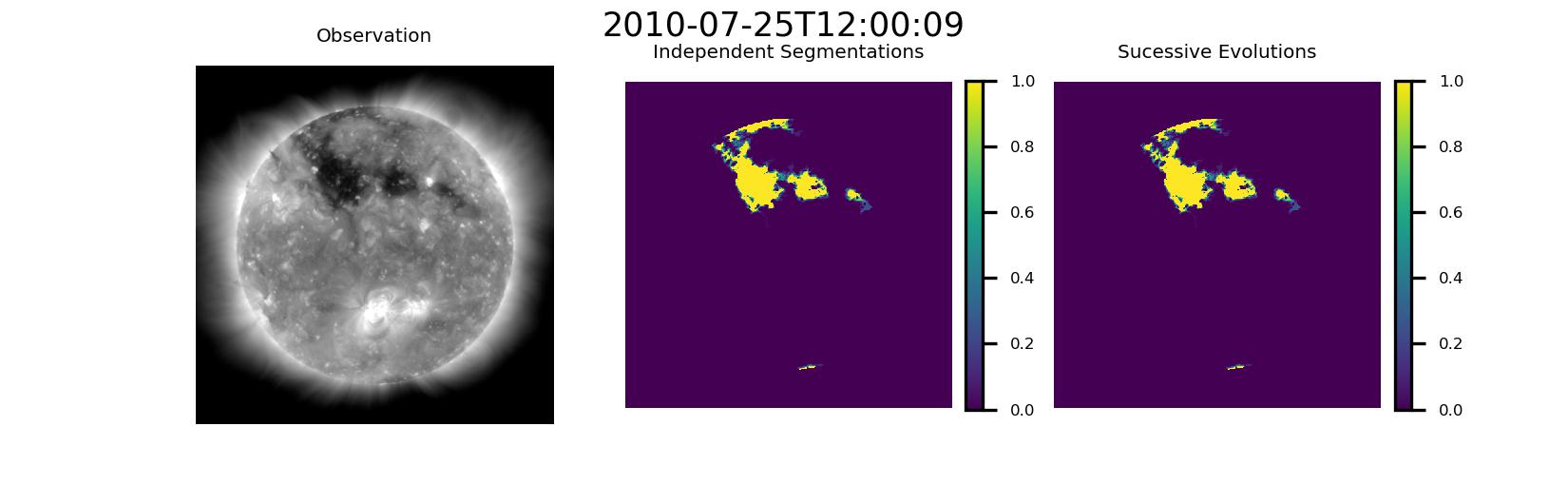}}\\
    \subfloat[Example of quickly varying segmentation versus homogeneity ratio for low homogeneity ratios.]{\includegraphics[trim=.25in 0in .25in 0in,clip,width=\textwidth]{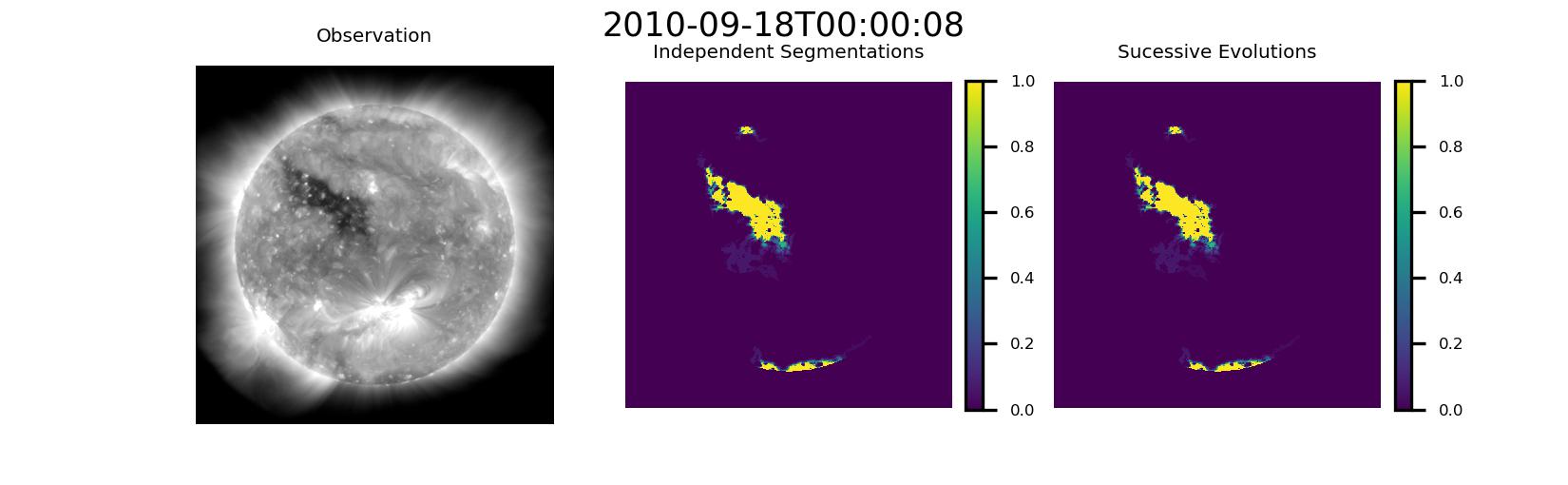}}\\
    \subfloat[Example of change of target at low homogeneity ratios.]{\includegraphics[trim=.25in 0in .25in 0in,clip,width=\textwidth]{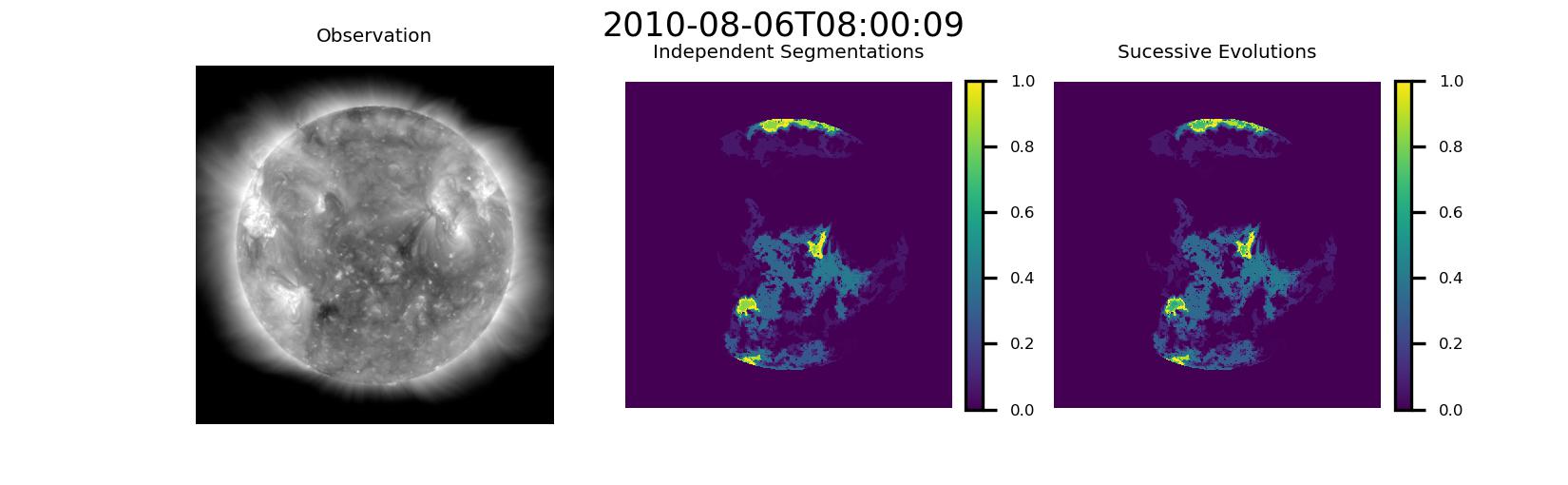}\label{fig:CompareChange}}\\
    \caption{Comparison of normalized summation of segmentations at homogeneity ratios $\lambda_i/\lambda_o\in[10,100]$ generated independently (center) and through the evolution of a single segmentation (right). The color bar to the right of each map shows the proportion of the segmentations in which a region in the original image was identified by the segmentation group as belonging to a CH. A value of 1 (yellow) indicates that all segmentations identified that region as a CH region. A value of 0 (purple) indicates that none of the segmentations identified the region as a CH region.}
    \label{fig:ConMapCompair}
\end{figure}

Figure \ref{fig:ConMapCompair} shows confidence maps generated from independently produced segmentations (center) and confidence maps generated from successive evolutions of ACWE (right) for the examples in Figure \ref{fig:ConMapOld}. These results show that maps generated via both methods are nearly identical in behavior, including in the case of change of target (see Figure \ref{fig:CompareChange}). It should be noted that, for all images within the dataset, the SSIM between confidence maps generated from independently produced segmentations and confidence maps generated from a successive evolutions of ACWE was 1. In addition, a comparison of maps generated via the two methods yielded high wIOU, low GCE, and low LCE. The mean and standard deviation for each metric, organized by Carrington Rotation, are presented in Table \ref{tab:ConnMapMethods}.

\begin{table}[t]
    \centering
    \caption{Comparison of confidence map methods, organized by Carrington Rotation}
    \label{tab:ConnMapMethods}
    \begin{tabular}{ccccc}
    \hline
    \textbf{CR} & \textbf{wIOU}  & \textbf{SSIM}  & \textbf{GCE}  & \textbf{LCE} \\\hline
    2099 & $0.9826\pm0.0207$ & $1.0000$ & $0.0096\pm0.0088$ & $0.0079\pm0.0065$\\
    2100 & $0.9890\pm0.0055$& $1.0000$ & $0.0073\pm0.0037$ & $0.0064\pm0.0033$\\
    2101 & $0.9867\pm0.0089$& $1.0000$ & $0.0065\pm0.0049$ & $0.0055\pm0.0042$\\
    2133 & $0.9879\pm0.0088$& $1.0000$ & $0.0083\pm0.0056$ & $0.0071\pm0.0048$\\\hline
    \end{tabular}
\end{table}

The differences between the independent and successive evolution confidence map as measured by wIOU, GCE, and LCE (see Table \ref{tab:ConnMapMethods}) can be explained by the fact that ACWE cannot define a boundary with sub-pixel accuracy. When the true boundary between foreground and background for a given $\lambda_i/\lambda_o$ ratio lies within a pixel instead of on a pixel boundary, this pixel will continually alternate between foreground and background as ACWE converges to the final solution. To account for this, a stopping criterion from \cite{boucheron2016segmentation} was used. This criterion dictates that ACWE will halt evolution and return a segmentation map whenever the only pixels that evolve between iterations are pixels along the boundary that alternate between foreground and background. 

By evolving from the previous segmentation instead of evolving from the initial seed, the optimized implementation changes (reduces) the number of iterations needed to converge to a solution, thus changing the point at which the stopping criteria is encountered. Over CR 2099 the successive evolutions method generated segmentation groups an average of $2.9209\pm0.2622$ times faster than the independent method. On the aforementioned Intel Core i7-8700K, this reduced computation time from anywhere between $136.431$ and $369.381$ s (with a mean of $194.597$~{s}) to between $45.266$ and $100.913$ s (with a mean of $66.527$~{s}) depending on the segmentation group.

\subsection{A Further Study of Change of Target}
\label{sec:filterconmaps}

Automated identification of change of target cases remains an area of open research. The current method of identifying change of target consists of calculating the change in the percent of the initial seed present in final segmentation as a function of $\lambda_i/\lambda_o$ ratio. If the drop in the percent of the seed present is $\geq5\%$ from one $\lambda_i/\lambda_o$ ratio to the next smallest $\lambda_i/\lambda_o$ ratio, then it is assumed that the segmentation with the smaller $\lambda_i/\lambda_o$ ratio exhibits signs of change of target.  When this occurs, that segmentation as well as all other segmentations with a smaller $\lambda_i/\lambda_o$ ratio, are omitted from the segmentation group during summation and normalization. 

\begin{figure}
    \centering
    \subfloat[AIA 193~{\AA} image (left) and initial seed (right).]{\includegraphics[trim=0in .2in 0in 0in,clip,width=.66\textwidth]{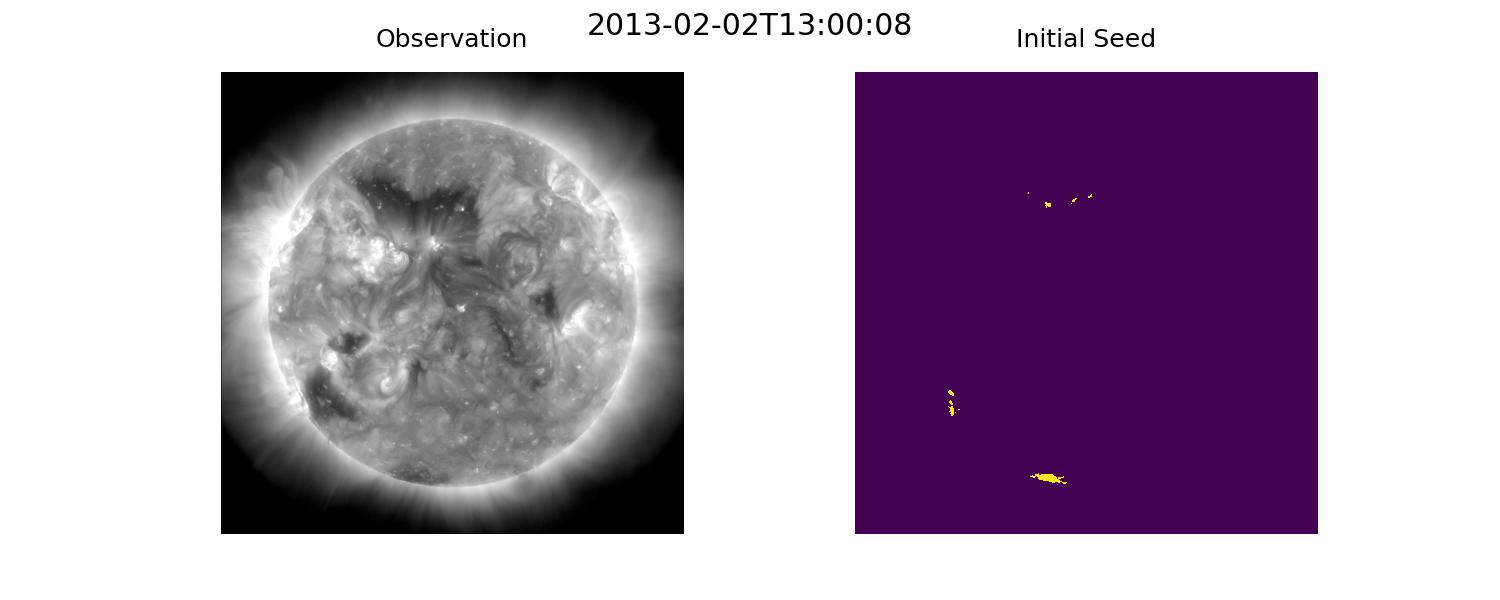}}\\
    \subfloat[Confidence map built from full ensemble (left) and using change of target criterion (right). Since change of target did not occur, both maps are identical.]{\includegraphics[trim=0in .2in 0in 0in,clip,width=.63\textwidth]{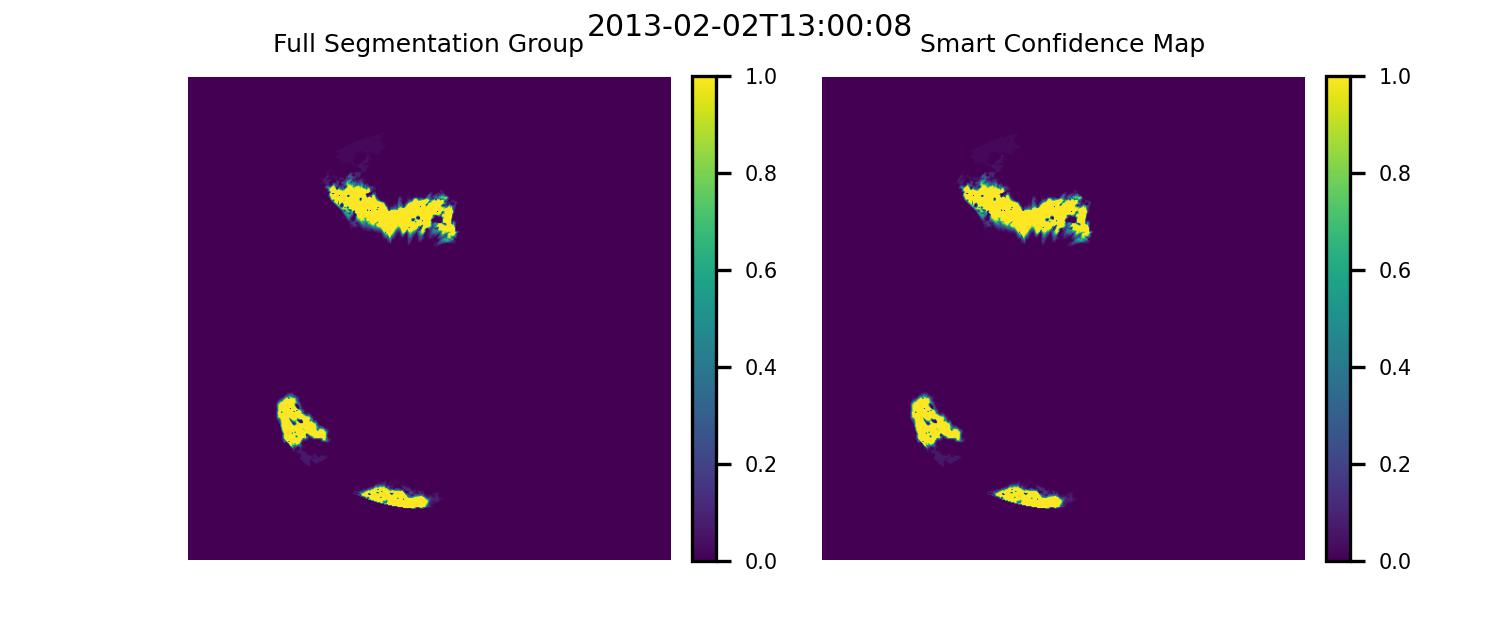}}\\
    \subfloat[Intensity of segmentations compared to intensity of initial seed, normalized by mean intensity of seed.  Note the smooth variation in the segmentation intensities versus homogeneity ratio, indicating the absence of a change of target.]{\includegraphics[trim=4in 0in 4in 0in,clip,width=0.86\textwidth]{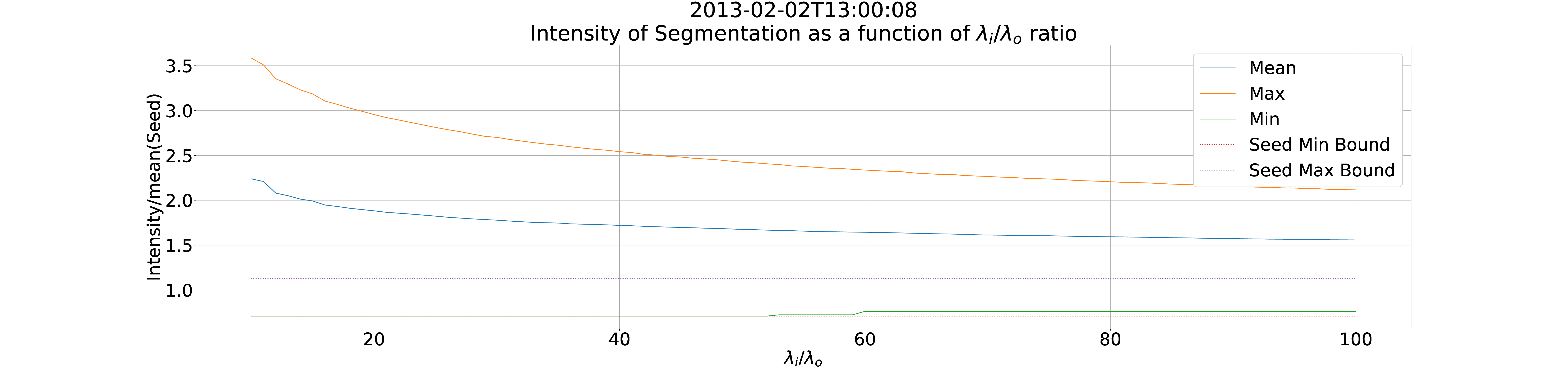}\label{fig:overfitIntenisty}}\\
    \subfloat[Area of initial seed retained in final segmentation.  A small portion of the initial seed is not included even for the highest homogeneity ratios, indicating overfitting, but not a change of target.]{\includegraphics[trim=4in 0in 4in 0in,clip,width=0.86\textwidth]{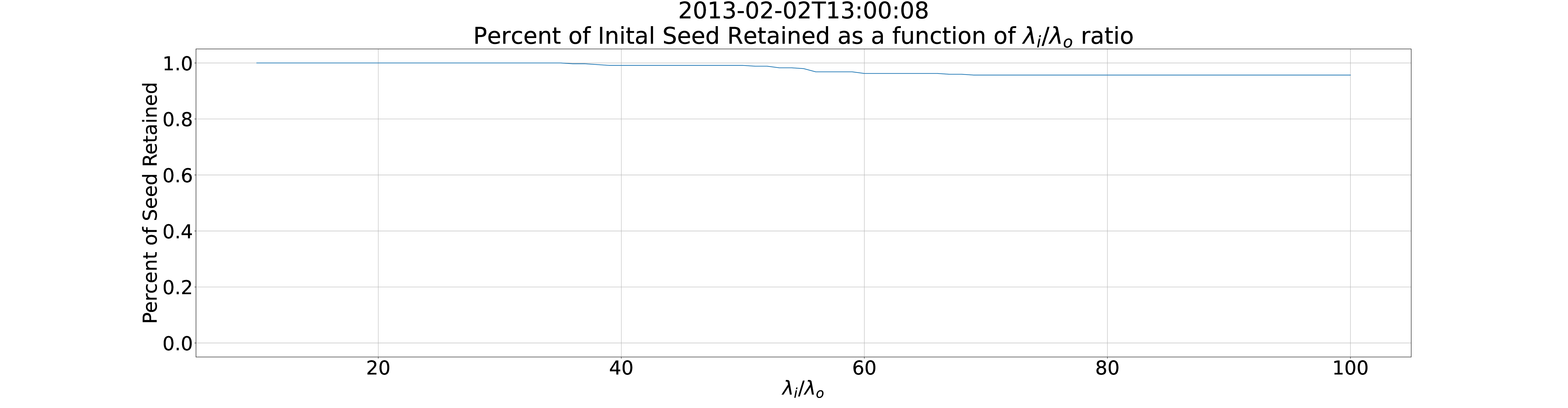}\label{fig:overfitArea}}
    \caption{Example of overfitting at high $\lambda_i/\lambda_o$. The two graphs show that absence of the darkest part of the initial seed is limited to higher homogeneity ratios, likely due to the $\lambda_i/\lambda_o$ ratio being too stringent for the image. Note that the proposed method for identifying change of target was able to identify that the absence of a portion of the initial seed was solely due to overfit and therefore did not exclude any segmentations when generating the confidence map.}
    \label{fig:overfit}
\end{figure}

\begin{figure}
    \centering
    \subfloat[AIA 193~{\AA} image (left) and initial seed (right).]{\includegraphics[trim=0in .2in 0in 0in,clip,width=.65\textwidth]{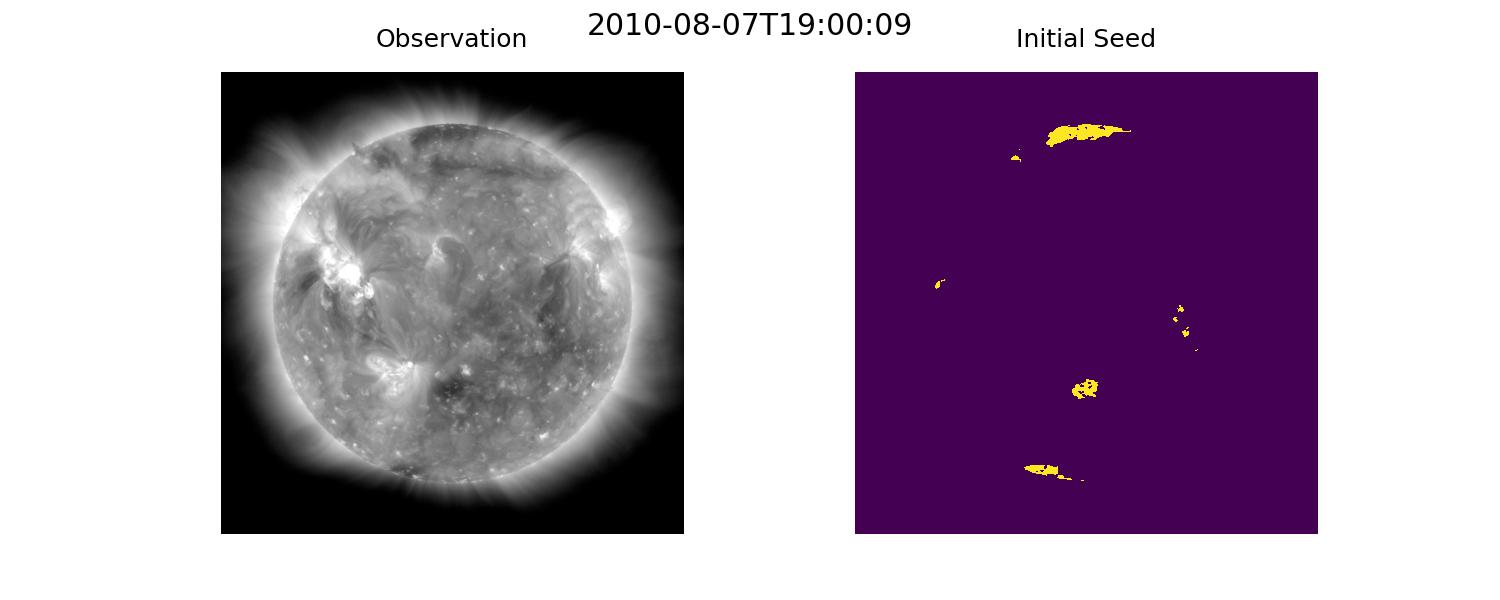}}\\
    \subfloat[Confidence map built from full ensemble (left) and confidence map built using change of target criterion (right).]{\includegraphics[trim=0in .2in 0in 0in,clip,width=.62\textwidth]{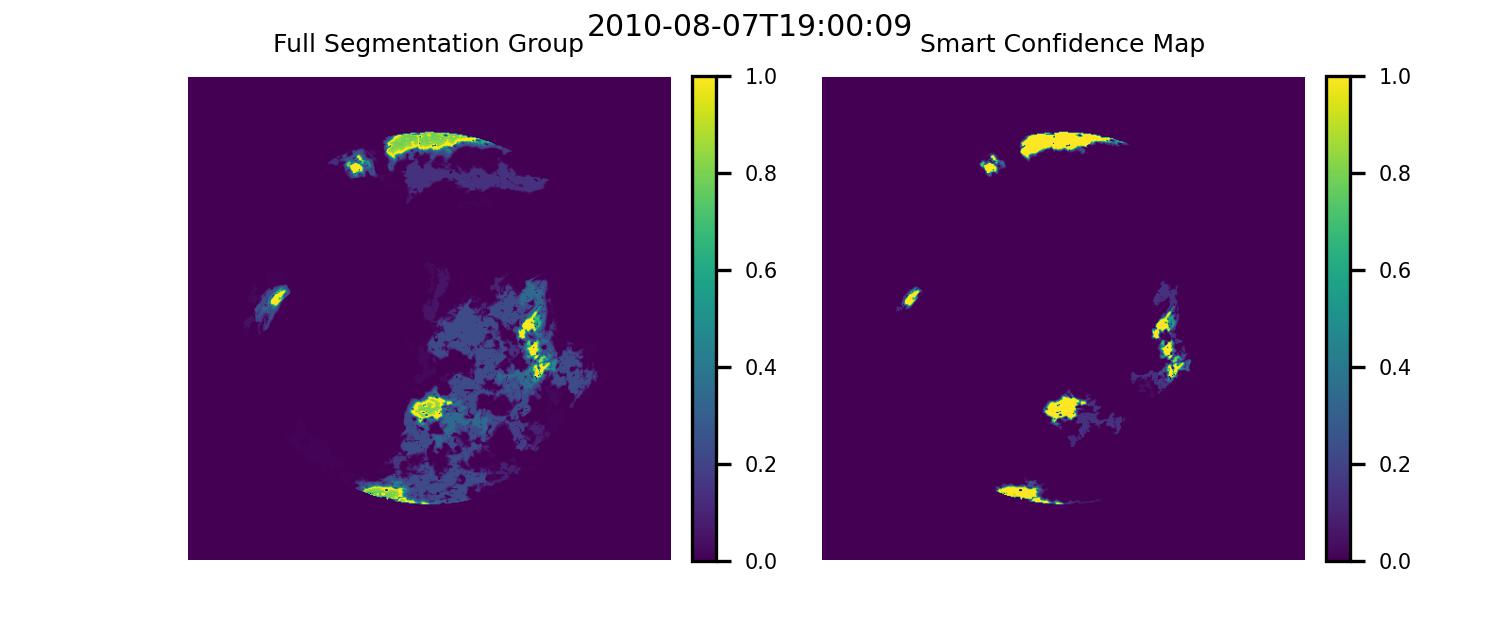}}\\
    \subfloat[Intensity of segmentations compared to intensity of initial seed, normalized by mean intensity of seed. Note the sudden jumps in minimum and mean intensities versus homogeneity ratio, a hallmark of change of target. Note also that the minimum intensity of the segmentation is always larger than the minimum intensity of the initial seed, indicating that overfitting resulted in the removal of the darkest part of the initial seed before change of target occurred.]{\includegraphics[trim=4in 0in 4in 0in,clip,width=0.85\textwidth]{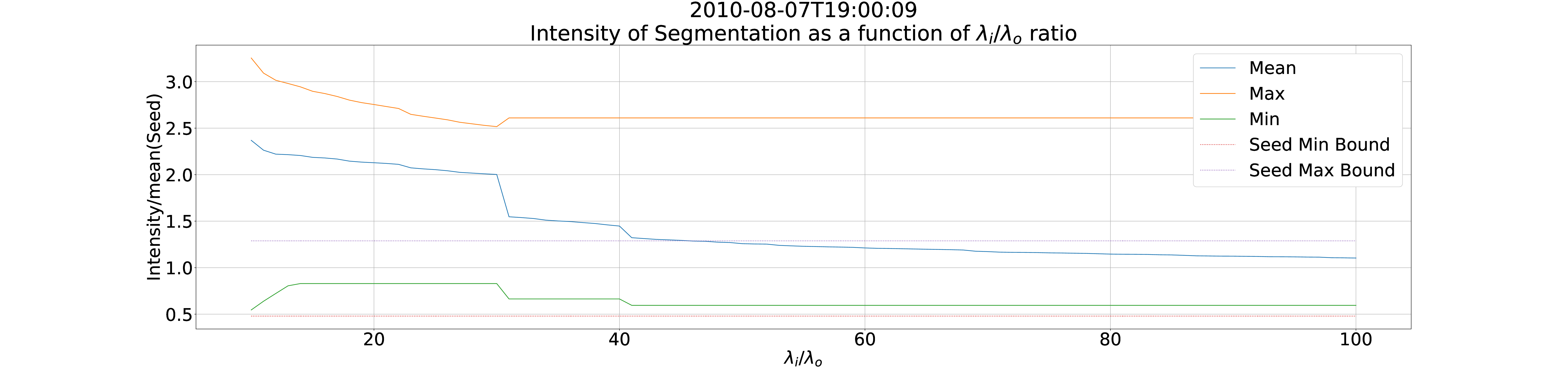}\label{fig:overfitChangeIntenisty}}\\
    \subfloat[Area of initial seed retained in final segmentation.  Note the sudden decrease in area with decreasing homogeneity ratio, a hallmark of change of target.]{\includegraphics[trim=4in 0in 4in 0in,clip,width=0.85\textwidth]{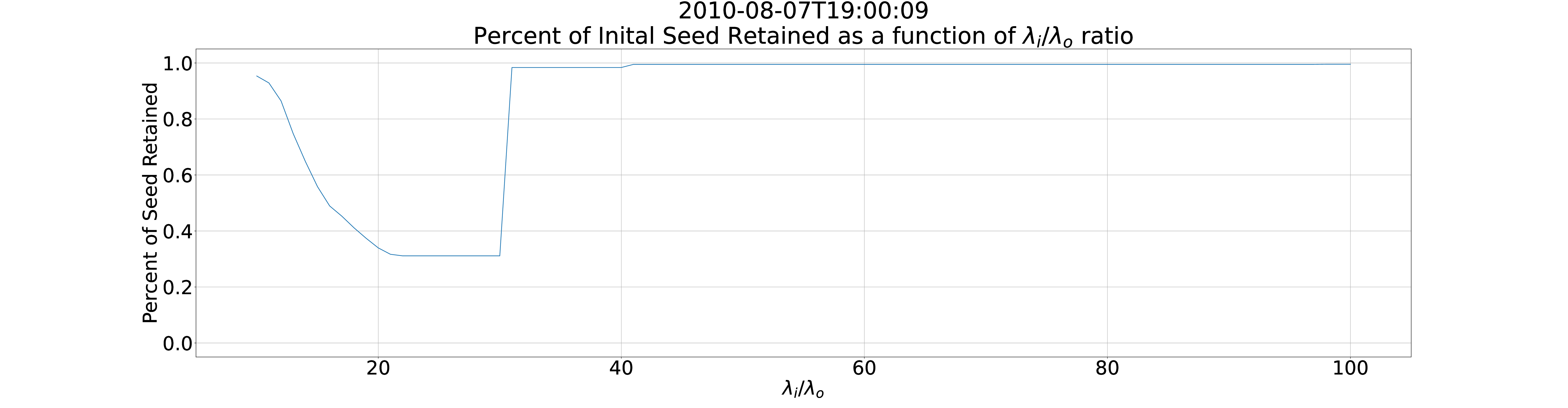}\label{fig:overfitChangeArea}}
    \caption{Example of overfitting and change of target within the same segmentation group. Note that the proposed method for identifying change of target can identify and retain the overfit region while removing the effects of change of target.}
    \label{fig:overfitAndChange}
\end{figure}

The percent of the initial seed present in the segmentation with the largest $\lambda_i/\lambda_o$ ratio is used as a starting value for this process. This ensures that in the case of overfitting (high $\lambda_i/\lambda_o$ ratios causing ACWE to open a hole in the darkest part of the initial seed, even when still detecting CHs), the absence of a portion of the initial seed does not result in an erroneous report of change of target.  Similarly, in the case that the seeding process includes a portion of the QS which is subsequently removed in the evolution of the contour, the absence of a portion of the initial seed does not result in an erroneous report of change of target.  Without this concession, the absence of a portion of the initial seed at the largest homogeneity ratio would result in the entire ensemble being dismissed, preventing the generation of a confidence map.  Figure~\ref{fig:overfit} illustrates an example of this scenario containing overfitting but no change of target.  On the other hand, Figure~\ref{fig:overfitAndChange} illustrates an example of both overfitting for high homogeneity ratios and also a change of target at lower homogeneity ratios.

Manual exploration of the full dataset revealed 125 cases of change of target, of which 96 cases were identified via this method. An additional 14 cases of change of target not identified by the human observer were also found via this method. The total number of cases of change of target identified via either method (139 cases) constitute $<5.63\%$ of the total dataset. The change of target cases identified by human observer and not identified by the automated method (29 cases) constitute $<1.22\%$ of the total dataset.  While a robust automatic identification of change of target is still under investigation, the proposed criterion appears to perform well for the datasets studied here.

\section{Validation of ACWE Boundaries}
\label{sec:val}

Validation of the confidence maps is performed by calculating the skew of the flux of the underlying magnetic field as a function of confidence value. For this process, evaluation is limited to CHs that appear near disk center to minimize the influence of projection effects \citep{leka2017}.

\subsection{Calculating Skew} 

In order to calculate the skew of the flux of the underling magnetic field, the HMI magnetogram image that corresponds to an AIA observation must be scaled and aligned to match the scale and orientation of the level 1.5 193~{\AA} image used to generate the segmentation group. In addition to this, the confidence map, which was generated at one-eighth spatial resolution, must be resized to match the resolution of the EUV image. The process of aligning the magnetogram is achieved by providing \verb+reproject.reproject_interp+ \citep{repoject_2020} with the HMI magnetogram and the Astropy \citep{astropy2013,astropy2018} world coordinate system (WCS) of the level 1.5 193~{\AA} EUV image (which is preserved in the metadata of the confidence map). To resize the confidence map, each segmentation within the group is upscaled from $512\times 512$ pixels to the original $4096\times4096$ pixel resolution using bi-linear interpolation. Once upscaled the segmentations are combined and normalized.

To calculate the skew of the underlying CH region, CHs are partitioned into CH groups. A CH group is defined as all eight-connected regions with a confidence value $V>0$ that are within 40 pixels (equivalent to 5 pixels at the original one-eighths resolution) of each other. The partitioning process was achieved by generating a binary mask of the upscaled confidence mask wherein all pixels with a confidence value $V>0$ are set to 1. Dilation is applied to this map using the scikit-image \citep{scikit-image} function \verb|skimage.morphology.dilation|, with a $40\times40$ pixel square. Once dilated this map is then labeled using \verb|skimage.measure.label|, and one binary mask is generated for each labeled region. These masks are converted into confidence maps of their respective CH groups by replacing the region of the binary mask with a value of 1 with the confidence values of the corresponding pixels in the original confidence map. 

Once the CH groups are generated, they are are then utilized as masks to identify the corresponding region within the matching HMI magnetogram. Skew is calculated as a function of confidence such that for a confidence value $V$ only regions of the magnetogram that correspond to a region within the CH group with a confidence $\geq V$ are included. The skew of the magnetic flux of the region is defined as 
\begin{equation}
    \gamma=\frac{1}{N}\sum^{N-1}_{i=0}\Bigg(\frac{\Phi_i-\Bar{\Phi}}{\sigma}\Bigg)^3
\end{equation}
where $N$ is the number of pixels within the region, $\Phi_i$ is the magnetic flux of pixel $i$, $\Bar{\Phi}$ is the mean flux of the region, and $\sigma$ is the standard deviation of the flux of the region.

Instead of attempting to mitigate the projection effects in HMI magnetograms by estimating the skew of each region, the skew of the underlying region is calculated twice for each confidence value, once directly on the magnetogram (called the “unweighted skew”), and a second time on a weighted magnetogram (“weighted skew”). For the weighted skew the weight for each pixel of the magnetogram is defined as 
\begin{equation}
    W_i=\Bigg(\frac{z_i}{\max(z)}\Bigg)^3
\end{equation}
where $z_i$ is the $z$ coordinate (Cartesian) of pixel $i$ within the heliocentric frame and $\max(z)$ is the $z$ coordinate of the pixel directly in line with the Sun center and observer. This weighting favors regions near disk center \citep{Weights}, where flux observations are least affected by projection effects \citep{leka2017}.

\subsection{Behavior of Confidence Maps}

\begin{figure}
    \centering
    \subfloat[CH at 02:00:09, in the graph a downward trend ($\downarrow$) indicates increasing unpopularity.]{\includegraphics[trim=0in .13in 0in 0in,clip,width=.8\textwidth]{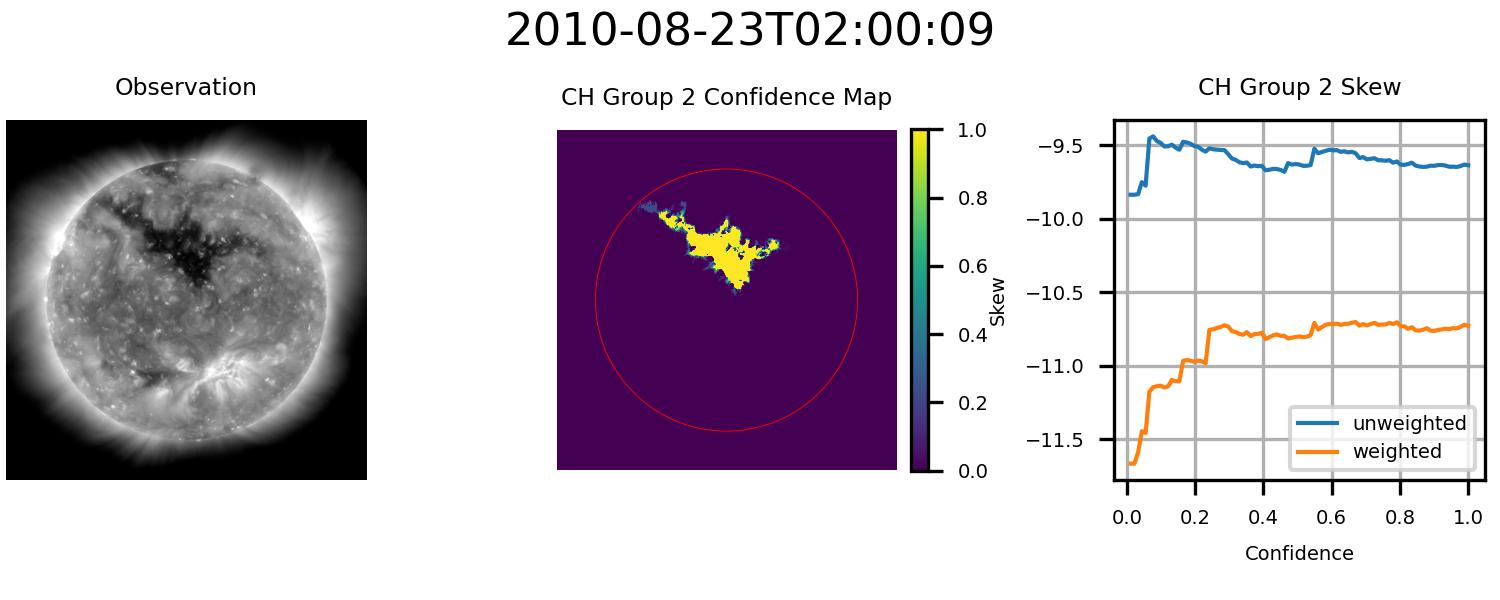}}\\
    \subfloat[CH at 03:00:09, in the graph a downward trend ($\downarrow$) indicates increasing unpopularity.]{\includegraphics[trim=0in 0.13in 0in 0in,clip,width=.8\textwidth]{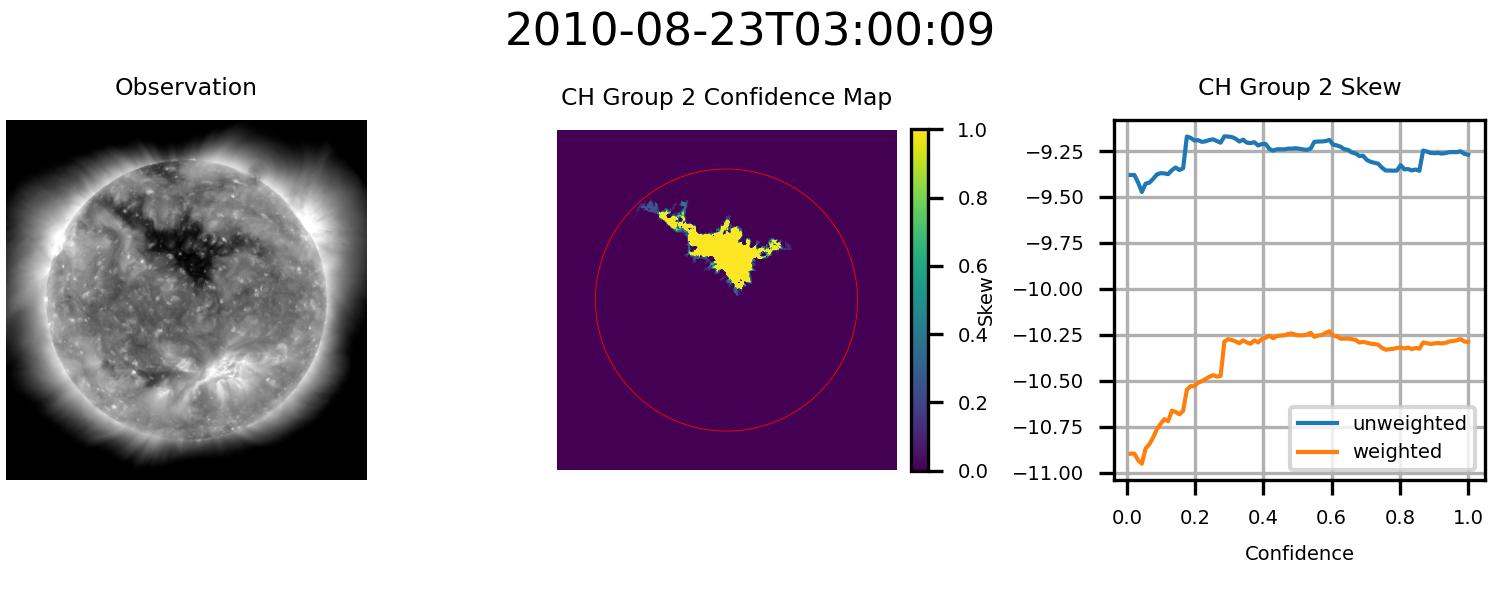}}\\
    \subfloat[CH at 04:00:09, in the graph a downward trend ($\downarrow$) indicates increasing unpopularity.]{\includegraphics[trim=0in 0.13in 0in 0in,clip,width=.8\textwidth]{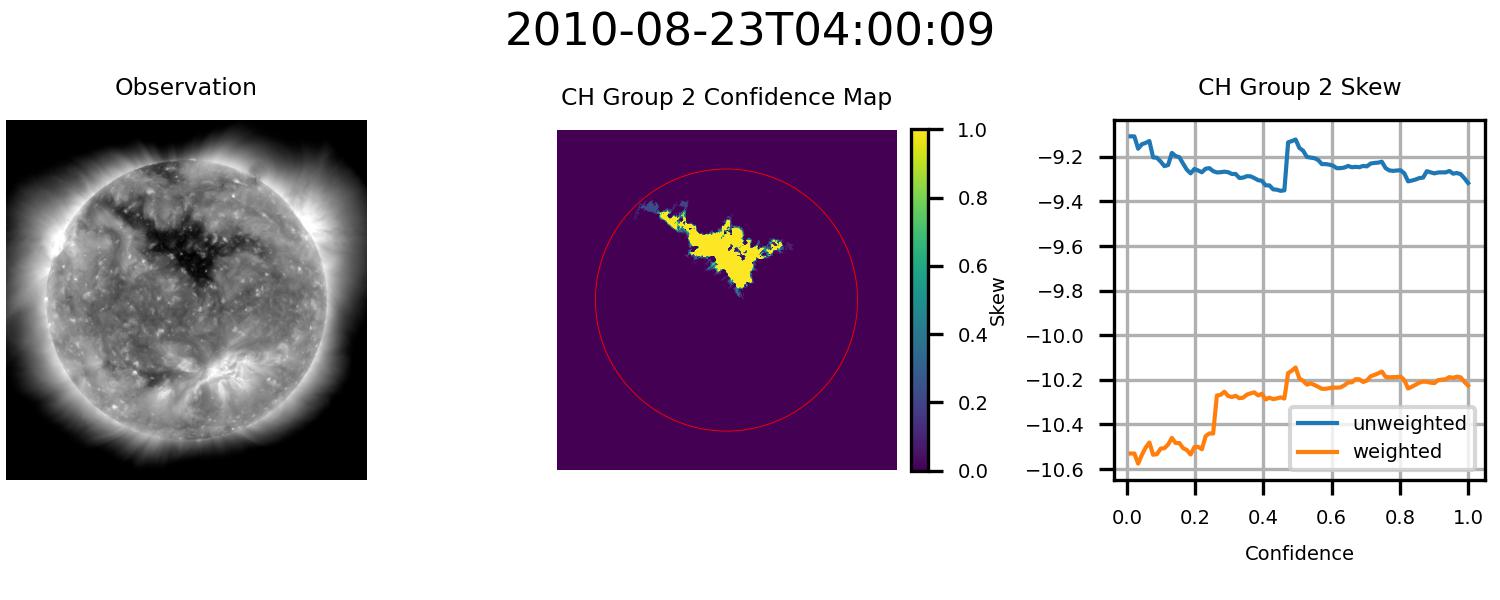}}
    \caption{Skew of the flux of the magnetic field of a centrally located CH in CR2100 for three consecutive hours. From left to right for (a)-(c): AIA 193~{\AA} observation, confidence map (on disk area encircled in red), graph of skew as a function of confidence level.  Note the relative consistency of skew values over the three observations.}
    \label{fig:hourDiff}
\end{figure}

The exact behavior of the skew of the flux of the underlying magnetic field as a function of confidence varied hour by hour as CH groups traversed the center of disk. Figure \ref{fig:hourDiff} provides an example of this variation, providing three observations of the same CH region at one hour intervals. 
As can be noted in Figure \ref{fig:hourDiff}, the range of the skew, both weighted and unweighted, is relatively consistent from one hour to the next.  This further validates the consistency of the ACWE CH segmentation algorithm and the accuracy of the confidence maps in delineating CHs.

\begin{figure}
    \centering
    \subfloat[CH observed in CR2099, in the graph a downward trend ($\downarrow$) indicates increasing unpopularity.]{\includegraphics[trim=0in 0.13in 0in 0in,clip,width=.74\textwidth]{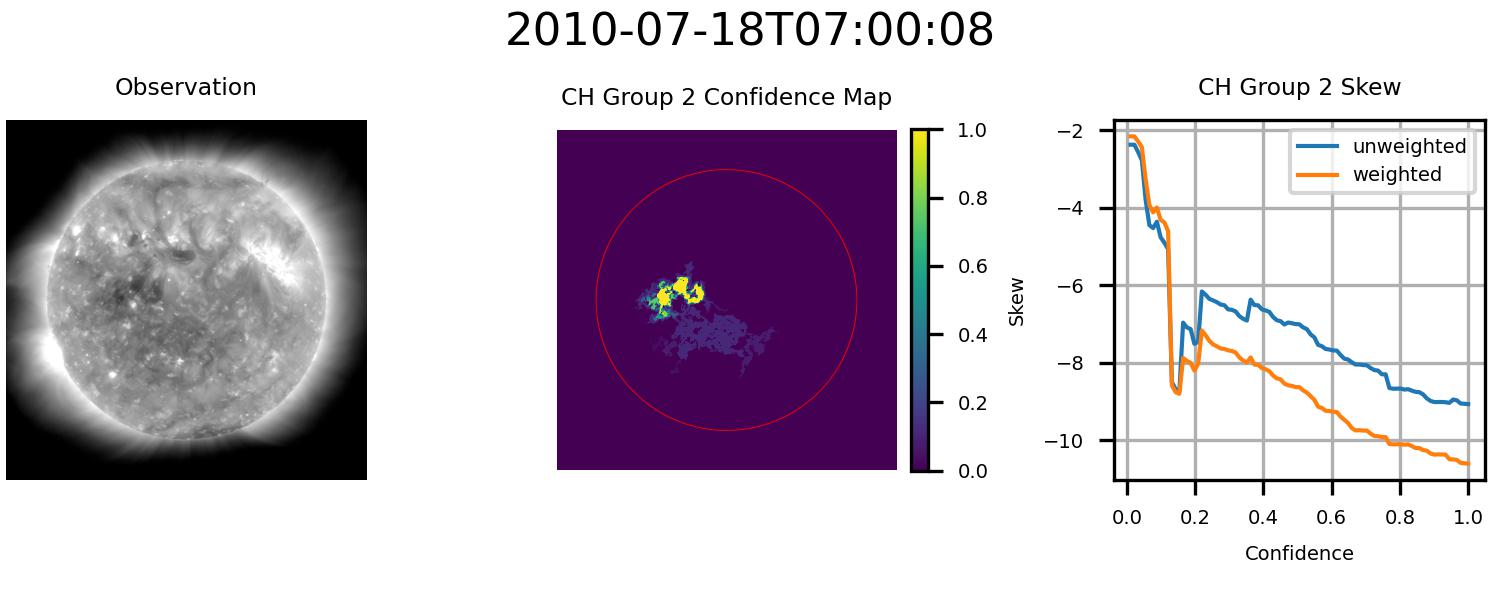}\label{fig:Formed}}\\
    \subfloat[CH observed in CR2100, in the graph an upward trend ($\uparrow$) indicates increasing unpopularity.]{\includegraphics[trim=0in 0.13in 0in 0in,clip,width=.74\textwidth]{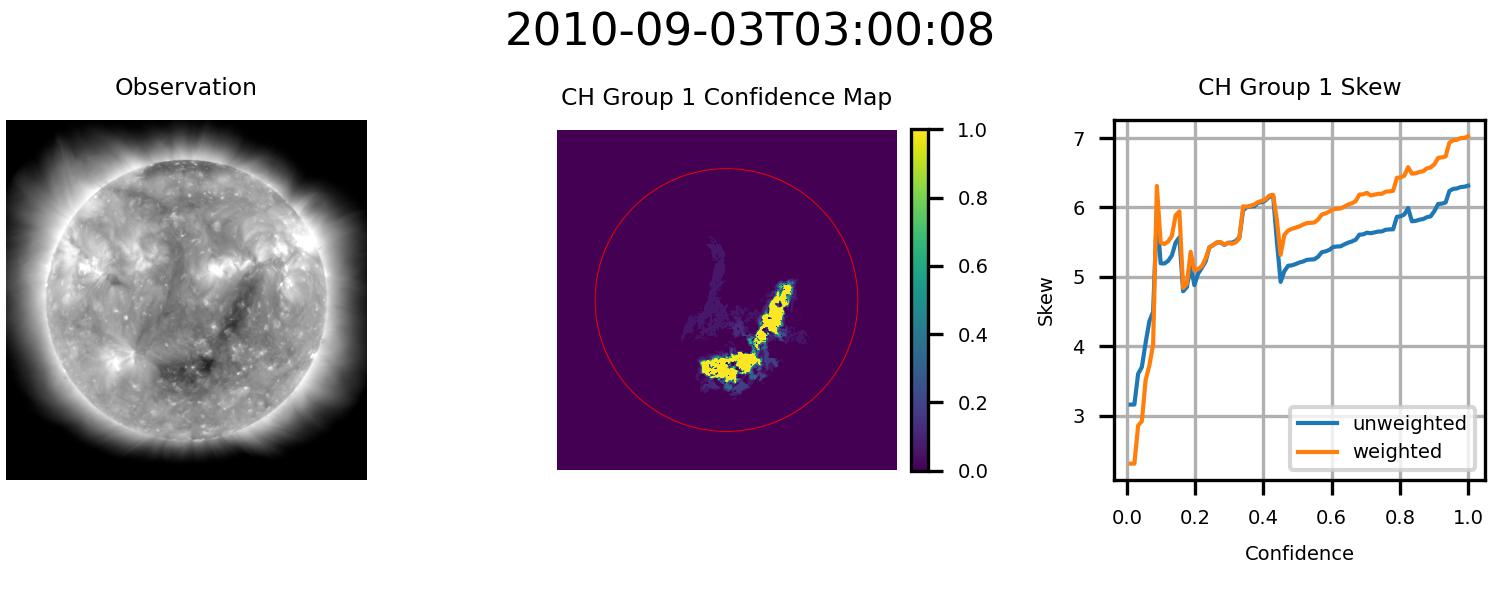}}\\
    \subfloat[CH observed in CR2101, in the graph a downward trend ($\downarrow$) indicates increasing unpopularity.]{\includegraphics[trim=0in 0.13in 0in 0in,clip,width=.74\textwidth]{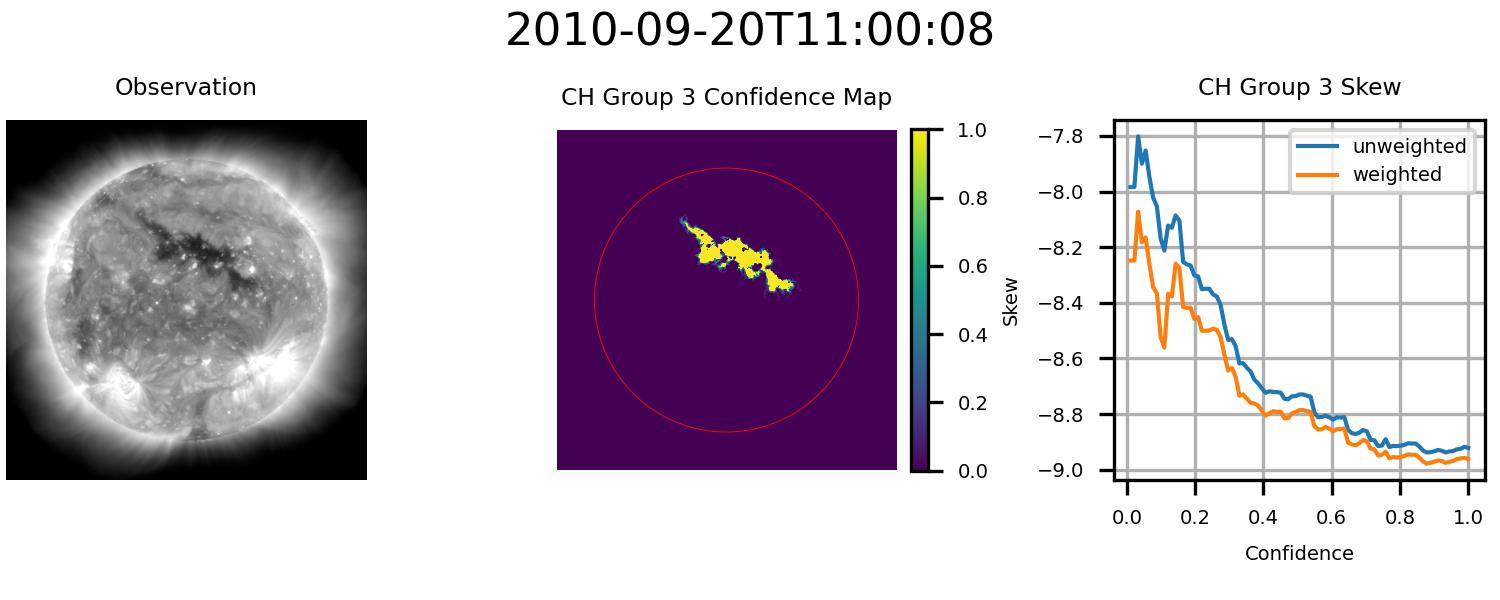}}\\
    \subfloat[CH observed in CR2133, in the graph a downward trend ($\downarrow$) indicates increasing unpopularity.]{\includegraphics[trim=0in 0.13in 0in 0in,clip,width=.74\textwidth]{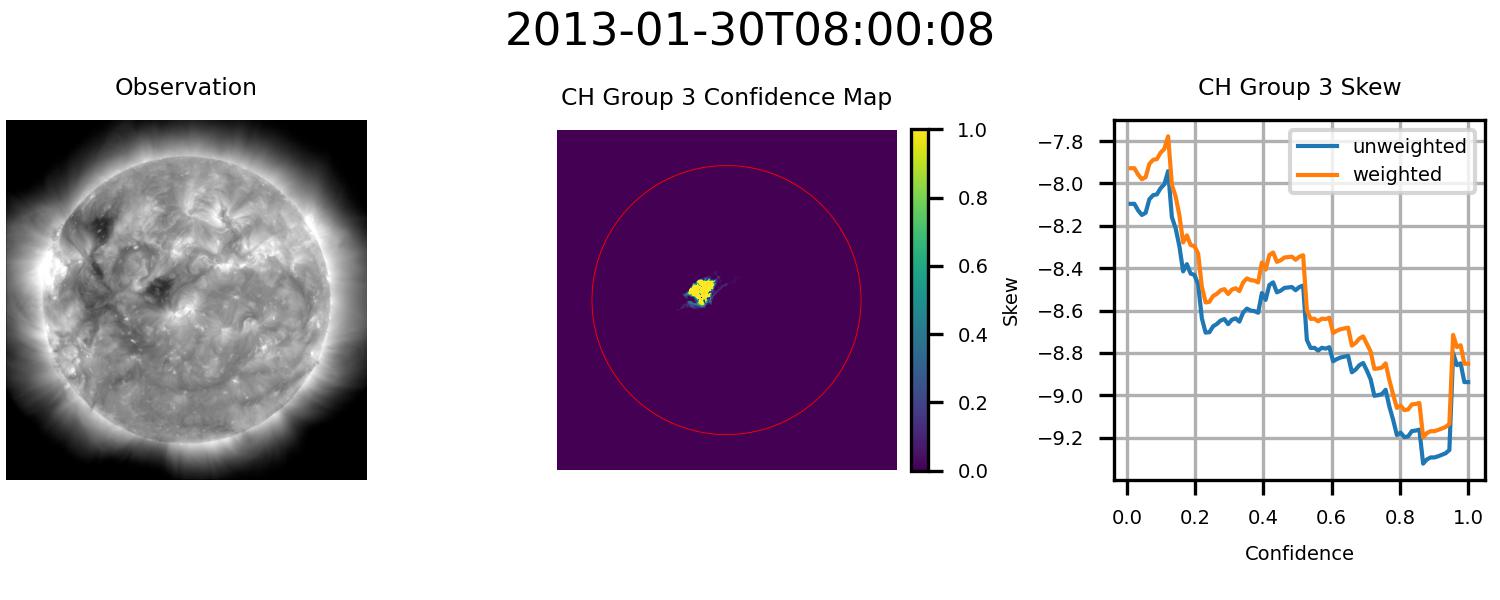}}\\
    \caption{Example CH groups where skew magnitude increased as a function of confidence, indicating that regions with high confidence are also more strongly unipolar.  From left to right for (a)-(d): AIA 193~{\AA} observation, confidence map (on disk area encircled in red), graph of skew as a function of confidence level.}
    \label{fig:grow}
\end{figure}

Three primary trends were noted within the data. The first trend, seen in Figure \ref{fig:grow}, consists of examples where the magnitude of the skew increased as a function of confidence. This trend was the most prevalent within the data, with nearly all cases where increasing confidence resulted in significantly decreased area showing this trend. This result suggests that higher confidence levels, as expressed by the ACWE confidence map, generally represent better delineations of CH regions in that the underlying magnetic field is more strongly unipolar.

\begin{figure}
    \centering
    \subfloat[CH observed in CR2099, in the graph a downward trend ($\downarrow$) indicates increasing unpopularity.]{\includegraphics[trim=0in 0.13in 0in 0in,clip,width=.8\textwidth]{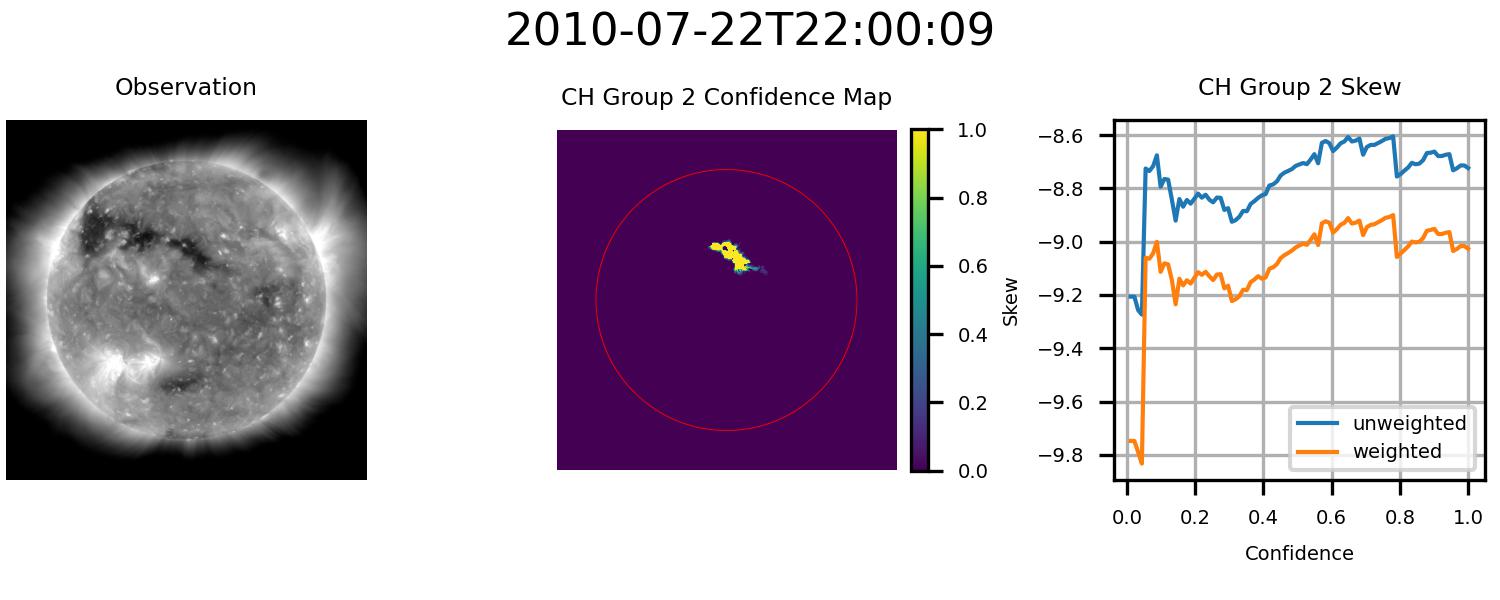}}\\
    \subfloat[CH observed in CR2101, in the graph a downward trend ($\downarrow$) indicates increasing unpopularity.]{\includegraphics[trim=0in 0.13in 0in 0in,clip,width=.8\textwidth]{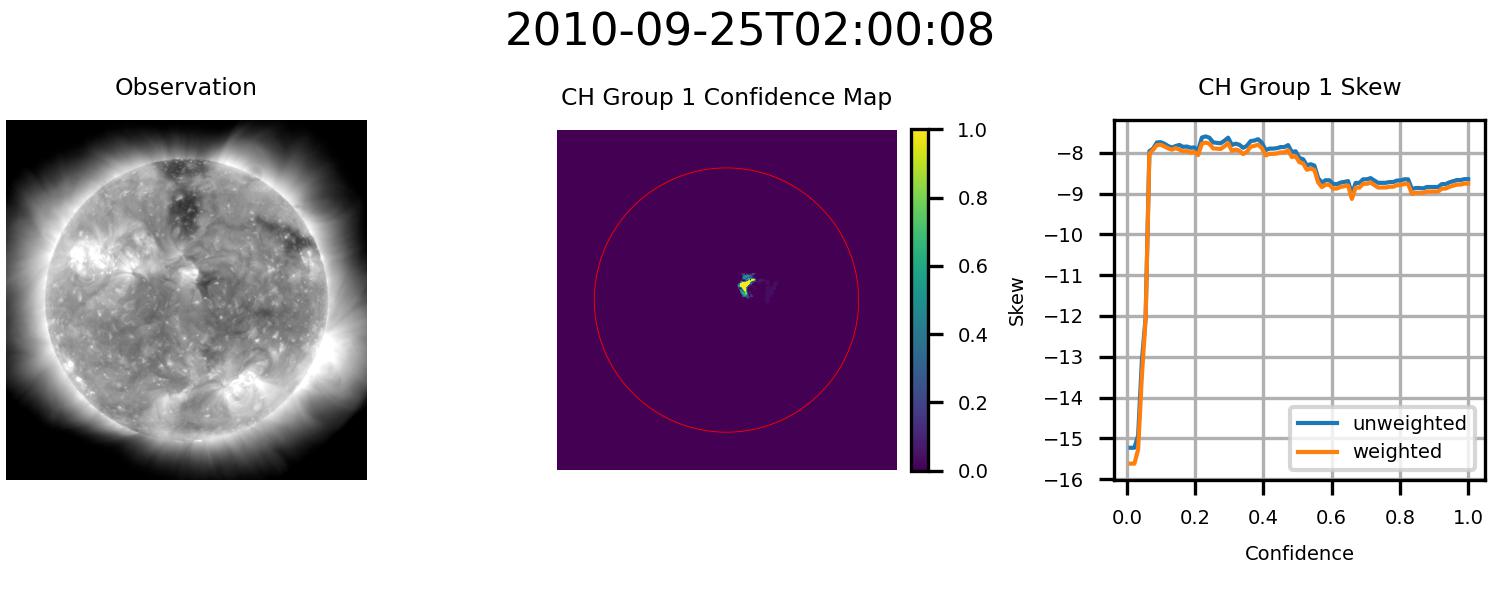}\label{fig:active}}\\
    \caption{Example CH groups where skew magnitude remained consistent for a majority of confidence levels.  From left to right for (a)-(b): AIA 193~{\AA} observation, confidence map (on disk area encircled in red), graph of skew as a function of confidence level.}
    \label{fig:Still}
\end{figure}

The second trend consisted of cases where confidence did not significantly affect skew magnitude for the majority of confidence levels. Figure \ref{fig:hourDiff} provides an example of this, with two additional cases presented in Figure \ref{fig:Still}. This indicates that a highly unipolar region is maintained throughout the suite of confidence levels, suggesting that these regions nonetheless represent accurate delineations of CH regions. 

\begin{figure}
    \centering
    \subfloat[CH observed in CR2100, in the graph a downward trend ($\downarrow$) indicates increasing unpopularity.]{\includegraphics[trim=0in 0.13in 0in 0in,clip,width=.8\textwidth]{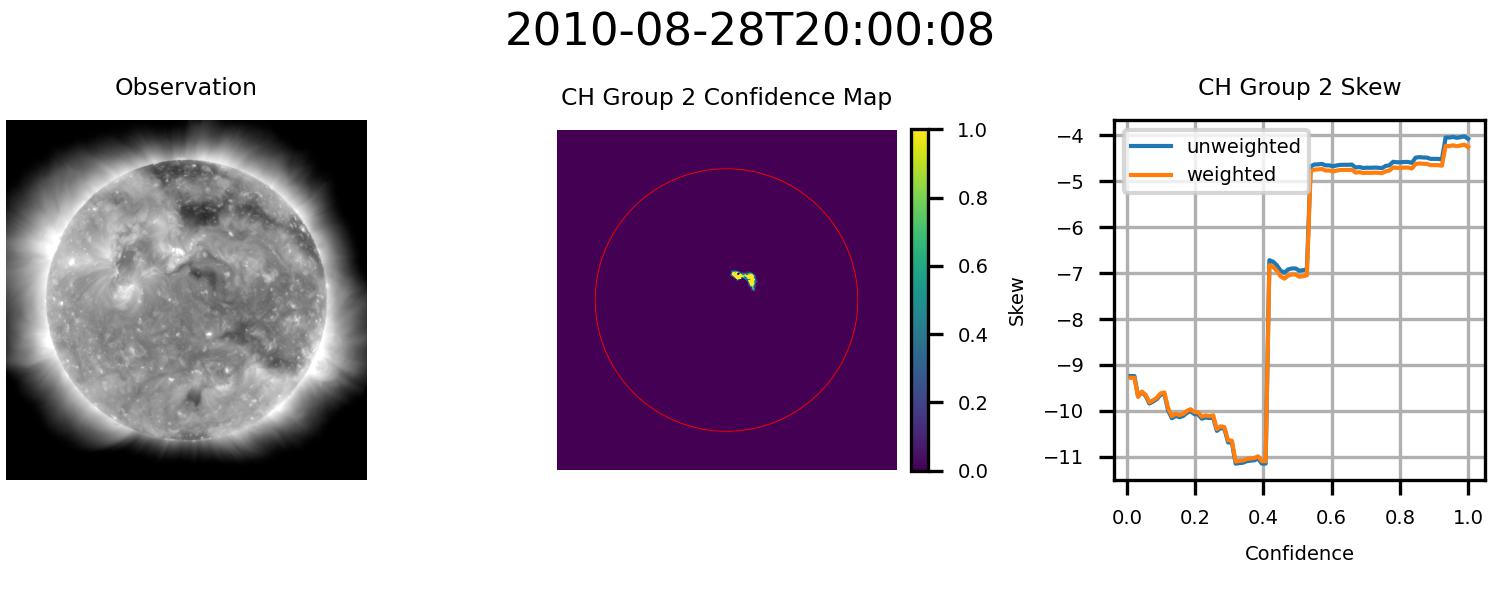}}\\
    \subfloat[CH observed in CR2101, in the graph an upward trend ($\uparrow$) indicates increasing unpopularity.]{\includegraphics[trim=0in 0.13in 0in 0in,clip,width=.8\textwidth]{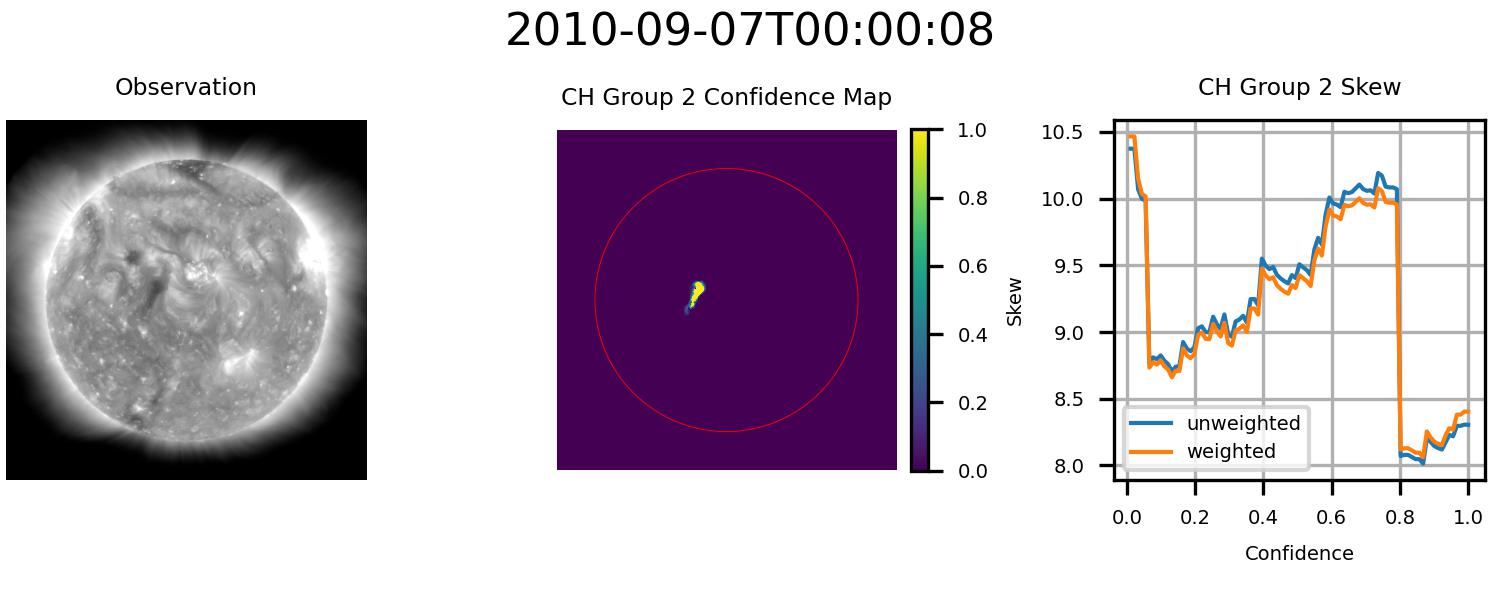}\label{fig:doubleDrop}}\\
    \caption{Example CH groups where skew magnitude dropped at higher confidence levels.  Skew magnitude remains high across all confidence levels, suggesting that these segmentations nonetheless represent good segmentations of the CH region.  From left to right for (a)-(b): AIA 193~{\AA} observation, confidence map (on disk area encircled in red), graph of skew as a function of confidence level.}
    \label{fig:drop}
\end{figure}

\begin{figure}
    \centering
    \includegraphics[trim=0in 0.13in 0in 0in,clip,width=.8\textwidth]{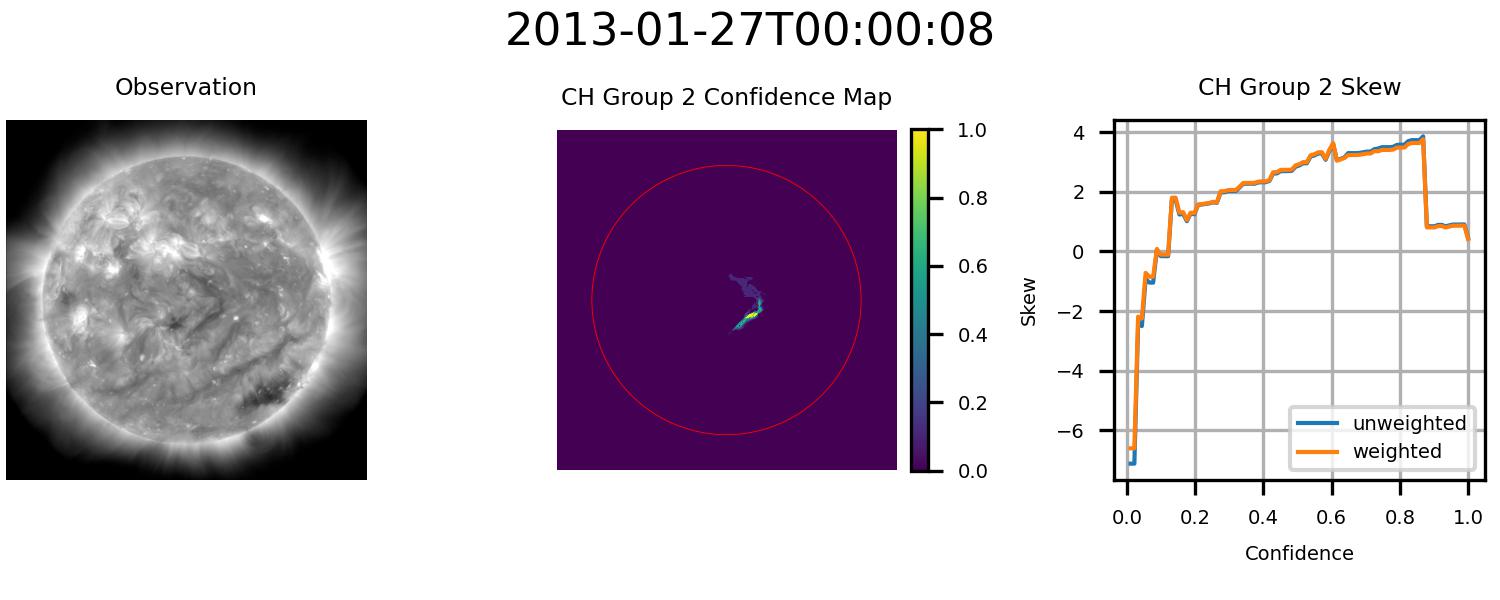}\\
    \caption{Filament captured by ACWE, based on the low absolute skewness values and the change from negative to positive skew as the confidence level is increased.  From left to right: AIA 193~{\AA} observation, confidence map (on disk area encircled in red), graph of skew as a function of confidence level. In the graph any trend away from 0 would indicate increasing unipolarity.}
    \label{fig:filament}
\end{figure}

\begin{figure}
    \centering
    \includegraphics[trim=0in 0.13in 0in 0in,clip,width=.8\textwidth]{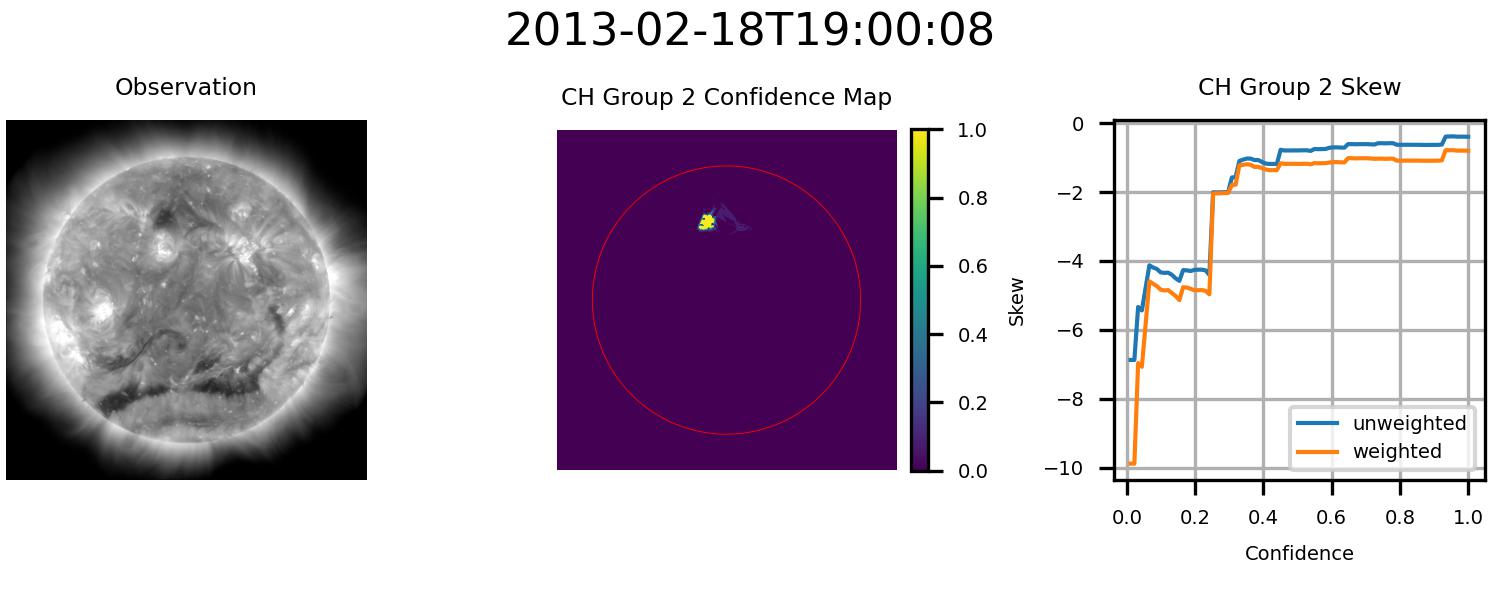}\\
    \caption{Example of CH with minimal skew at high confidence, possibly due to the nearby active region.  From left to right: AIA 193~{\AA} observation, confidence map (on disk area encircled in red), graph of skew as a function of confidence level. In the graph a downward trend ($\downarrow$) would indicate increasing unpopularity}
    \label{fig:oddDrop}
\end{figure}

The decrease in skew magnitude present in the early, low confidence regions in both cases in Figure \ref{fig:Still} is the third notable trend. This result is not isolated to lower confidence levels. Figure \ref{fig:drop} presents two examples of drops in skew magnitude at larger confidence levels. The cause of these drops in skew may vary from CH to CH, with the presence of nearby spurious bright regions providing potential explanations in the case of Figure \ref{fig:active} and the two examples in Figure \ref{fig:drop}. Drops in confidence may also be the result of overfitting, potentially explaining the second drop in skew magnitude seen in Figure \ref{fig:doubleDrop}. Despite this, skew magnitude remains high across all confidence levels, suggesting that these segmentations nonetheless represent accurate segmentations of the CH region.

Based on the skew data provided by HMI, it appears that in addition to the CHs identified by ACWE, at least two additional regions, both from CR2133, were also identified. The first region, seen in Figure \ref{fig:filament}, appears to be a filament based on the low absolute skewness values and the change from negative to positive skew as the confidence level is increased. This indicates that ACWE will misidentify sufficiently dark regions as CHs. This result may highlight the need to integrate data related to the magnetic field into ACWE to provide better segmentations.  The second group, shown in Figure \ref{fig:oddDrop}, consists of what appears to be a CH region with very low skew at high confidence. This result may the the result of the nearby active region, similar to the effects in Figure \ref{fig:active}.

\subsection{Formation of a CH}

\begin{figure}
    \centering
    \subfloat[First observation.]{\includegraphics[trim=0in 0.13in 0in 0in,clip,width=.57\textwidth]{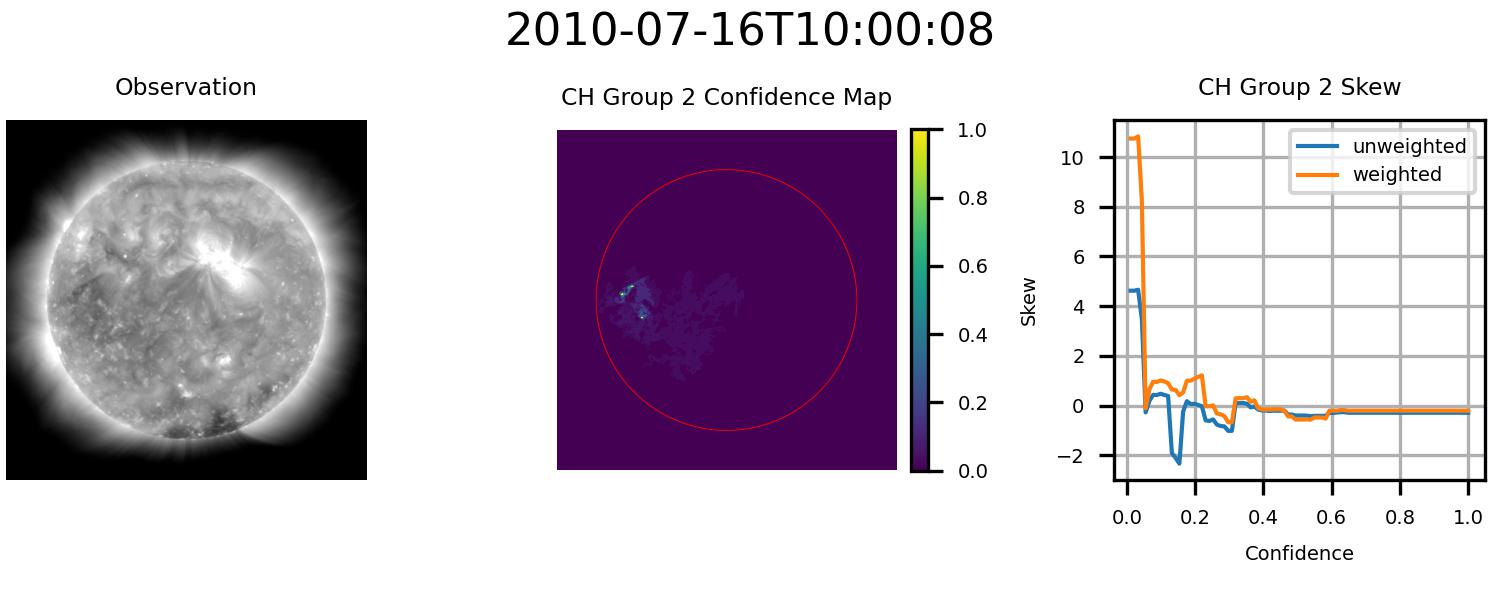}}\\
    \subfloat[Six hours into formation.]{\includegraphics[trim=0in 0.13in 0in 0in,clip,width=.57\textwidth]{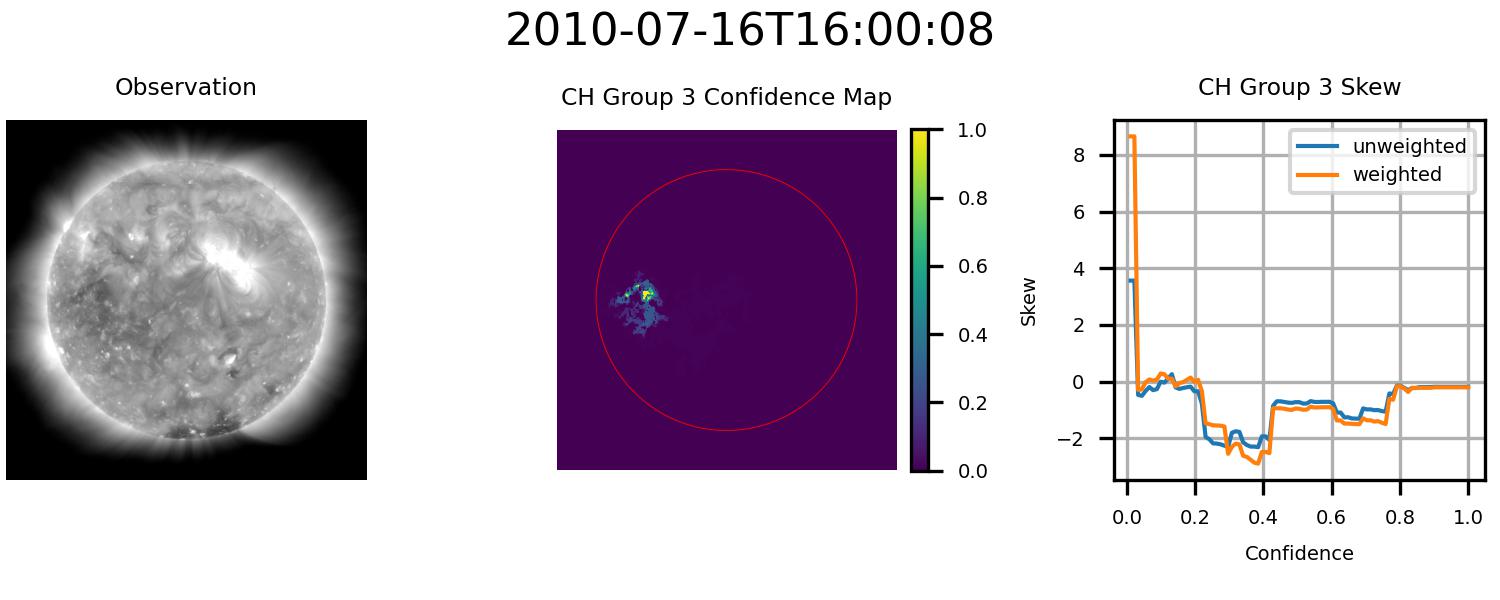}}\\
    \subfloat[Twelve hours into formation.]{\includegraphics[trim=0in 0.13in 0in 0in,clip,width=.57\textwidth]{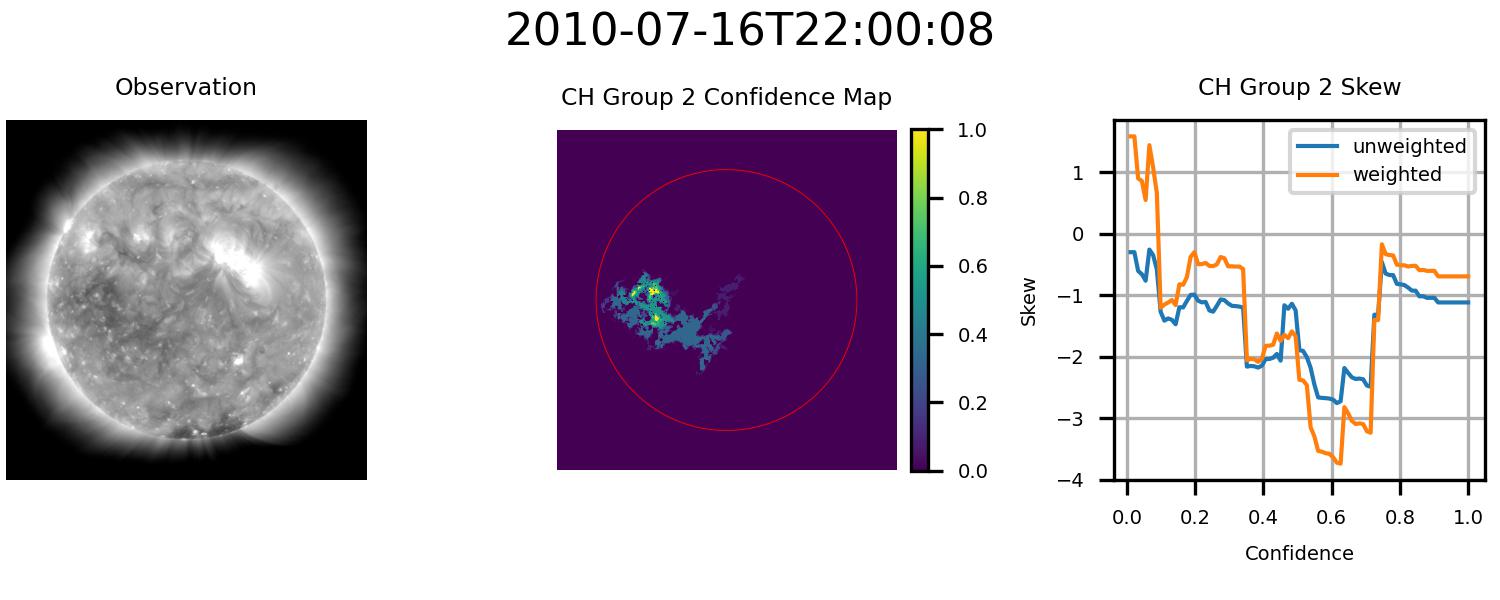}}\\
    \subfloat[Fifteen hours into formation.]{\includegraphics[trim=0in 0.13in 0in 0in,clip,width=.57\textwidth]{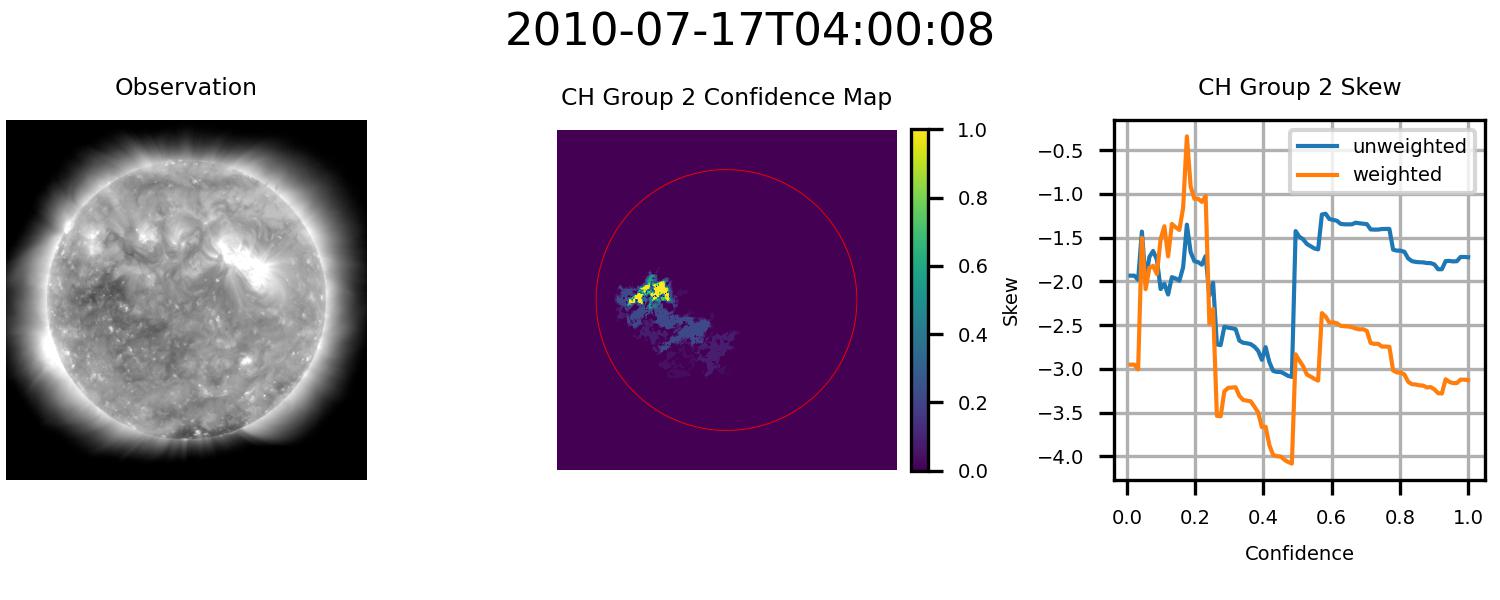}}\\
    \subfloat[Twenty-one hours into formation.]{\includegraphics[trim=0in 0.13in 0in 0in,clip,width=.57\textwidth]{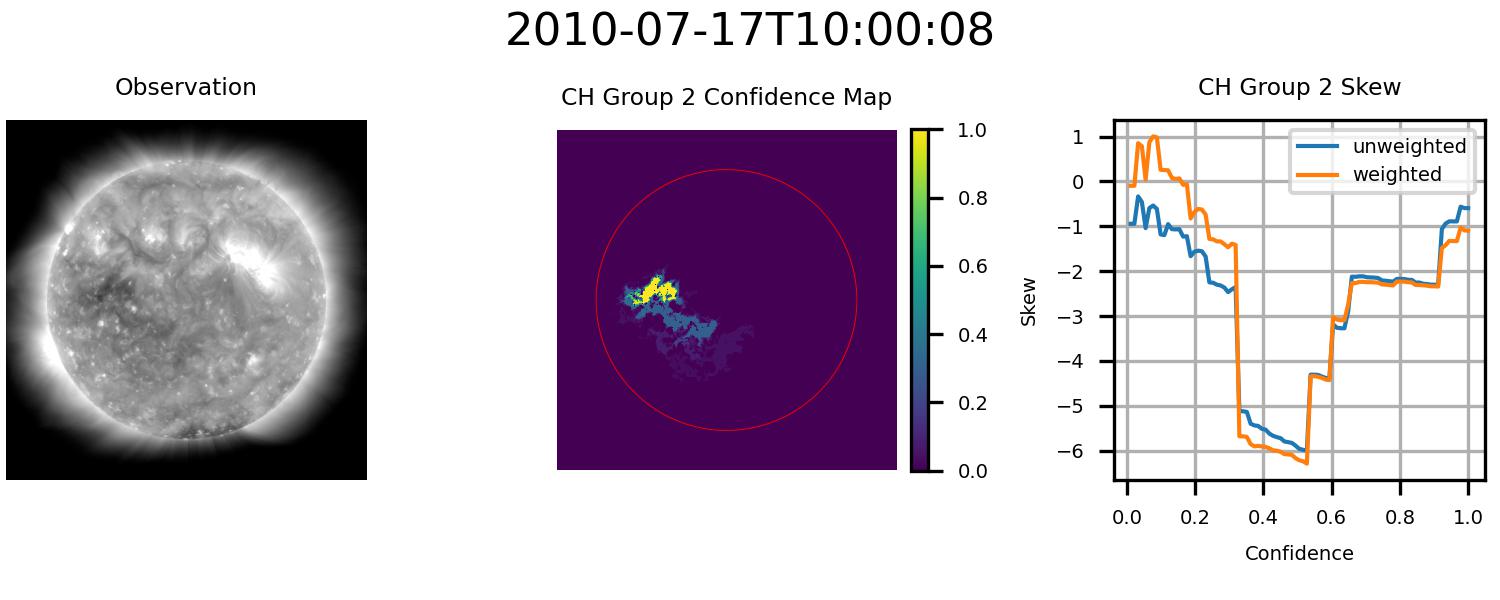}}\\
    \caption{Example of formation of a CH identified by ACWE.  From left to right for (a)-(e): AIA 193~{\AA} observation, confidence map (on disk area encircled in red), graph of skew as a function of confidence level.}
    \label{fig:form}
\end{figure}

As a final interesting result stemming from the validation of the ACWE confidence maps, ACWE was able to identify the formation of a CH in CR 2099. This region initially appeared as a low confidence region with large changes in segmentation area from one confidence level to the next and minimal skew on 2010-07-16 at 10:00:00 before forming into a CH as it entered disk center. As the region formed, the area of confidence increased, and difference in area from one confidence level to the next at higher confidence levels decreased until this observation matched other CHs within the dataset. Figure \ref{fig:form} shows the formation of this CH at six hour intervals. This is the same CH as seen in Figure \ref{fig:Formed}. The frequency with which ACWE can identify and catalog the formation of CH regions remains an open area of research.

\section{Conclusion and Future Work}
\label{sec:conclustion}

This work demonstrates the robustness and consistency of coronal hole segmentation via the Active Contours Without Edges (ACWE) algorithm and furthers this by introducing an ensemble method that leverages the ability of ACWE to define region boundaries on the basis of homogeneity. The ACWE algorithm was found to preserve general structures in the CH segmentations, even when operating on images decimated by $8\times$.  Additionally, the ACWE algorithm was found to be relatively insensitive to aggressive intensity scaling (e.g., from float64 to uint8) provided that the relative difference between CH and quiet Sun intensity is maintained.  These results have implications for cross-instrument application of the ACWE algorithm (or any other algorithm that depends on intensities, homogeneity of intensities, or dynamic range) as they indicate a strong effect of changing the dynamic range and distribution of intensities within that dynamic range. It would thus be expected that instruments with a reduced dynamic range of intensities may cause issues in obtaining accurate segmentations of CHs.  On the other hand, these results suggest that an accurate delineation of CH boundaries can be determined with significantly reduced spatial resolution, including the $512\times512$ pixel resolution employed by \cite{Jarolim2021}.

The ACWE algorithm was found to provide consistent segmentations across short timescales (i.e., shorter than the expected timsecales for CH evolution).  This highlights the importance of relying on image-derived features rather than hard thresholds to segment CHs.  The ACWE segmentation method relies on image features not only in initial seeding (using an estimate of the quiet Sun intensity), but also in defining the final CH regions as the initial seed is evolved to maximize the homogeneity of the CH regions with respect to the remaining on disk area.

In order to better leverage the region refinement capabilities of ACWE, an intraensemble method was developed that correlates likelihood that a region is part of a CH with the similarity of that region to the core region of that CH. It was found that the magnitude of magnetic skew increased with increase in confidence of the ensemble, suggesting that the higher confidence regions represent delineations of CH regions with a more strongly unipolar magnetic field. 

These confidence maps introduce two additional contributions to the field. First, by developing this intraensemble method, ACWE is able to directly catalog and report on ambiguity that exists along CH boundaries, which are often difficult to define not only due to effects of limb brightening and stray light from nearby bright regions, but also due to surrounding QS structures that may obscure this boundary \citep{Caplan2016}. Second, these results demonstrate that classical segmentation methods can be leveraged to generate confidence maps that provide strong correlations between reported likelihood that a region belongs to a CH and the actual probability that this is the case as measured by the underlying unipolarity of the region.

These results, however, also introduce additional considerations, constituting avenues for future work. 
First, it was found that some segmentations began targeting quiet Sun rather than CH regions, likely a result of relative homogeneity of the non-CH regions on disk. More robust methods to identify these change of target cases are needed, along with methods to identifying overfit wherein inner portions of CH regions are excluded for stringent homogeneity ratios.  Second, ACWE is vulnerable, as are many CH detection algorithms, to false detections caused by other dark regions such as filaments.  To mitigate this, ACWE will require additional data, such as from the magnetic field of the underlying regions.  In both of these avenues of future work, it is important that any developed methods be able to preserve any regions where ACWE identifies CH formation.

Additional improvements to this method may also come in the form of introducing data from other wavelengths captured by AIA or other instruments with an overlapping field of view. These data could be used to both aid in the seeding of the algorithm, and also to better define the CH region. Introduction of other data could be performed either independently, generating additional segmentations to be considered within the ensemble, or jointly through a vector-valued implementation of the ACWE algorithm \citep{chan2000130}. Finally, adaptation of ACWE to operate on additional observations, such as those produced by the Solar Terrestrial Relations Observatory (STEREO) \citep{kaiser2008stereo} can help to develop a more complete picture of CH activity by allowing for the development of synchronic maps to aid in space weather applications.

%
\appendix
\section{Intensity and Dynamic Range Effects}
\label{sec:intensity}

In addition to evaluating the effects of spatial resolution on segmentation, the effects of intensity on segmentation were also evaluated.  These evaluations were designed to study the sensitivity of ACWE segmentation of CHs with the hypothesis that intensity rescaling, e.g., a log scaling commonly used to visualize AIA images, may provide an advantage.  It was found, however, that direct representation of EUV intensities using a log scale hindered the segmentation process.  Since the use of an intensity rescaled image is not used for subsequent processing (e.g., to compute a confidence map) these results are not included in the main text.  Since these results could be of interest in application of ACWE for segmentation of CHs on other data with other intensity distributions (e.g., cross-instrument applications), the results are included here as an appendix.  

For the process of evaluating the effects of intensity scaling on segmentation, the Level 1.5 EUV images were remapped to a reduced intensity range of 256 intensity levels (the dynamic range of popular image formats such as PNG and JPEG). The data from CRs 2099 through 2101 originally contained intensities in the range $[-128,16383]$, while the data from CR 2133 contained intensities in the range $[-16,16383]$. Eight remapping schemes were tested:
\begin{itemize}
    \item \textbf{Linear Full:} The EUV image is linearly rescaled so that the maximum intensity is 255 and the minimum intensity is 0. Each intensity is then then rounded to the nearest integer value producing an image with 256 discrete intensity levels. The minimum and maximum intensity present within the original EUV image are separately recorded for use in restoring the native intensity range.
    \item \textbf{Linear 0 to Max:} The image is clipped so that all negative intensity values are replaced with a value of 0. This clipped image is then rescaled. As before, the intensity of each pixel is rounded to the nearest integer, and the minimum (0) and maximum intensity (image dependent) are separately recorded.
    \item \textbf{Linear Solar Limits:} The image is clipped to the range of on-disk intensities prior to remapping the image.
    \item \textbf{Linear 20 to 2500:} The image is clipped to the range of 20 to 2500 prior to remapping the image.
    \item \textbf{Log10 Full:} An offset is applied to the intensities of all pixels to ensure that the minimum intensity is remapped to 1 and the range of intensities are preserved. The resulting image is remapped by taking the $\log_{10}$ of each pixel, consistent with common practice in visualizing AIA images. This remapped image is then rescaled. The original minimum, original maximum, and the offset are separately recorded for use in restoring the native intensity range.
    \item \textbf{Log10 Compress 0 to Max:} The image is clipped so that all negative intensity values are replaced with a value of 0. An offset of 1 is then applied to all pixels within the image. The resulting image is remapped by taking the $\log_{10}$ of each pixel. This reduced and remapped image is then rescaled. The minimum (0), offset (1), and the original maximum are separately recorded for use in restoring the native intensity range.
    \item \textbf{Log10 Solar Limits:} The image is clipped to the range of on-disk intensities and, if needed, offset to ensure the minimum intensity is $>0$. The resulting image is remapped by taking the $\log_{10}$ of each pixel. This reduced and remapped image is then rescaled.
    \item \textbf{Log10 20 to 2500:} The image is clipped to the range of 20 to 2500, the resulting image is remapped by taking the $\log_{10}$ of each pixel. This reduced and remapped image is then rescaled.
\end{itemize}

For each of the intensity compression schemes, ACWE is performed both on the compressed image and on a “restored image.” For linear remapping schemes, the restoration process is achieved by assigning to all pixels of intensity $b$ the intensity value
\begin{equation}
    v = \frac{b}{255}\times(\max(I)-\min(I)) + \min(I),
    \label{eq:v}
\end{equation}
where $\max(I)$ is the maximum intensity of the original EUV image and $\min(I)$ is the minimum intensity of the original image. For $\log_{10}$ remapping schemes, the restoration process is completed in two steps. First the intensity range of the compressed image in the log domain is restored by assigning each pixel the intensity value
\begin{equation}
  \begin{aligned}
    v_{l10} = &\frac{b}{255}\times(\log_{10}(\max(I)+O)-\log_{10}(\min(I)+O))\\
    &+\log_{10}(\min(I)+O),
  \end{aligned}
  \label{eq:vl10}
\end{equation}
where the offset $O$ is defined as
\begin{equation}
    O = \begin{cases}
        0, & \text{for } \min(I) > 0\\
        1-\min(I), & \text{else}.
        \end{cases}
\end{equation}
Next the original intensity value of each pixel is approximated using
\begin{equation}
    v = 10^{v_{l10}} - O.
\end{equation}
In all cases (both restored and unrestored), decimation to $512\times512$ pixels (one-eighth spatial resolution), correction for limb brightening, and seeding are performed, following the method outlined in \cite{boucheron2016segmentation}, using the altered EUV image as if it was the original. ACWE is then performed, also using the altered EUV image. These results are compared against the one-eighth scale segmentations generated in Section~\ref{sec:spatial}.

\begin{table}[t]
\centering
    \caption{Number of cases where an empty segmentation was returned.  The remapping scheme is noted along with whether the image was segmented using the $[0,255]$ range or the restored range (Rest.).}
    \label{tab:intensityEmpty}
    \begin{tabular}{cccccc}
    \hline
    \multicolumn{2}{c}{\textbf{Remapping Scheme}} & \textbf{CR 2099}  & \textbf{CR 2100}  & \textbf{CR 2101}  & \textbf{CR 2133} \\
    \hline
    \multirow{2}{*}{Linear Full} & [0,255] & 0 & 1 (0.2\%) & 0 & 2 (0.3\%)\\
    & Rest. & 1 (0.2\%) & 0 & 0 & 1 (0.2\%)\\\hline
    \multirow{2}{*}{Linear 0 to Max} & [0,255] & 0 & 0 & 0 & 0\\
    & Rest. & 0 & 0 & 0 & 0\\\hline
    \multirow{2}{*}{Linear Solar Limits} & [0,255] & 0 & 1 (0.2\%) & 0 & 0\\
    & Rest. & 0 & 0 & 0 & 0\\\hline
    \multirow{2}{*}{Linear 20 to 2500} & [0,255] & 0 & 0 & 0 & 0\\
    & Rest. & 0 & 0 & 0 & 1 (0.16\%)\\\hline
    \multirow{2}{*}{Log10} & [0,255] & 562 (100\%) & 619 (100\%) & 582 (100\%) & 618 (100\%)\\
    & Rest. & 0 & 0 & 0 & 0\\\hline
    \multirow{2}{*}{Log10 0 to Max} & [0,255] & 562 (100\%) & 619 (100\%) & 582 (100\%) & 618 (100\%)\\
    & Rest. & 0 & 0 & 0 & 0\\\hline
    \multirow{2}{*}{Log10 Solar Limits} & [0,255] & 4 (0.7\%) & 4 (0.6\%) & 3 (0.5\%) & 6 (1.0\%)\\
    & Rest. & 0 & 0 & 0 & 0\\\hline
    \multirow{2}{*}{Log10 20 to 2500} & [0,255] & 202 (35.9\%) & 309 (49.9\%) & 419 (72.0\%) & 247 (40.0\%)\\
    & Rest.  & 0 & 0 & 0 & 1 (0.2\%)\\\hline
    \end{tabular}
\end{table}

Table \ref{tab:intensityEmpty} summarizes the number of times where ACWE returned a segmentation wherein no regions were identified as belonging to CHs. These results indicate that ACWE, when applied to CH detection using the method outlined in \cite{boucheron2016segmentation}, has difficulty with data that has been remapped via the the $\log_{10}$ operator unless attempts are made to restore the original dynamic range of the data. This result may be, in part, due to the fact that remapping the intensities via a $\log$ transform minimizes the dynamic range of higher intensities in order to maximize the dynamic range of lower intensities. By maximizing the dynamic range of lower intensities, this remapping scheme is artificially decreasing the homogeneity of the CH region while simultaneously increasing the homogeneity of the non-CH region.

\begin{figure}
    \centering
    \subfloat[Intersection Over Union]{\includegraphics[width=.72\textwidth]{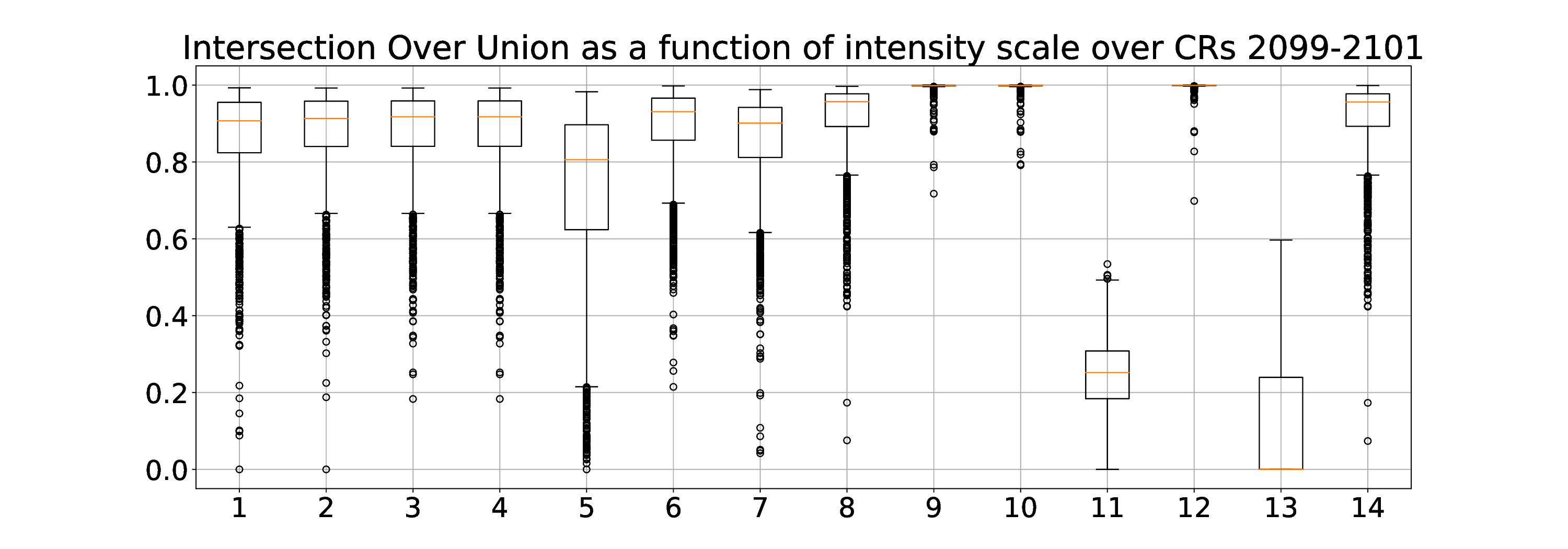}}\\
    \subfloat[Structural Similarity]{\includegraphics[width=.72\textwidth]{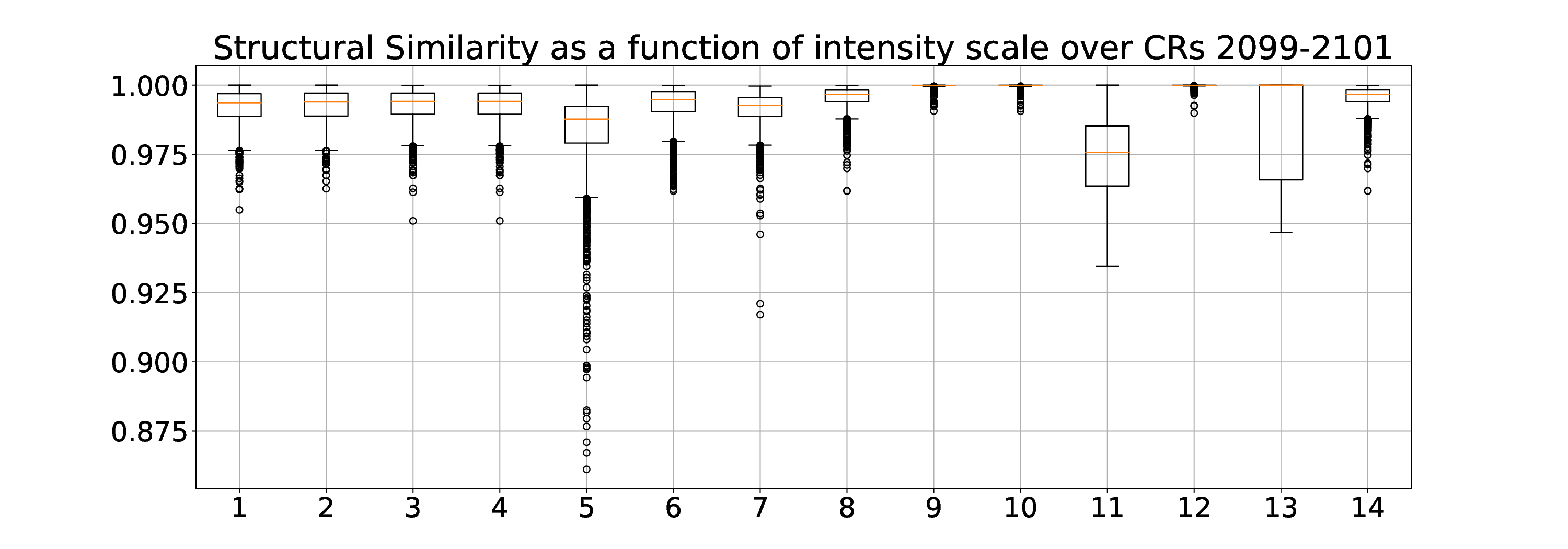}}\\
    \subfloat[Global Consistency Error]{\includegraphics[width=.72\textwidth]{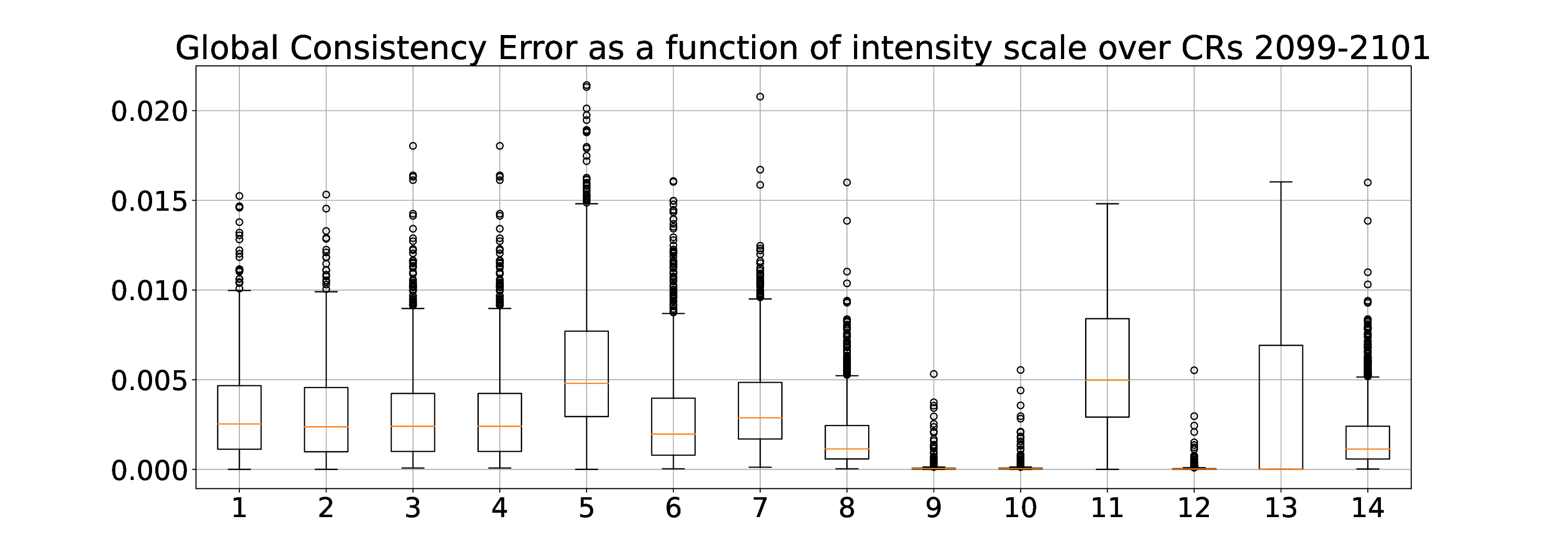}}\\
    \subfloat[Local Consistency Error]{\includegraphics[width=.72\textwidth]{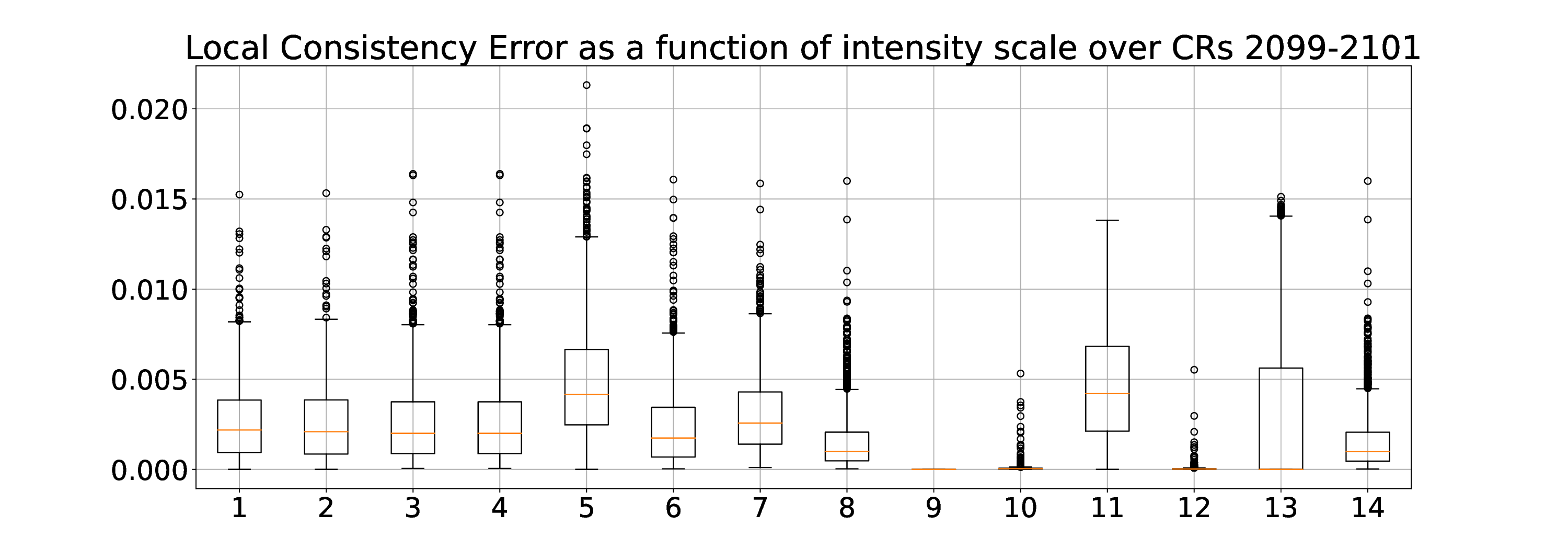}}\\
    \caption{Similarity of segmentation on intensity rescaled image to segmentation on original, unscaled data for CRs 2099, 2100, and 2101. The compression schemes presented are 1 \& 2: Linear Full (using the $[0,255]$ range and restored range, respectively), 3 \& 4: Linear 0 to Max, 5 \& 6: Linear Solar Limits, 7 \& 8: Linear 20 to 2500, 9: Log10 Linear Restored (the $[0,255]$ range is omitted as no valid maps were returned), 10: Log10 0 to Max Restored, 11 \& 12: Log10 Solar Limits (using the $[0,255]$ range and restored range, respectively), and 13 \& 14: Log10 20 to 2500. The orange line within each plot is the median value, the box represents the range between the first (Q1) and third (Q3) quartile, the whiskers represent 1.5 times the interquartile range above Q3 or below Q1, and circles represent outliers.}
    \label{fig:Intensity_CR2099_2101}
\end{figure}

\begin{figure}
    \centering
    \subfloat[Intersection Over Union]{\includegraphics[width=.72\textwidth]{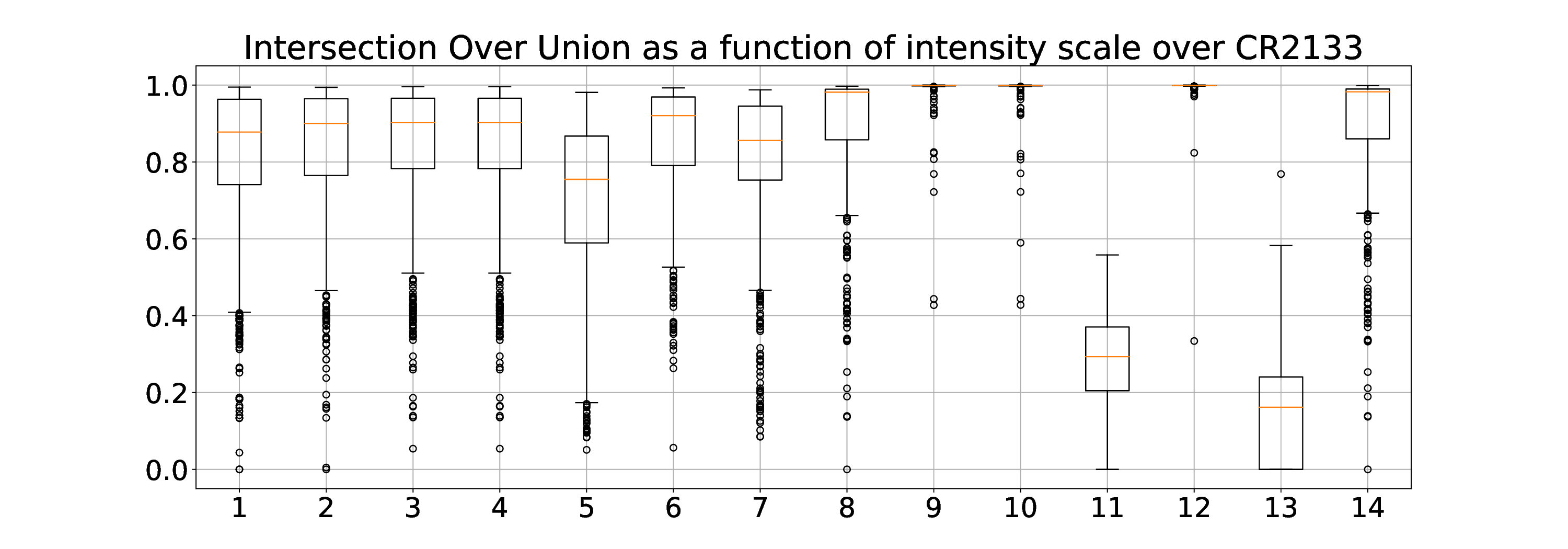}}\\
    \subfloat[Structural Similarity]{\includegraphics[width=.72\textwidth]{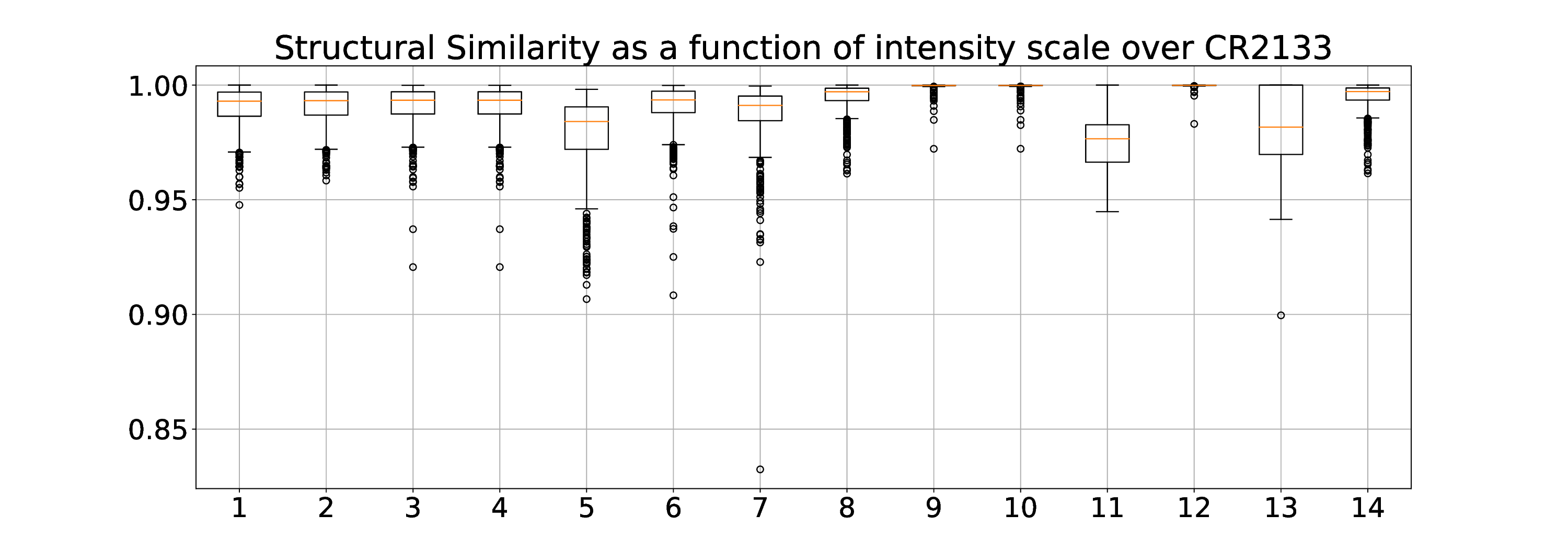}}\\
    \subfloat[Global Consistency Error]{\includegraphics[width=.72\textwidth]{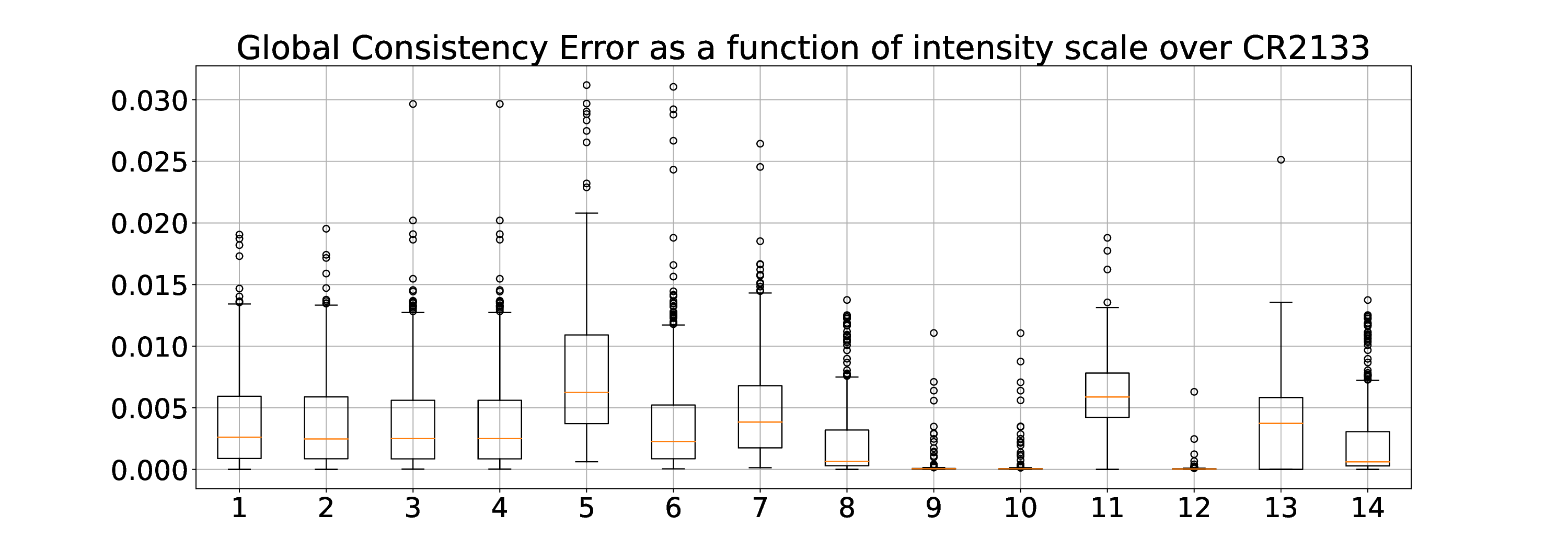}}\\
    \subfloat[Local Consistency Error]{\includegraphics[width=.72\textwidth]{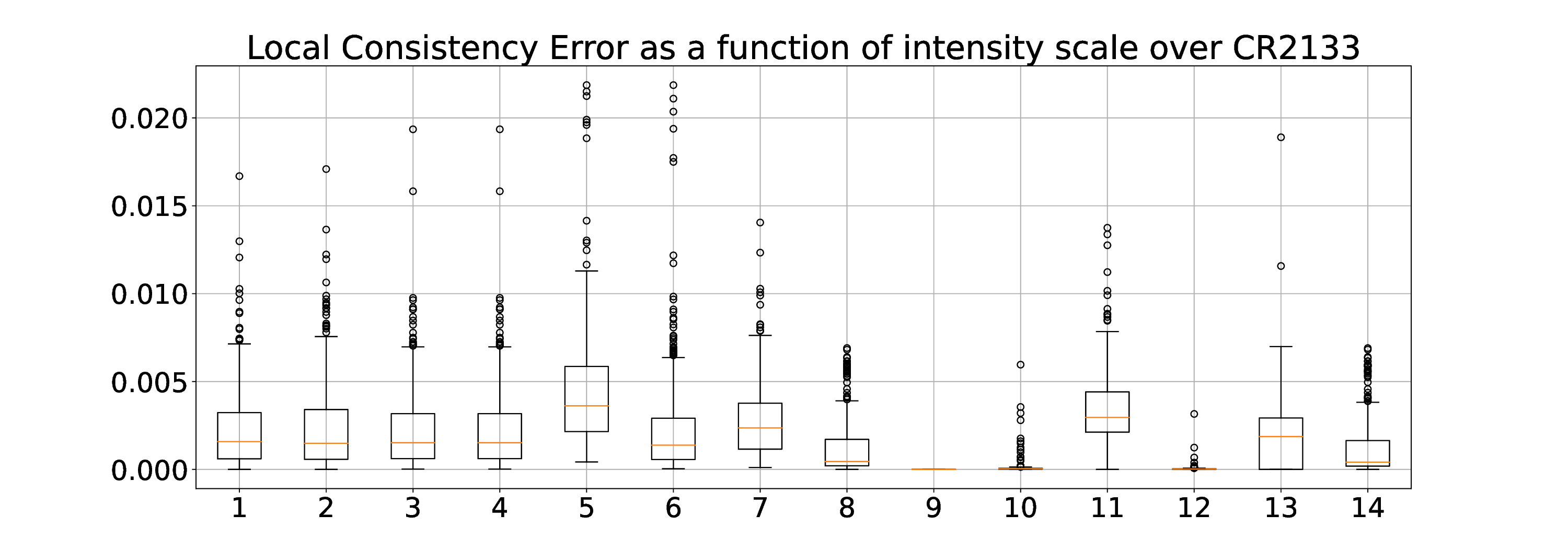}}\\
    \caption{Similarity of segmentation on intensity rescaled image to segmentation on original, unscaled data for CR 2133. The compression schemes presented are 1 \& 2: Linear Full (using the $[0,255]$ range and restored range, respectively), 3 \& 4: Linear 0 to Max, 5 \& 6: Linear Solar Limits, 7 \& 8: Linear 20 to 2500, 9: Log10 Linear Restored (the $[0,255]$ range is omitted as no valid maps were returned), 10: Log10 0 to Max Restored, 11 \& 12: Log10 Solar Limits (using the $[0,255]$ range and restored range, respectively), and 13 \& 14: Log10 20 to 2500. The orange line within each plot is the median value, the box represents the range between the first (Q1) and third (Q3) quartile, the whiskers represent 1.5 times the interquartile range above Q3 or below Q1, and circles represent outliers.}
    \label{fig:Intensity_CR2133}
\end{figure} 

\begin{figure}
    \centering
    \includegraphics[trim=0in 0.5in 0in 0in,clip,width=\textwidth]{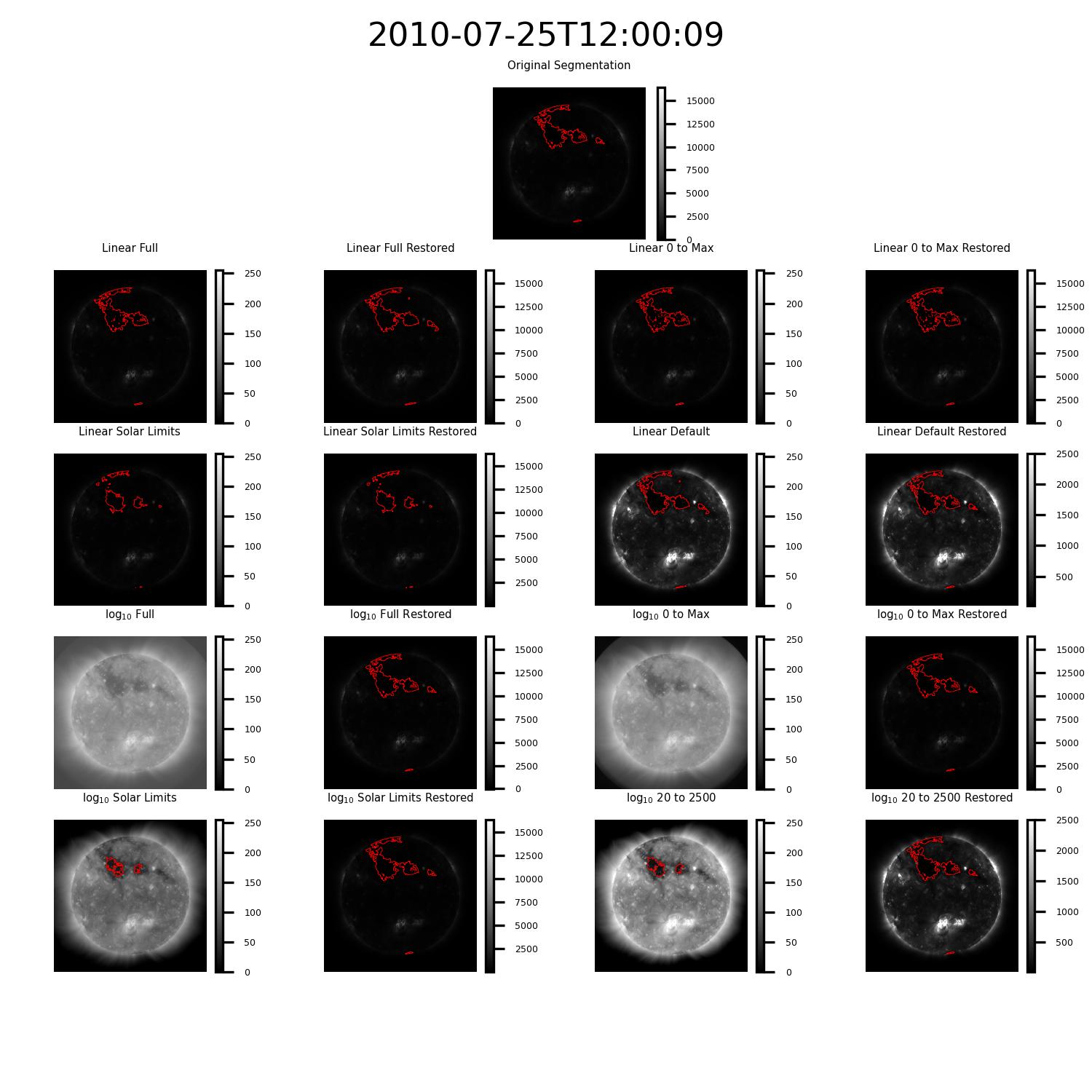}
    \caption{Comparison of segmentations of the same observation performed using the various intensity remapping schemes overlaid onto the remapped image. It should be noted that the topmost image is the original EUV image, presented here without any remapping of intensities, and the segmentation is the same segmentation seen in Figure \ref{fig:ScaleSample} at one-eighth scale resolution. Each image is accompanied by a color bar showing the range of intensities in the final image.}
    \label{fig:IntensitySample}
\end{figure}

Figure \ref{fig:Intensity_CR2099_2101} shows the IOU, SSIM, GCE and LCE between the segmentations generated from the intensity remapped images and segmentations generated from the original EUV image in CRs 2099, 2100, and 2101. Figure \ref{fig:Intensity_CR2133} shows the results for CR 2133. It should be noted the restoration process had little effect on the linear remapping schemes which preserved the majority of the dynamic range (such as the full and 0 to max mapping schemes), and minimal effect on the remaining linear schemes. By contrast, restoring the $\log_{10}$ images significantly improved segmentation similarity across all four metrics. Based on these metrics, the restoration processed resulted in nearly identical segmentations for the restored $\log_{10}$ schemes that used the full on-disk range, 0 to max intensity range, and solar intensity range. This result is consistent with observations of the resulting segmentation, a sample of which is provided in Figure \ref{fig:IntensitySample}. As noted above, the process of remapping the intensities via a $\log$ transform minimizes the dynamic range of higher intensities in order to maximize the dynamic range of lower intensities. This means that a larger portion of the intensities in the compressed image (variable $b$ in Equations (\ref{eq:v}) and (\ref{eq:vl10})) were used to represent the lower intensity levels, allowing for a finer representation of the structure in CH regions. The low GCE and LCE for the un-restored $\log_{10}$ images (i.e., those images in the range $[0,255]$) represent a known problem with these metrics. \cite{martin2001} note that both GCE and LCE fail to characterize the quality of a segmentation when one segmentation is empty as both metrics treat any segmentation as a refinement of an empty segmentation. For that reason, the presence of a large number of empty segmentations in the un-restored $\log_{10}$ examples reduced the reported error of the ensemble. These results have implications for cross-instrument application of the ACWE algorithm (or any other algorithm that depends on intensities, homogeneity of intensities, or dynamic range) as they indicate a strong effect of changing the dynamic range and distribution of intensities within that dynamic range. It would thus be expected that instruments with a reduced dynamic range of intensities may cause issues in obtaining accurate segmentations of CHs.

%
\section*{Acknowledgements}
The authors gratefully acknowledge the support of NASA grant 80NSSC20K0517 and NSF grants 1945705 and 1649052 which helped support this work.



\section*{Data and Code Availability} 
The data used in this work are publicly available.  The GitHub repository at \url{https://github.com/DuckDuckPig/CH-ACWE} contains code to download the dataset used in this paper, the base ACWE segmentation code, and code to replicate all experiments described herein.




%
%
\bibliographystyle{plainnat}
\bibliography{References}  
%
%
%
%
\end{document}